\documentclass[useAMS,usenatbib,usegraphicx]{mn2e}
\usepackage{graphicx}
\usepackage[active]{srcltx}
\usepackage{times}
\def\kms{$\mbox{km s}^{-1}$}
\newcommand{\sauron}{{\texttt {SAURON}}}

\newcounter{subfigure}

\newcommand{\hb}{H$\beta$}
\newcommand{\mgb}{Mg\,$b$} 
\newcommand{\fes}{Fe5270$_{\rmn{S}}$}

\newcommand{\mgfefzp}{[MgFe50]$^\prime$}
\newcommand{\mgfeftp}{[MgFe52]$^\prime$}
\newcommand{\re}{$(R_{\rmn{e}}/8)$}
\newcommand{\reb}{$R_{\rmn{e}}/8$}
\newcommand{\mvir}{$M_\rmn{dyn}$}

\newcommand{\lesssim}{\mathrel{\hbox{\rlap{\hbox{\lower4pt\hbox{$\sim$}}}\hbox{$<$}}}}
\newcommand{\gtrsim}{\mathrel{\hbox{\rlap{\hbox{\lower4pt\hbox{$\sim$}}}\hbox{$>$}}}}

\title[The \sauron\ project - XVII] {The SAURON project - XVII.  Stellar
  population analysis of the absorption line strength maps of 48
  early-type galaxies}

\author[Kuntschner et al.] {Harald Kuntschner,$^{1}$\thanks{Email: hkuntsch@eso.org} Eric Emsellem,$^{2,3}$ Roland Bacon,$^{3}$ Michele Cappellari,$^{4}$ \newauthor 
Roger L.\ Davies,$^{4}$ P.~Tim de Zeeuw,$^{2,12}$ Jes\'us Falc\'on-Barroso,$^{5}$ Davor Krajnovi\'c,$^{2}$ \newauthor
Richard M.\ McDermid,$^{6}$ Reynier F.\ Peletier,$^{7}$ Marc Sarzi,$^{8}$ Kristen L. Shapiro,$^{9}$ \newauthor 
Remco C. E. van den Bosch,$^{10}$ Glenn van de Ven$^{11}$ \\
$^{1}$Space Telescope European Coordinating Facility, European Southern Observatory, Karl-Schwarzschild-Str. 2, 85748 Garching, Germany\\
$^{2}$European Southern Observatory, Karl-Schwarzschild-Str~2, 85748
Garching, Germany\\
$^{3}$Centre de Recherche Astronomique de Lyon, 9~Avenue Charles 
Andr\'e, 69230 Saint Genis Laval, France\\
$^{4}$Sub-department of Astrophysics, University of Oxford, Denys Wilkinson Building, Keble Road, Oxford OX1 3RH, United Kingdom \\
$^{5}$Instituto de Astrofsica de Canarias, Canarias, Via Lactea s/n, 38700 
La Laguna, Tenerife, Spain\\ 
$^{6}$Gemini Observatory, Northern Opertations Center, 670 N. Aohoku 
Place, Hilo, HI 96720, USA\\
$^{7}$Kapteyn Astronomical Institute, Rijksuniversiteit Groningen, Postbus 800, 9700 AV 
Groningen, The Netherlands\\
$^{8}$Centre for Astrophysics Research, University of Hertfordshire, 
College Lane, Hatfield, Herts, AL10 9AB, UK\\
$^{9}$UC Berkeley Department of Astronomy, Berkeley, CA 94720, USA\\
$^{10}$Department of Astronomy, University of Texas, Austin, TX 78712, 
USA\\
$^{11}$Max Planck Institute for Astronomy, D-69117 Heidelberg, Germany\\ 
$^{12}$Sterrewacht Leiden, Universiteit Leiden, Postbus 9513, 2300 RA Leiden, The Netherlands\\}

\begin{document}

\maketitle

\clearpage

%
%
\begin{abstract}

  We present a stellar population analysis of the absorption line
  strength maps for $48$ early-type galaxies from the \sauron\/
  sample. Using the line strength index maps of \hb, Fe5015, and \mgb,
  measured in the Lick/IDS system and spatially binned to a constant
  signal-to-noise, together with predictions from up-to-date stellar
  population models, we estimate the simple stellar
  population-equivalent (SSP-equivalent) age, metallicity and
  abundance ratio [$\alpha$/Fe] over a two-dimensional field extending
  up to approximately one effective radius.  A discussion of
  calibrations and differences between model predictions is given.
  Maps of SSP-equivalent age, metallicity and abundance ratio
  [$\alpha$/Fe] are presented for each galaxy.  We find a large range
  of SSP-equivalent ages in our sample, of which $\sim$40 per cent of
  the galaxies show signs of a contribution from a young stellar
  population.  The most extreme cases of post-starburst galaxies, with
  SSP-equivalent ages of $\le 3$\,Gyr observed over the full
  field-of-view, and sometimes even showing signs of residual
  star-formation, are restricted to low mass systems ($\sigma_{\rm e}
  \le 100$\,\kms \/ or $\sim$2 $\times 10^{10} M_{\odot}$).  Spatially
  restricted cases of young stellar populations in circumnuclear
  regions can almost exclusively be linked to the presence of
  star-formation in a thin, dusty disk/ring, also seen in the near-UV
  or mid-IR on top of an older underlying stellar population.

  The flattened components with disk-like kinematics previously
  identified in all fast rotators (\citeauthor{kraj08}) are shown to
  be connected to regions of distinct stellar populations. These range
  from the young, still star-forming circumnuclear disks and rings
  with increased metallicity preferentially found in intermediate-mass
  fast rotators, to apparently old structures with extended disk-like
  kinematics, which are observed to have an increased metallicity and
  mildly depressed [$\alpha$/Fe] ratio compared to the main body of
  the galaxy. The slow rotators, often harboring kinematically
  decoupled components in their central regions, generally show no
  stellar population signatures over and above the well known
  metallicity gradients in early-type galaxies and are largely
  consistent with old ($\ge$10\,Gyr) stellar populations.

  Using radially averaged stellar population gradients we find in
  agreement with Spolaor et al. a mass-metallicity gradient relation
  where low mass fast rotators form a sequence of increasing
  metallicity gradient with increasing mass. For more massive systems
  (above $\sim$3.5 $\times 10^{10} M_{\odot}$) there is an overall
  downturn such that metallicity gradients become shallower with
  increased scatter at a given mass leading to the most massive
  systems being slow-rotators with relatively shallow metallicity
  gradients. The observed shallower metallicity gradients and
  increased scatter could be a consequence of the competition between
  different star-formation and assembly scenarios following a general
  trend of diminishing gas fractions and more equal mass mergers with
  increasing mass, leading to the most massive systems being devoid of
  ordered motion and signs of recent star-formation.
 
\end{abstract}


%
%
\section{INTRODUCTION}
\label{sec:intro}

In hierarchical galaxy formation models with a $\Lambda$-dominated
cold dark matter cosmology ($\Lambda$CDM), early-type galaxies are
predicted to have formed through the merging of smaller galaxies and
accretion events over time
\citep[e.g.,][]{som99,col00,deLuc06,bow06}. Large galaxy surveys in
the local universe as well as at medium redshift have shown that
galaxies follow a strong bimodality in colour magnitude space, forming
a 'red sequence' and a 'blue cloud' with an intermediate region
sparsely populated by galaxies sometimes dubbed the 'green valley'
\citep[e.g.,][]{bal04,bell04,will06}. The hierarchical galaxy
formation scenario has received renewed observational support by the
recent findings of the COMBO-17 and DEEP2 surveys that the stellar
mass of galaxies on the red sequence has increased by a factor of 2-5
since redshift $z \sim 1$ (\citealt{bell04,fab07}; but see
\citealt{mar09}). This growth of mass can arise from two main sources
(1) the growth of objects already on the red-sequence (in the case of
cold gas accretion they may move from the red sequence into the 'green
valley' during the period of star-formation) and (2) the addition of
galaxies from the blue cloud once star-formation is shut off either by
consumption or by the removal of the cold gas reservoir \citep{fab07}.

There is observational evidence that the detailed growth of mass on
the red sequence is a function of luminosity (here used as a proxy for
total galaxy mass) as well as a function of environment. While most
massive galaxies have already joined the red sequence at $z \sim 1$
\citep{treu05,bun06,cim06}, the low mass end of the red sequence seems
to be subject to the continued arrival and re-rejuvenation of galaxies
up to the present day (\citealt{treu05,scha07,san08,smi09,sha2010},
but see \citealt{tra08}). Taken together, a scenario emerges where low
mass early-type galaxies (on the red sequence), irrespective of their
environment, are predicted to show younger stellar population ages
than their more massive counterparts. In addition, galaxies residing
in high-density environments are older and show less spread in stellar
age than those in less dense regions \citep[see also][]{tra08}.

Many authors have investigated the scenarios summarized above by means
of a stellar population analysis of nearby early-type galaxies based
on optical line strength indices often measured in the Lick/IDS
system.  In most of the studies galaxies are treated as spatially
unresolved sources \citep[e.g.,][]{ber06,yam06,scha07,tra08,tho2010}
allowing access to either large samples of galaxies, as for example in
the case of the Sloan Digital Sky Survey \citep{yor00}, or the
analysis of extremely high signal-to-noise ratio (S/N) data. Only a
limited number of studies have investigated line strength gradients
obtained with long slits \citep[e.g.,][]{dav93,san07,spo09} or
radially averaging the data from integral-field-units \citep{raw08};
spatially resolved stellar population maps of early-type galaxies
remain the exception \citep[e.g.,][]{mcd06,chi09,prac09,bla09}.

With the advent of high throughput integral field units on medium size
telescopes, the study of two-dimensional stellar populations maps has
become feasible. We have conducted a survey of 72 representative
early-type (E, S0, and Sa) galaxies using the integral-field
spectrograph \sauron\/ mounted on the William Herschel Telescope in La
Palma \citep[][]{bac01,deZ02}. This allowed us to map the stellar and
gas kinematics as well as a number of stellar absorption line indices
up to about one effective radius $R_{\rmn{e}}$ for most of the
galaxies in the sample.  Analyzing the global properties of the
velocity fields of the 48 E and S0 galaxies \citet[][hereafter
Paper~IX]{ems07} characterized the population of early-type galaxy
with two classes of slow and fast rotators, according to their
specific (projected) angular momentum measured within one effective
radius, $\lambda_R$.  \citet[][hereafter Paper~XII]{kraj08} further
analyzed the velocity and velocity dispersion maps using kinemetry, a
generalization of surface photometry to the higher order moments of
the line-of-sight velocity distribution \citep{kraj06} and found that
74 per cent of Es and 92 per cent of S0s in the \sauron\/ sample
exhibit rotation, which can be well-described by that of a simple,
inclined disc.  \citet{sco09} explore the surprisingly tight
correlation between the local escape velocity $V_{\rmn{esc}}$ and
stellar population parameters.

In this paper, we present for the 48 E and S0 galaxies in the
\sauron\/ sample, estimates of the mean simple stellar
population-equivalent \citep[hereafter SSP-equivalent; see ][]{tra08}
stellar population parameters, age, metallicity and abundance ratio
[$\alpha$/Fe] for each bin in our maps. The study is based on the line
strength measurements presented in \citet[][hereafter
Paper~VI]{kun06}, although we provide a short description of an
improved data-reduction scheme and the resulting updated line strength
measurements in this paper. Compared to many other line strength based
stellar population studies, our work rests on a limited number (3-4)
of line strength indices. However, our study has the benefit of
covering a contiguous region of the galaxies, often reaching out to
approximately one effective radius $R_{\rmn{e}}$. Furthermore, it
complements and extends the \sauron\/ survey studies of the {\em
  GALEX} near-UV (hereafter NUV) imaging \citep[][hereafter
Paper~XIII]{jeo09}, {\em Spitzer} mid-IR imaging \citep[][hereafter
Paper~XV]{sha2010} and ionized gas \citep[][hereafter
Paper~XVI]{sar2010}.

This paper is organized as follows. In Section~\ref{sec:revised}, we
present the details of our revised data-reduction leading to improved
line strength measurements. In Section~\ref{sec:pops}, we describe our
method to estimate stellar population parameters from the available
set of Lick/IDS indices along with a discussion on the reliability of
the results. The two-dimensional stellar population maps are presented
in Section~\ref{sec:results}, where we first discuss in detail the
findings and also analyze overall stellar population gradients and
scaling relations of the aperture measurements with central velocity
dispersion and dynamical mass of the galaxies. The conclusions follow
in Section~\ref{sec:conclusions}. We present a revised table of
circular aperture line strength measurements within \reb\/ and one
$R_{\rmn{e}}$ in the Appendix.

\section{Revised Data reduction}
\label{sec:revised}
The results presented here are based on the observations and data
reduction described in \citet[][hereafter Paper~III]{em04}, Paper~VI
and \citet[][hereafter Paper~V]{sar06}.  However, our spectral
template library which is used to derive the kinematics and the gas
emission was changed from the Jones library \citep{jon99} to the
stars from the recently published MILES stellar library
\citep{san06}. This new library provides better fits to our spectra,
therefore minimizing template mismatch \citep[see also][]{cap07}. The
main consequence of this are small, but significant changes to the
kinematics; namely, reduced $\sigma$ and $h_4$ values where the change
is most significant in the central regions of our sample galaxies and
over a larger area for the most massive systems. These changes in the
kinematics carry through to the fully corrected line strength
measurements via the internal velocity broadening corrections and
result in slightly reduced values for \mgb, Fe5015 and \fes\/ in the
affected regions of the maps (see Paper VI).

In an attempt to further improve the emission-line corrections to our
absorption line strength measurements, we also re-extracted the
nebular emission fluxes and kinematics using the MILES stellar
library.  In the case of emission-line measurements, however, it is
crucial to provide an underlying fit to the stellar continuum that is
as physically motivated as possible in order to avoid spurious
results. For this reason, rather than using a subset of stellar
spectra from the MILES library for each separate galaxy, we used for
the entire sample a single template library consisting of 48 simple
stellar population models from \citet[][based on MILES stellar
spectra]{vaz10} to which we added a number of stellar templates
obtained by matching \sauron\/ spectra devoid of emission.
Specifically, based on the maps of Paper~V, we extracted 50 high-S/N
spectra from circular apertures in regions without significant
emission.  We fitted these with the MILES stars over the {\it entire}
wavelength range (i.e. without excluding regions prone to emission),
excluding only the stellar templates with exceedingly low
($<0.9$\,\AA) values of the H$\beta$ absorption-line strength. The
optimal combination of the stellar spectra that best matched each of
our emission-free aperture spectra constitute each of the empirical
templates in our new template library. Care was taken to extract
\sauron\/ spectra covering the range of absorption-line strengths
observed in Paper~VI. In this respect, our empirical templates
approximate as closely as possible the spectra of real early-type
galaxies, were they unaffected by kinematic broadening.

Although adopting the new template library does not lead to a
dramatically better sensitivity in the detection of nebular emission,
the use of such a physically-motivated set of templates leads to more
robust emission-line measurements, and therefore a better correction
of the absorption-line strength indices affected by emission (mostly
\hb\/ and Fe5015 in our survey). For further discussion of the new
emission-line measurements see Paper~XVI.

In summary, the use of the new spectral template library significantly
improved the measurements of the line strength indices. Main results
presented in earlier papers, however, are not affected. Specifically,
in Paper~VI we presented a comparison between the isoindex contours of
\mgb\/ and the isophotes of the surface brightness finding that the
\mgb\/ contours appear to be flatter for about 40 per cent of our
galaxies without significant dust features. Using the revised index
measurements, we find that ellipticity measurements for the \mgb\/
contours have slightly changed, but agree within the one $\sigma$
errors with the previous measurements. Furthermore, the stellar M/L
ratios derived in \citet{cap06} remain the same within the quoted
errors.

In Table~\ref{tab:LSnew} of the Appendix we present the updated line
strength measurements for circular apertures within \reb\/ and one
$R_{\rmn{e}}$. These values supersede the ones presented in
Paper~VI. Aperture line strength measurements changed by less than 17
per cent with robustly measured mean offsets of $0.12$, $-0.37$,
$-0.02$, and $0.00$~\AA\/ for the \hb, Fe5015, \mgb\/ and \fes\/
indices over the \re\/ aperture, respectively. For the one $R_{\rmn
  e}$ aperture, the offsets are $0.12$, $-0.31$, $-0.01$ for the \hb,
Fe5015 and \mgb\/ indices, respectively. Note that we also made use of
revised estimates of the effective radii $R_{\rmn{e}}$ (see
Table~\ref{tab:LSnew}) which were derived with a $R^{1/n}$ growth
curve analysis from our wide-field MDM (1.3m) imaging survey of the
\sauron\/ sample (Falc{\'o}n-Barroso in preparation).

\section{Estimating stellar population parameters}
\label{sec:pops}
Estimates of luminosity-weighted stellar population parameters such as
age, metallicity and abundance ratios can be inferred from a set of
line strength indices measured on integrated spectra in combination
with the predictions from stellar population models. In this section
we begin by presenting a first order overview of the ages and
metallicities of our sample galaxies using classical age/metallicity
diagnostic diagrams and then describe the $\chi^2$ technique used for
the full analysis of our data including the abundance ratio
variations.

\subsection{A first look with classical techniques}
The classical problem of the degeneracy between age and metallicity
changes in an old ($\ge$2\,Gyr) integrated stellar population can be
partially overcome by plotting a combination of metallicity-sensitive
indices against an age-sensitive index \citep[e.g.,][]{wor94}. This
method has been used successfully in the past and we will present an
overview of our data with this traditional method before we employ
state-of-the-art techniques for the full analysis of our line
strengths maps (see Section~\ref{sec:chi2}).

Due to our limited wavelength range ($4800 - 5380$\,\AA) we can only
use \hb\/ as an age-sensitive index. In order to minimize the
influence of abundance ratio variations on our first-order age
estimates derived from the classical index {\em vs} index diagrams we
define composite indices. These indices are constructed such that a
mean metallicity is measured rather than a metallicity biased to one
species. The \hb\/ index shows only a mild dependence on abundance
ratios \citep[e.g., ][]{TMB03} and can be safely used in a first order
analysis. We emphasize that these composite indices are largely used
for presentation purposes when we want to show our measurements in the
observational plane of index {\em vs} index diagrams.

Within the \sauron\/ wavelength range we measure three different
metallicity indicators: Fe5015, \mgb, an \fes\/ (for details see
Paper~VI).  These indices are not pure metallicity indicators, but are
also sensitive to age changes.  Furthermore, non-solar abundance
ratios can severely affect the interpretation of these indices if not
taken into account \citep[e.g., ][]{kun01,TMB03}. In practice it is
sometimes convenient to define metallicity indicators which are
insensitive to abundance ratio variations. The first
index\footnote{$\rmn{[MgFe]} = \sqrt{\rmn{Mg}\,b\, \times
    \frac{(\rmn{Fe5270} + \rmn{Fe5335})}{2}}$} of this sort was
defined by \citet{gon93} and further improved by
\citet{TMB03}\footnote{ $\rmn{[MgFe]}^\prime = \sqrt{\rmn{Mg}\,b\,
    \times (0.72 \times \rmn{Fe5270} + 0.28 \times \rmn{Fe5335})}$}.
However, the \sauron\/ wavelength range does not allow us to measure
Fe5335. Therefore, we define new, abundance-ratio-insensitive index
combinations for the available indices. We follow the approach by
\citet{TMB03} and also use their models to derive the index
definitions. The subset of the model library we use to define the new
indices spans ages from 1 to 15\,Gyr, metallicities from $\rmn{[Z/H]}
=-1.35 ~\rmn{to}~ 0.35$ and abundance ratios of
$\rmn{[}\alpha/\rmn{Fe]} = 0.0 ~\rmn{to}~ 0.5$.  This covers most of
the range of stellar population parameters which we probe in this
survey.

The new abundance ratio insensitive indices are defined as:

\begin{equation}
  \label{eq:mgfe52}
  \rmn{[MgFe52]}^\prime =  \frac{0.64 \times \rmn{Mg}\,b + \rmn{Fe5270}}{2}
\end{equation}

\begin{equation}
  \label{eq:mgfe50}
  \rmn{[MgFe50]}^\prime =  \frac{0.69 \times \rmn{Mg}\,b + \rmn{Fe5015}}{2}
\end{equation}

\begin{figure}
 \includegraphics[width=84mm]{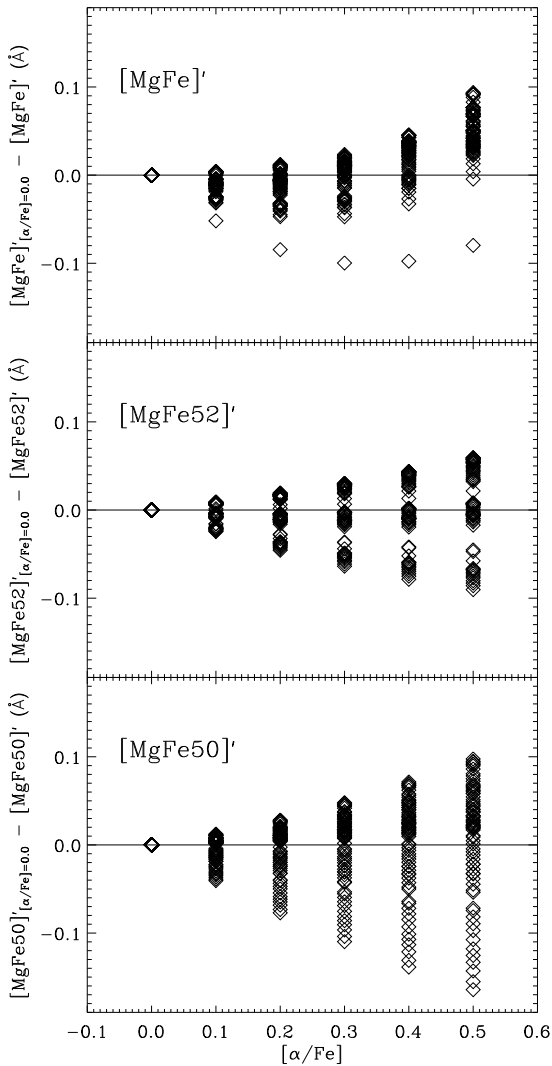} 
 \caption{The dependence of [MgFe]$^{\prime}$ \citep{TMB03},
   \mgfeftp\/ and \mgfefzp\/ on abundance ratio [$\alpha$/Fe]. Shown
   is the difference between the model predictions \citep{TMB03} at
   solar abundance ratio and the model predictions for non-solar
   abundance ratios. The model library used here spans ages from 1 to
   15\,Gyr, metallicities from $\rmn{[Z/H]} =-1.35 ~\rmn{to}~ 0.35$
   and abundance ratios of $\rmn{[}\alpha/\rmn{Fe]} = 0.0 ~\rmn{to}~
   0.5$. }
 \label{fig:mgfe52new}
\end{figure}

\noindent The scaling factor for the \mgb\/ index was optimized so
that the mean difference between solar and non-solar models is zero
for this index.  Figure~\ref{fig:mgfe52new} demonstrates that, similar
to the [MgFe]$^{\prime}$ index defined by \citet{TMB03}, the new
composite indices are rather insensitive to abundance ratio changes
over the range probed by the model predictions.

\begin{figure}
 \includegraphics[width=84mm]{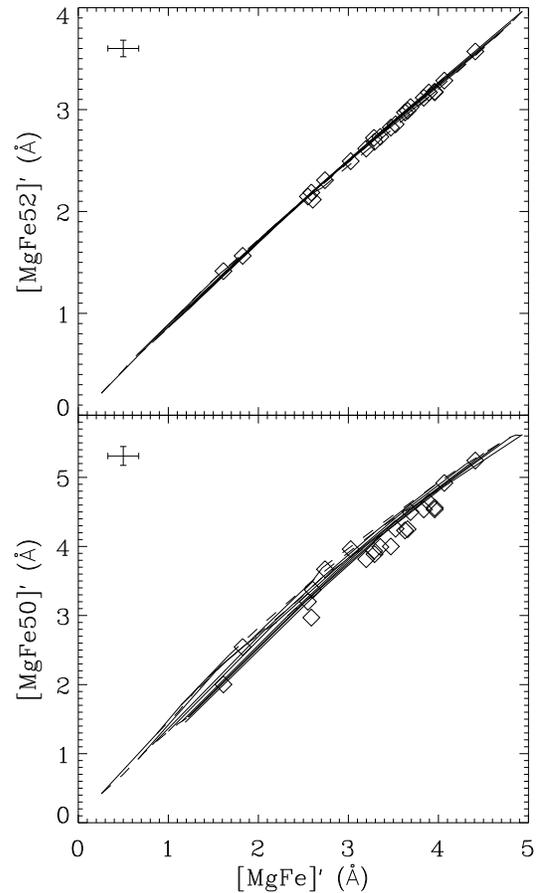}
 \caption{\mgfeftp\/ and \mgfefzp\/ {\em versus}\/ [MgFe]$^{\prime}$
   for a sample of Fornax galaxies taken from \citet{kun00} are shown
   as open diamonds.  Overplotted are stellar population models by
   \citet{TMB03} at solar abundance ratio. Average error bars for the
   Fornax data are shown in the upper left corner. Note, however, that
   the errors are correlated due to some indices being used in both
   axis.}
 \label{fig:mgfe}
\end{figure}

In Figure~\ref{fig:mgfe} we show the relation between \mgfeftp\/ and
\mgfefzp\/ {\em versus}\/ [MgFe]$^{\prime}$ for a sample of well
calibrated Fornax early-type galaxies taken from \citet{kun00}. There
is a good linear correlation between the new composite indices and
[MgFe]$^\prime$ as expected if they are good indicators of mean
metallicity.
 
\begin{figure*}
 \includegraphics[width=177mm]{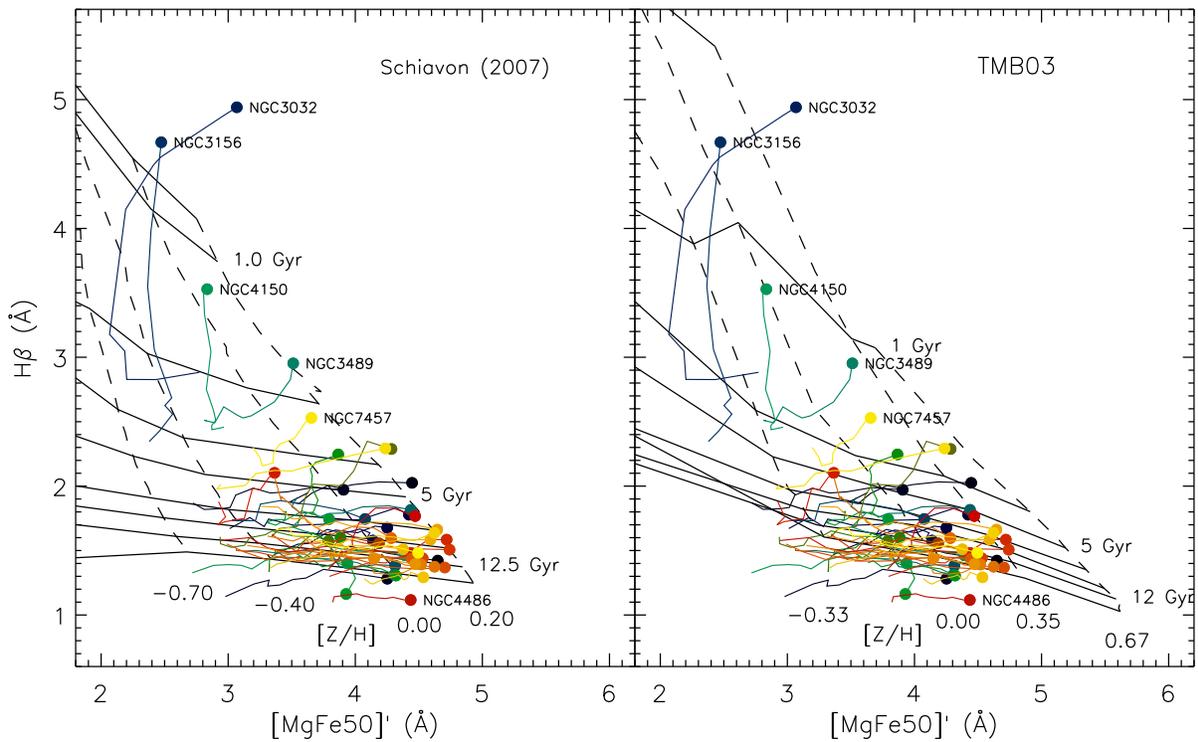}
 \caption{Radial line strength gradients averaged along isophotes for
   the 48 early-type galaxies in the \sauron\/ sample. The center of
   each galaxy is indicated by a filled circle and different colours
   are used for each galaxy. For galaxies with a strong contribution
   from a young stellar population (\hb\, $> 2.5$\,\AA) we give the
   galaxy name next to the central data-point. We also indicate the
   location of NGC\,4486 which exhibits \hb\/ absorption strength
   substantially below the models.  Overplotted are stellar population
   models by \citet[][left]{schia07} and \citet[][right; hereafter
   TMB03]{TMB03} for solar abundance ratios. Note that the TMB03
   models extend to $\rmn{[Z/H]} = +0.67$, whereas Schiavon models
   reach only $\rmn{[Z/H]} = +0.2$.  }
 \label{fig:hb_mgfe50_all}
\end{figure*}

Throughout this paper we use index measurements calibrated to the
Lick/IDS system (see Paper~VI and Section~\ref{sec:revised}). There
are various stellar population models available for this system
\citep[e.g., ][]{wor94,vaz97,TMB03,lee05,schia07} which agree to first
order in their predictions of index strength as function of age,
metallicity and abundance ratio, but differ in details. We demonstrate
this in Figure~\ref{fig:hb_mgfe50_all} where we show radial gradients
averaged along lines of constant surface brightness in the \hb\/
versus [MgFe50]$^\prime$ plane for the 48 early-type galaxies in our
sample.  The center of each galaxy is indicated by a filled
circle. Overplotted are model predictions from
\citet{schia07}\footnote{We use here the ``base'' models from Schiavon
  (2007) with solar abundance ratios, where the abundance ratio bias
  of the stellar library has been removed.  The model predictions were
  transformed to the Lick system by using the offsets described in
  Table~1 of Schiavon 2007.} for $\rmn{[}\alpha/\rmn{Fe]} = 0.0$ on the
left and model predictions by \citet{TMB03} for
$\rmn{[}\alpha/\rmn{Fe]} = 0.0$ on the right. This diagram provides a
convenient overview of our line strength measurements while at the
same time showing graphically the relation between the stellar
population parameters, luminosity-weighted age and metallicity, as
predicted by the models.

It is evident from the plot and consistent for both stellar population
models that the galaxies span a large range in luminosity weighted age
with the majority of galaxies clustering at old ages ($\sim$12\,Gyr).
However, taking a $\sim$9\,Gyr isochrone as a dividing line, about 40
per cent of our sample galaxies show signs for the presence of young
stellar populations. Excluding the very youngest objects, radial line
strength gradients translate into a decline in metallicity with
increasing radius.

Examining the diagram in more detail one can see that most galaxies
that are dominated by old stellar populations (e.g., age $>$ 9\,Gyr)
exhibit roughly constant \hb\/ absorption strength with radius (see
also Paper~VI). Taken at face value, the \citet{TMB03} models predict
for those galaxies an age gradient such that the center is younger
than the outskirts. Moreover, a significant fraction of galaxies shows
\hb\/ strengths at large radii which are below the model predictions
for an age of 15\,Gyr. Using the \citet{schia07} models both effects
are mitigated in the sense that there is little or no evidence for age
gradients for old galaxies and that the majority of line strength
measurements lie within the range of model predictions.

The TMB03 and Schiavon models both use the same index sensitivity
functions from \citet{kor05} to correct for abundance ratio patterns,
however, they use different stellar libraries; namely the
\citet{jon99} library in the Schiavon models and the Lick/IDS library
\citep{wor94b} in TMB03.

Although the \citet{schia07} models appear to better fit the
observational data of the \sauron\/ survey, we emphasize that stellar
population models are not considered perfect by us and calibration
uncertainties between the models and observed data remain an issue
\citep[see also][]{grav08}. For the remainder of this paper we will
use the models of \citet{schia07} for our stellar population analysis.
However, we will make specific reference to \citet{TMB03} at key
points to demonstrate the model independence of our results.

\subsection{The $\chi^2$ technique}
\label{sec:chi2}
In the classical index-index diagrams described above one does
typically not make use of all available index measurements and
furthermore it requires an iterative scheme if one is interested in
abundance ratio estimates. A more elegant way of determining stellar
population parameters is the $\chi^2$ technique \citep[e.g.,
][]{proc04}. Here the best fitting stellar population model to a given
set of index measurements is found by minimizing the differences
between the measured index values and the stellar population model
predictions in a $\chi^2$ sense. This approach was already used by our
team in \citet[][Paper~VIII]{mcd06}, and will be employed for the
remainder of this paper to derive the estimates of age, metallicity
and abundance ratio for a given set of observed indices. To allow for
a reasonable accuracy in the estimates we interpolate the Schiavon
models to a fine grid providing 340681 individual models, spanning
[Z/H]~$= -1.19$~to~$0.57$ (in steps of 0.02 dex), age~$=
0.89$~to~$17.7$\,Gyr (logarithmically, in steps of 0.02 dex), and
[$\alpha$/Fe]~$= -0.2$~to~0.5 (in steps of 0.01 dex)\footnote{The
  Schiavon (2007) models are provided as function of [Fe/H] and were
  converted by us to [Z/H] as used throughout this paper by a
  transformation kindly provided by R. Schiavon. As a result of the
  transformation the models do not cover the full metallicity range
  given above at each [$\alpha$/Fe].}. The above model parameter space
covers the majority of our measurements for the sample of 48
galaxies. For index measurements which lie outside of the stellar
population model parameter space, we use the best fitting values at
the boundary of the interpolated models. However, observed index
values are generally covered by the model parameter space and only
individual data points may cross the boundaries. We estimate
confidence levels on the derived parameters using the $\Delta \chi^2$
values \citep[for details see][]{press92}.

We use the $\chi^2$ technique to derive SSP-equivalent stellar
population parameters for each bin in our maps. We also use this
technique to derive stellar population estimates over circular
apertures of one effective radius $R_{\rmn{e}}$ and of radius \reb,
where all the spectra in these apertures have been summed up and line
strength values re-derived. For 28/48 galaxies the \sauron\/ data
cover less then one effective radius $R_{\rmn{e}}$ and we applied
aperture corrections to the line strength using the average
corrections derived in Paper~VI where necessary (see Table~4 therein).
Corrections for the \hb\/ index are negligible and are relatively small
for the Fe5015 and \mgb\/ indices with maximum corrections of 8 per
cent.  The fully corrected set of indices is given in
Table~\ref{tab:LSnew} of the Appendix.

\begin{figure*}
  \includegraphics[width=177mm]{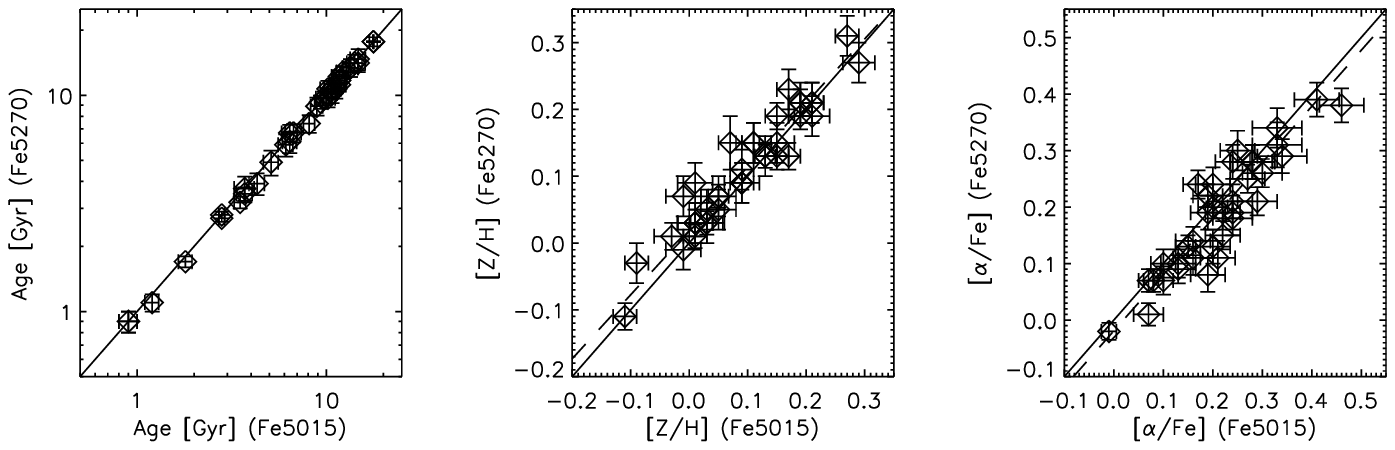}
  \caption{Comparison of stellar population parameters derived from
    \reb\/ circular aperture line strength measurements (see also
    Table~\ref{tab:LSnew}) by using the indices \hb, Fe5015 and \mgb\/
    or the indices \hb, \mgb\/ and Fe5270 with the $\chi^2$ method and
    the [Ti/Fe] adjusted \citet{schia07} models as described in
    Section~\ref{sec:chi2}. The axis labels indicate which one of the
    Fe5270 or F5015 indices is used. The one-to-one relation is
    indicated by a solid line in each panel whereas the dashed line
    shows linear fits to the data taking into account errors in both
    variables.}
  \label{fig:model_comp2} 
\end{figure*}

Due to our limited wavelength range, only the \hb\/, Fe5015, and
\mgb\/ indices are available for the full field of view. The fourth
index, \fes, typically covers a smaller fraction of the field-of-view
(see Paper VI).  However, for a circular aperture of \reb\/ we can
derive the \fes\/ values for 37 out of the 48 sample galaxies (see
also Table~\ref{tab:LSnew}). This allows us to compare the stellar
population parameters derived by using the Fe5015 index or the more
widely utilized Fe5270 index\footnote{We convert our \fes\/
  measurements to Fe5270 by using $\rmn{Fe5270} = 1.26 \times
  \rmn{Fe5270}_{\rmn{S}} + 0.06$ \citep[for details
  see][]{kun06}.}. The results of the comparison are shown in
Figure~\ref{fig:model_comp2}.

Because the \hb\/ index is the primary age indicator and is used in
both derivations, the age estimates show excellent agreement. Overall
we also find good agreement for the metallicity and abundance ratio
estimates with maximum differences of 0.08\,dex and 0.11\,dex for
[Z/H] and [$\alpha$/Fe], respectively. Linear fits taking into account
errors in both variables are close to the one-to-one relation (see
dashed lines in Figure~\ref{fig:model_comp2}). However, in order to
obtain a good agreement for the abundance ratio estimates we used a
custom built version of the Schiavon models (Schiavon, private
communication) where $\rmn{[Ti/Fe]} = 0.5 \times \rmn{[Mg/Fe]}$. More
specifically abundance ratios of [C/Fe], [N/Fe], and [Cr/Fe] are fixed
to solar ratios, whereas [O/Fe], [Mg/Fe], [Ca/Fe], [Na/Fe], [Si/Fe]
are varied with $\rmn{[Ti/Fe]} = 0.5 \times \rmn{[Mg/Fe]}$. Using the
standard version of the Schiavon models with $\rmn{[Ti/Fe]} =
\rmn{[Mg/Fe]}$ together with the Fe5015 index leads to somewhat
underestimated [$\alpha$/Fe] ratios for the most massive
galaxies. From our data alone it is not clear whether the Fe5015 index
shows a slightly incorrect calibration in the fitting functions or if
indeed the [Ti/Fe] ratio is not tracking the [Mg/Fe] ratio. Further
calibrations with e.g., globular cluster data are needed to settle
this particular issue. Throughout this paper we employ the [Ti/Fe]
adjusted models in our estimates of the stellar population parameters.

\begin{figure*} 
  \includegraphics[width=177mm]{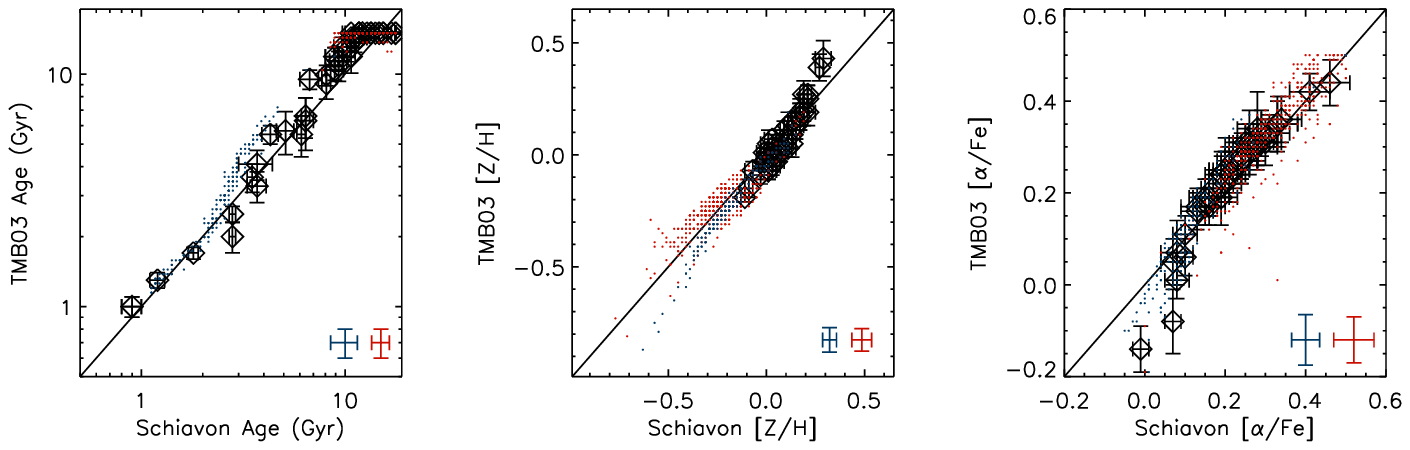}
  \caption{Comparison of stellar population parameters derived from
    the stellar population models of \citet{schia07} and
    \citet[][labeled TMB03]{TMB03} using the indices \hb, Fe5015 and
    \mgb. The black diamonds with error bars represent the \reb\/
    circular aperture values of our sample of galaxies. The results
    from the index maps of NGC\,4660 and NGC\,4150 are shown as red
    and blue points, respectively.  Representative error bars are
    shown in the lower right corner of each panel for typical stellar
    population values for each galaxy.}
  \label{fig:model_comp}
\end{figure*}

In Figure~\ref{fig:model_comp} we carry out a comparison of the
stellar population parameters derived with our standard set of indices
(\hb, Fe5015 and \mgb) between the \citet[][]{schia07} and
\citet{TMB03} models. In order to cover the parameter space probed in
this paper we use the line strength measurements of the \reb\/
apertures as well as the index maps of a galaxy with old stellar
populations (NGC\,4660) and one where young stellar populations
dominate (NGC\,4150).  Overall we find reasonable to good agreement
between the models with the known `saturation' effects of the
\citet{TMB03} model at old ages (15\,Gyr; see also
Figure~\ref{fig:hb_mgfe50_all}). For estimates of the metallicities
[Z/H] and abundance ratios [$\alpha$/Fe], we find good agreement
between the models over a large parameter space.  One difference to
note though is that for the most metal rich populations found in the
center of early-type galaxies, the Thomas et al. models predict higher
metallicities and younger ages compared to Schiavon. This effect is
mainly caused by the different slopes the models predict for \hb\/ as
function of metallicity (see also Figure~\ref{fig:hb_mgfe50_all}).
\citet{grav08} carry out a more detailed comparison between the
predictions of the \citeauthor{schia07} and \citeauthor{TMB03} models
using the sample of galaxies from \citet{tho05} and also find overall
agreement, albeit finding more significant offsets for metallicity and
age estimates \citep[see also][]{smi09}.

In summary, we conclude that our stellar population estimates derived
from \hb, Fe5015 and \mgb\/ and the Schiavon models, which we will use
for all maps in this paper, agree well with the estimates derived from
the more widely used index combination of \hb, \mgb\/ and Fe5270 and
the predictions from the stellar population models of \citet{TMB03}.

\begin{figure*}
  \includegraphics[width=177mm]{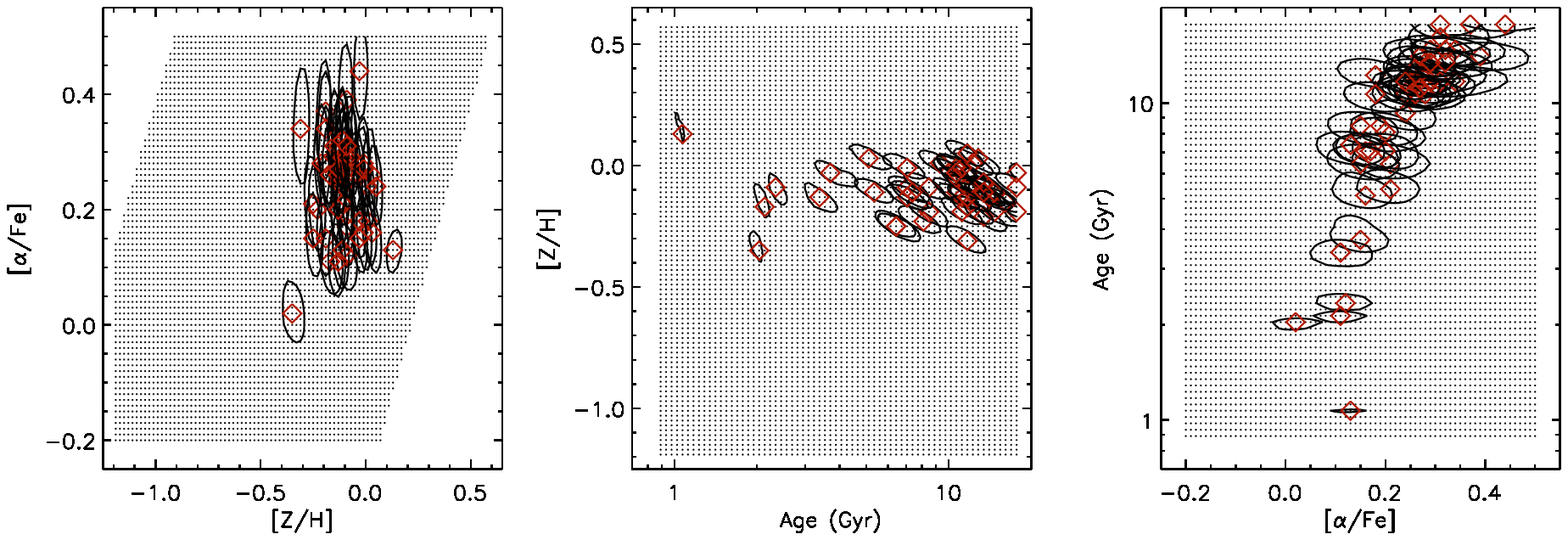}
  \caption{Examples of $\Delta \chi^2$ space for finding the
    best-fitting stellar population parameters (SSP-equivalent age,
    metallicity [Z/H] and [$\alpha$/Fe]; Schiavon 2007 models) to the
    line strengths of all 48 early-type galaxies as measured within a
    circular aperture of one $R_{\rmn{e}}$.  The best-fitting model is
    found by minimizing the difference between the model predictions
    and the three Lick/IDS indices \hb, Fe5015 and \mgb.  The three
    panels show different projections of the three-dimensional
    solution space, with one of the parameters fixed at the best
    fitting value.  The contours indicate the 1$\sigma$ confidence
    levels based on $\Delta \chi^2$ for two degrees of freedom. The
    small black dots show all grid points where $\chi^2$ is
    evaluated.}
  \label{fig:central_pop_dia}
\end{figure*}

As final demonstration of the $\chi^2$ technique, we show in
Figure~\ref{fig:central_pop_dia} how the one $R_{\rmn{e}}$ aperture
index values of our sample of 48 early-type galaxies project onto the
model grids in planes of [$\alpha$/Fe] vs [Z/H], [Z/H] vs Age and Age
vs [$\alpha$/Fe]. Error ellipses are drawn for each galaxy and
demonstrate graphically the well known error correlation between [Z/H]
and Age \citep[e.g.,][]{tra00b,kun01}. In Table~\ref{tab:stellpop}, we
present our estimates of SSP-equivalent age, metallicity and
[$\alpha$/Fe] for circular apertures within \reb\/ and one $R_{\rmn
  e}$.

\begin{table*}
  \begin{center} 
    \begin{minipage}{175mm} 
      \caption{List of SSP-equivalent age, metallicity [Z/H] and
        [$\alpha$/Fe] estimates within a circular aperture of $R_{\rm
          e}/8$ and $R_{\rm e}$.}
      \label{tab:stellpop} 
	\begin{center}
      \begin{tabular}{lrrrrrr} \hline
          Name      & Age  (Gyr)& Metallicity & [$\alpha$/Fe]  & Age (Gyr)         & Metallicity  & [$\alpha$/Fe]  \\
          & \reb & \reb        &  \reb          & $R_{\rmn{e}}$ & $R_{\rmn{e}}$ & $R_{\rmn{e}}$   \\
          (1)      & (2)  &  (3)        & (4)            & (5)          & (6)          & (7)            \\
          \hline
  NGC\,474 & $ 9.3^{+ 0.4}_{-1.2}$ & $ 0.05\pm0.04$ & $ 0.16\pm0.05$ & $ 7.1^{+ 0.7}_{-0.6}$ & $-0.09\pm0.04$ & $ 0.17\pm0.04$ \\
  NGC\,524 & $15.4^{+ 1.5}_{-1.4}$ & $ 0.03\pm0.04$ & $ 0.20\pm0.04$ & $12.2^{+ 1.2}_{-1.1}$ & $ 0.01\pm0.04$ & $ 0.18\pm0.05$ \\
  NGC\,821 & $11.2^{+ 1.1}_{-1.0}$ & $ 0.07\pm0.02$ & $ 0.22\pm0.04$ & $11.2^{+ 0.5}_{-1.0}$ & $-0.13\pm0.02$ & $ 0.25\pm0.04$ \\
 NGC\,1023 & $11.7^{+ 0.6}_{-1.0}$ & $ 0.09\pm0.02$ & $ 0.19\pm0.04$ & $11.7^{+ 1.7}_{-0.5}$ & $-0.03\pm0.04$ & $ 0.26\pm0.04$ \\
 NGC\,2549 & $ 3.7^{+ 0.4}_{-0.3}$ & $ 0.27\pm0.02$ & $ 0.15\pm0.03$ & $ 5.1^{+ 0.5}_{-0.2}$ & $ 0.03\pm0.02$ & $ 0.16\pm0.04$ \\
 NGC\,2685 & $ 5.1^{+ 0.5}_{-0.4}$ & $-0.01\pm0.04$ & $ 0.19\pm0.04$ & $ 6.4^{+ 1.0}_{-0.6}$ & $-0.25\pm0.04$ & $ 0.21\pm0.05$ \\
 NGC\,2695 & $16.9^{+ 0.8}_{-1.5}$ & $-0.01\pm0.04$ & $ 0.31\pm0.05$ & $17.7^{+ 0.1}_{-0.8}$ & $-0.19\pm0.02$ & $ 0.37\pm0.06$ \\
 NGC\,2699 & $ 6.4^{+ 0.6}_{-0.6}$ & $ 0.21\pm0.02$ & $ 0.16\pm0.04$ & $ 8.5^{+ 0.4}_{-1.1}$ & $-0.09\pm0.04$ & $ 0.19\pm0.05$ \\
 NGC\,2768 & $ 8.9^{+ 0.9}_{-0.8}$ & $ 0.01\pm0.04$ & $ 0.26\pm0.04$ & $11.2^{+ 1.1}_{-1.0}$ & $-0.19\pm0.02$ & $ 0.28\pm0.06$ \\
 NGC\,2974 & $ 8.9^{+ 0.9}_{-0.4}$ & $ 0.11\pm0.04$ & $ 0.29\pm0.04$ & $ 9.3^{+ 0.9}_{-0.8}$ & $ 0.01\pm0.02$ & $ 0.24\pm0.04$ \\
 NGC\,3032 & $ 0.9^{+ 0.1}_{-0.1}$ & $ 0.03\pm0.04$ & $ 0.07\pm0.02$ & $ 1.1^{+ 0.1}_{-0.1}$ & $ 0.13\pm0.06$ & $ 0.13\pm0.02$ \\
 NGC\,3156 & $ 0.9^{+ 0.1}_{-0.1}$ & $ 0.05\pm0.02$ & $-0.01\pm0.02$ & $ 2.0^{+ 0.1}_{-0.1}$ & $-0.35\pm0.06$ & $ 0.02\pm0.04$ \\
 NGC\,3377 & $ 6.7^{+ 0.7}_{-0.6}$ & $ 0.01\pm0.04$ & $ 0.24\pm0.04$ & $ 8.1^{+ 0.4}_{-1.0}$ & $-0.23\pm0.04$ & $ 0.20\pm0.05$ \\
 NGC\,3379 & $14.7^{+ 0.7}_{-1.9}$ & $ 0.01\pm0.04$ & $ 0.28\pm0.05$ & $14.7^{+ 1.4}_{-1.3}$ & $-0.11\pm0.02$ & $ 0.29\pm0.05$ \\
 NGC\,3384 & $ 6.1^{+ 0.3}_{-0.8}$ & $ 0.19\pm0.02$ & $ 0.13\pm0.04$ & $ 7.1^{+ 0.7}_{-0.6}$ & $-0.01\pm0.04$ & $ 0.16\pm0.04$ \\
 NGC\,3414 & $13.4^{+ 1.3}_{-1.2}$ & $ 0.03\pm0.04$ & $ 0.34\pm0.05$ & $13.4^{+ 1.3}_{-1.2}$ & $-0.21\pm0.04$ & $ 0.28\pm0.06$ \\
 NGC\,3489 & $ 1.8^{+ 0.1}_{-0.2}$ & $ 0.15\pm0.02$ & $ 0.10\pm0.02$ & $ 2.3^{+ 0.1}_{-0.1}$ & $-0.09\pm0.04$ & $ 0.12\pm0.03$ \\
 NGC\,3608 & $11.7^{+ 0.6}_{-1.0}$ & $ 0.05\pm0.04$ & $ 0.24\pm0.04$ & $10.7^{+ 0.5}_{-0.9}$ & $-0.11\pm0.02$ & $ 0.24\pm0.04$ \\
 NGC\,4150 & $ 1.2^{+ 0.1}_{-0.1}$ & $ 0.07\pm0.02$ & $ 0.08\pm0.03$ & $ 2.1^{+ 0.1}_{-0.1}$ & $-0.17\pm0.06$ & $ 0.11\pm0.04$ \\
 NGC\,4262 & $14.1^{+ 1.4}_{-1.2}$ & $-0.01\pm0.04$ & $ 0.33\pm0.05$ & $14.7^{+ 1.4}_{-1.3}$ & $-0.19\pm0.04$ & $ 0.34\pm0.06$ \\
 NGC\,4270 & $ 8.1^{+ 0.4}_{-1.0}$ & $-0.07\pm0.04$ & $ 0.08\pm0.04$ & $ 8.5^{+ 0.8}_{-0.7}$ & $-0.19\pm0.04$ & $ 0.15\pm0.05$ \\
 NGC\,4278 & $17.7^{+ 0.1}_{-1.6}$ & $ 0.01\pm0.02$ & $ 0.46\pm0.05$ & $14.1^{+ 1.4}_{-1.2}$ & $-0.09\pm0.04$ & $ 0.39\pm0.06$ \\
 NGC\,4374 & $14.7^{+ 1.4}_{-1.3}$ & $ 0.01\pm0.04$ & $ 0.30\pm0.04$ & $16.1^{+ 1.6}_{-2.1}$ & $-0.15\pm0.04$ & $ 0.31\pm0.06$ \\
 NGC\,4382 & $ 3.7^{+ 0.7}_{-0.2}$ & $ 0.05\pm0.04$ & $ 0.16\pm0.04$ & $ 5.4^{+ 0.5}_{-0.5}$ & $-0.11\pm0.04$ & $ 0.21\pm0.04$ \\
 NGC\,4387 & $10.2^{+ 0.5}_{-1.3}$ & $-0.03\pm0.04$ & $ 0.20\pm0.04$ & $11.7^{+ 1.7}_{-0.5}$ & $-0.17\pm0.04$ & $ 0.26\pm0.06$ \\
 NGC\,4458 & $10.7^{+ 1.0}_{-0.9}$ & $-0.11\pm0.02$ & $ 0.28\pm0.05$ & $11.7^{+ 1.1}_{-1.0}$ & $-0.31\pm0.04$ & $ 0.34\pm0.06$ \\
 NGC\,4459 & $ 3.5^{+ 0.2}_{-0.3}$ & $ 0.17\pm0.02$ & $ 0.22\pm0.04$ & $ 7.1^{+ 0.7}_{-0.6}$ & $-0.13\pm0.02$ & $ 0.20\pm0.04$ \\
 NGC\,4473 & $11.7^{+ 0.6}_{-1.0}$ & $ 0.09\pm0.02$ & $ 0.24\pm0.04$ & $12.2^{+ 1.8}_{-0.6}$ & $-0.09\pm0.04$ & $ 0.27\pm0.04$ \\
 NGC\,4477 & $ 9.7^{+ 0.9}_{-0.9}$ & $ 0.03\pm0.04$ & $ 0.24\pm0.04$ & $11.7^{+ 1.1}_{-1.0}$ & $-0.13\pm0.02$ & $ 0.26\pm0.04$ \\
 NGC\,4486 & $17.7^{+ 0.1}_{-0.1}$ & $ 0.13\pm0.02$ & $ 0.41\pm0.05$ & $17.7^{+ 0.1}_{-0.1}$ & $-0.03\pm0.02$ & $ 0.44\pm0.05$ \\
 NGC\,4526 & $ 6.4^{+ 0.3}_{-0.8}$ & $ 0.19\pm0.02$ & $ 0.22\pm0.04$ & $10.7^{+ 1.0}_{-0.9}$ & $-0.01\pm0.02$ & $ 0.28\pm0.05$ \\
 NGC\,4546 & $10.7^{+ 1.0}_{-0.9}$ & $ 0.13\pm0.02$ & $ 0.27\pm0.04$ & $11.7^{+ 1.1}_{-1.0}$ & $-0.13\pm0.02$ & $ 0.31\pm0.05$ \\
 NGC\,4550 & $ 4.3^{+ 0.4}_{-0.4}$ & $-0.09\pm0.02$ & $ 0.21\pm0.04$ & $ 6.4^{+ 0.6}_{-0.8}$ & $-0.25\pm0.04$ & $ 0.15\pm0.05$ \\
 NGC\,4552 & $10.2^{+ 1.0}_{-0.9}$ & $ 0.21\pm0.04$ & $ 0.25\pm0.04$ & $12.8^{+ 1.2}_{-1.1}$ & $ 0.03\pm0.04$ & $ 0.26\pm0.05$ \\
 NGC\,4564 & $10.7^{+ 0.5}_{-0.9}$ & $ 0.21\pm0.02$ & $ 0.19\pm0.03$ & $10.7^{+ 1.6}_{-0.5}$ & $-0.05\pm0.04$ & $ 0.26\pm0.04$ \\
 NGC\,4570 & $11.7^{+ 1.1}_{-1.0}$ & $ 0.17\pm0.02$ & $ 0.17\pm0.03$ & $14.1^{+ 1.4}_{-1.2}$ & $-0.13\pm0.04$ & $ 0.27\pm0.04$ \\
 NGC\,4621 & $13.4^{+ 1.3}_{-1.2}$ & $ 0.09\pm0.04$ & $ 0.28\pm0.05$ & $14.7^{+ 0.7}_{-1.9}$ & $-0.11\pm0.04$ & $ 0.32\pm0.06$ \\
 NGC\,4660 & $12.2^{+ 1.2}_{-0.6}$ & $ 0.15\pm0.02$ & $ 0.24\pm0.04$ & $13.4^{+ 1.3}_{-1.2}$ & $-0.11\pm0.02$ & $ 0.29\pm0.05$ \\
 NGC\,5198 & $12.2^{+ 1.2}_{-1.1}$ & $ 0.13\pm0.04$ & $ 0.26\pm0.04$ & $11.7^{+ 1.1}_{-1.0}$ & $-0.09\pm0.02$ & $ 0.29\pm0.04$ \\
 NGC\,5308 & $12.8^{+ 0.6}_{-1.1}$ & $ 0.05\pm0.04$ & $ 0.24\pm0.04$ & $13.4^{+ 1.3}_{-1.2}$ & $-0.07\pm0.04$ & $ 0.29\pm0.05$ \\
 NGC\,5813 & $11.7^{+ 1.7}_{-0.5}$ & $ 0.05\pm0.04$ & $ 0.33\pm0.05$ & $13.4^{+ 0.6}_{-1.7}$ & $-0.11\pm0.04$ & $ 0.32\pm0.06$ \\
 NGC\,5831 & $ 8.1^{+ 0.4}_{-0.7}$ & $ 0.11\pm0.02$ & $ 0.13\pm0.04$ & $ 7.4^{+ 0.7}_{-0.7}$ & $-0.11\pm0.04$ & $ 0.13\pm0.04$ \\
 NGC\,5838 & $ 9.3^{+ 0.9}_{-0.4}$ & $ 0.19\pm0.02$ & $ 0.20\pm0.03$ & $11.2^{+ 0.5}_{-1.4}$ & $-0.01\pm0.04$ & $ 0.27\pm0.04$ \\
 NGC\,5845 & $11.2^{+ 0.5}_{-1.0}$ & $ 0.15\pm0.04$ & $ 0.20\pm0.04$ & $11.7^{+ 0.6}_{-1.0}$ & $ 0.05\pm0.02$ & $ 0.24\pm0.04$ \\
 NGC\,5846 & $15.4^{+ 0.7}_{-1.4}$ & $ 0.09\pm0.02$ & $ 0.28\pm0.04$ & $17.7^{+ 0.1}_{-1.6}$ & $-0.09\pm0.04$ & $ 0.31\pm0.05$ \\
 NGC\,5982 & $ 8.9^{+ 0.9}_{-0.8}$ & $ 0.17\pm0.02$ & $ 0.15\pm0.03$ & $10.7^{+ 0.5}_{-0.9}$ & $-0.03\pm0.02$ & $ 0.18\pm0.04$ \\
 NGC\,7332 & $ 2.8^{+ 0.1}_{-0.1}$ & $ 0.29\pm0.04$ & $ 0.10\pm0.03$ & $ 3.7^{+ 0.5}_{-0.2}$ & $-0.03\pm0.02$ & $ 0.15\pm0.04$ \\
 NGC\,7457 & $ 2.8^{+ 0.1}_{-0.1}$ & $ 0.01\pm0.04$ & $ 0.07\pm0.03$ & $ 3.4^{+ 0.2}_{-0.3}$ & $-0.13\pm0.04$ & $ 0.11\pm0.04$ \\
          \hline 
        \end{tabular} 
	\end{center}

        {\em Notes:}\/ (1) NGC number. (2)~-~(4) SSP-equivalent
        estimates of age, metallicity [Z/H] and [$\alpha$/Fe] within a
        circular aperture of \reb.  (5)~-~(7) SSP-equivalent estimates
        of age, metallicity [Z/H] and [$\alpha$/Fe] within a circular
        aperture of one $R_{\rmn{e}}$. For galaxies with less than one
        $R_{\rmn{e}}$ coverage we applied the aperture corrections given
        in Paper~VI to the line strength indices and then derived the
        stellar population estimates.

    \end{minipage} 
      
  \end{center}
\end{table*}

\subsection{Interpretation of derived stellar population parameters}
\label{sec:interpretation}
A word of caution is necessary when making use of the stellar
population models described above to interpret the observed line
strength indices.  Each of the stellar population models for a given
age, metallicity and abundance ratio is made of a set of stars that
all have the same age, metallicity and abundance ratio. Strictly
speaking, these models can then only be applied to objects, or
specific regions within objects for which this is also true. Since we
observe the integrated light along the line-of-sight for our target
galaxies we clearly violate this condition even if only metallicity
gradients are considered. The problem is even larger for galaxies with
signs of multiple star-formation periods. In practice, the derived
stellar parameters, from any of the methods described above, are
interpreted as {\em SSP-equivalent}. \citet{ser07} investigated the
SSP-equivalent age and chemical composition measured from Lick/IDS
line strength indices of a large set of two-population (old plus
young), composite model stellar populations. Their conclusions are
that SSP-equivalent estimates are biased such that the derived SSP age
depends primarily on the age of the young population, while the
derived SSP chemical composition is weighted more towards that of the
old population.  Furthermore, in this context, the age of a young
stellar component and the mass fraction between old and young stars
are degenerate.

Changes in observed line strength measurements due to interstellar
extinction, most likely to be present for early-type galaxies with
signs of recent star-formation, are negligible. Basic simulations show
that with a standard extinction curve \citep{fit99} and E(B-V) = 0.7,
the \hb, Fe5015, \mgb, and Fe5270 indices change by less than 0.01,
0.05, 0.01, and 0.02 \AA, respectively, compared to dust free models.

Given the biases and degeneracies described above, we find that the
broad conclusions on the ages and chemical composition of our sample
as described here are still valid, but individual determinations for a
given spectrum can be biased in the presence of composite stellar
populations. Specifically, the SSP-equivalent age estimate is strongly
biased towards the last star-formation event rather than tracing the
age of the majority of the stars contributing to the integrated
spectrum.

\section{RESULTS}
\label{sec:results} 
Figures~\ref{fig:maps1}--l below present maps of SSP-equivalent age,
metallicity [Z/H] and abundance ratio [$\alpha$/Fe] of the 48
early-type galaxies in the \sauron\/ sample, ordered by increasing NGC
number. For each galaxy, we show the total intensity reconstructed
from the full wavelength range of the \sauron\ spectra (see also
Paper~III), and the two-dimensional distributions of age, [Z/H] and
[$\alpha$/Fe] overplotted with isophotes of the reconstructed image
spaced by single magnitude steps.  We also include the stellar
velocity maps previously shown in Paper~III\footnote{The kinematics
  were re-derived using the new stellar template library as described
  in Section~\ref{sec:revised}.  However, there are only very minor
  ($\Delta v \le 10$\,\kms) changes in the recession velocities thus
  the visual impression of the maps is identical to Paper~III. } in
the bottom panel for each galaxy, used as reference for our
discussion. The maps are all plotted with the same spatial scale, and
oriented with respect to the \sauron\ field for presentation
purposes. The relative directions of north and east are indicated by
the orientation arrow next to the galaxy title. The maximum and
minimum of the plotting range is given in the tab attached to each
parameter map, and the colour bar indicates the colour table used. In
order to allow for an easy comparison between galaxies, the plotting
range of the age maps is fixed to the range 0.4 to 18\,Gyr, where the
colour bars reflect a linear scaling. The colours are adjusted such
that blue to dark green shades correspond to old stellar populations
($10 - 18$\,Gyr), while younger stellar populations are represented by
red and yellow shades. Similarly, the plotting range of the
[$\alpha$/Fe] maps is fixed to the range $-0.2$ to +0.5 for all
galaxies. For the [Z/H] maps we use an independent plotting range for
each galaxy in order to better visualize the metallicity distribution
across the field-of-view.

In Paper~VI we found non-physical values of the \hb, Fe5015 and \mgb\
line strengths in the outer regions of some galaxies where many
individual lenslets are averaged to achieve the target S/N of 60.
Additionally, we removed regions where a simple stellar population
interpretation is not meaningful; for example the central AGN
contaminated region and the jet of NGC\,4486 or the regions affected
by neighbour galaxies of NGC\,5846 and the regions affected by
foreground stars. These bins are indicated with grey colour in the
maps presented in Figures~\ref{fig:maps1}--l.

Whenever we make reference to the dynamical mass of a galaxy, we have
estimated it with the scalar virial relation

\begin{equation}
	M_{\rm dyn} = 5.0\times R_{\rmn e}\sigma_{\rm e}^2/G,
  	\label{equ:mvir}
\end{equation}

\noindent where $\sigma_{\rm e}$ is the luminosity-weighted second
velocity moment within one effective (half light) radius $R_{\rmn e}$
and the factor 5.0 was calibrated using detailed dynamical models in
Paper~IV. This mass represents

\begin{equation}
	M_{\rm dyn}\approx2\times M_{1/2}, 
\end{equation}

\noindent where $M_{1/2}$ is the total dynamical mass within a sphere
containing half of the galaxy light. It should not be confused with
the much larger total galaxy mass within the virial radius.

\subsection{Overview of stellar population maps} 
\label{sec:maps} 
The stellar population maps in Figs~\ref{fig:maps1}--l show a wealth
of structures. Some general trends are apparent and we discuss these
in the following.

\subsubsection{Age maps} 
\label{sec:agemaps} 
\begin{figure*}
  \includegraphics[width=177mm]{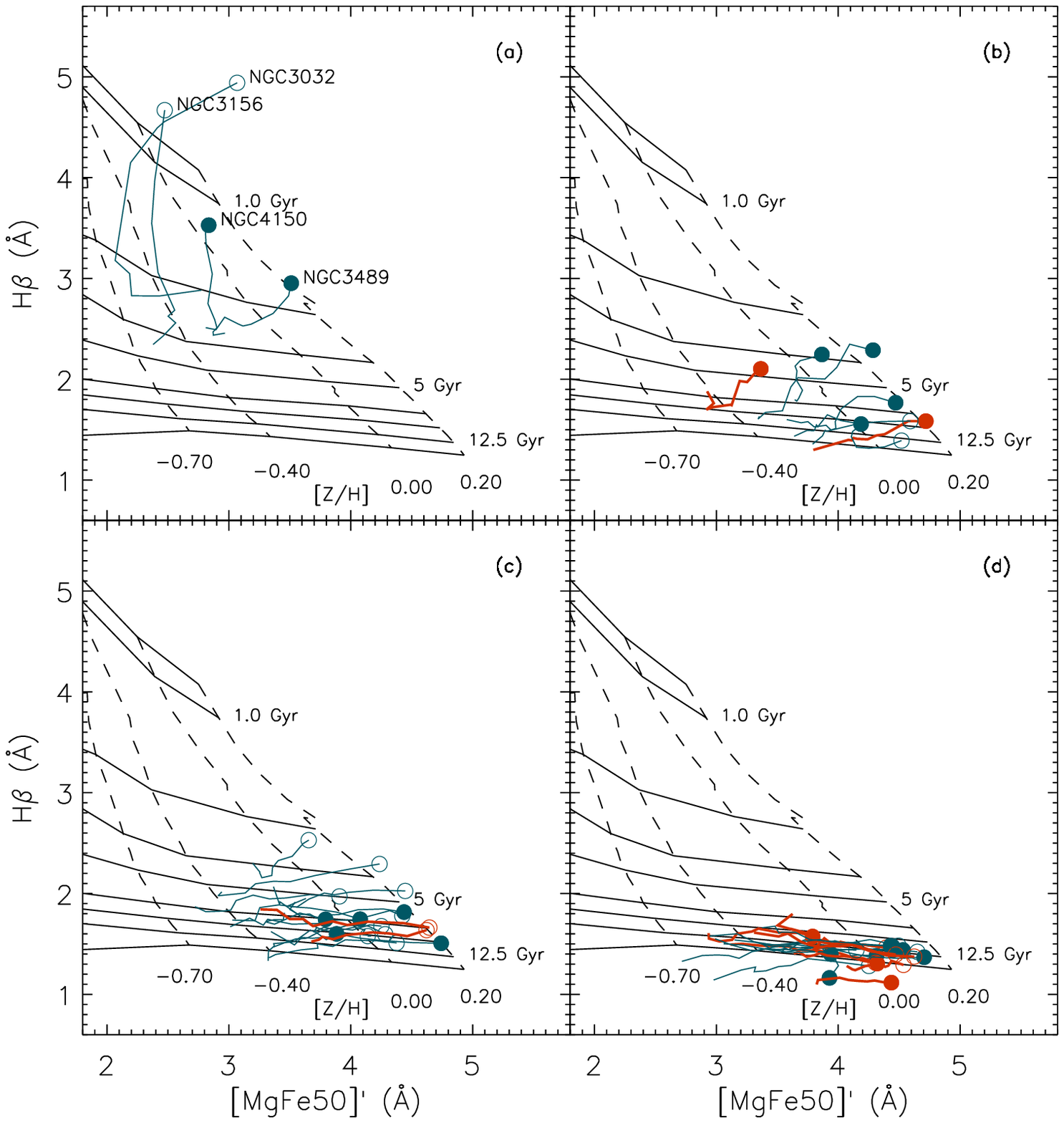}
  \caption{Age/metallicity diagnostic diagrams for our sample of the
    early-type galaxies in the \sauron\/ survey. For each galaxy a
    radial gradient averaged along isophotes is shown. The center of
    each galaxy is indicated by a filled circle and open circle for
    cluster and field galaxies, respectively. The colour indicates
    whether a galaxy belongs to the fast rotators (blue) or slow
    rotators (red). Overplotted are stellar population models by
    \citet{schia07}. (a) Galaxies with an extended, recent
    star-formation episode; (b) galaxies with signs of a central,
    spatially constrained, region of younger stellar populations; (c)
    galaxies with milder but spatially extended signs of a younger
    stellar population; and (d) galaxies consistent with overall old
    stellar populations.}
  \label{fig:pop_young} 
\end{figure*}

To aid the discussion in this section and to provide a simple overview
of the different age map categories, we show in
Figure~\ref{fig:pop_young} radial line strength gradients averaged
along isophotes with the predictions of stellar population models
overplotted. This is similar to Figure~\ref{fig:hb_mgfe50_all}.
However, here we have separated the galaxies into four groups according
to their radial age gradients. For each galaxy we indicate by colour
whether it belongs to the fast rotators (blue) or slow rotators (red;
see Paper~IX).  An overview of the various star-formation indicators
used in the \sauron\/ survey is given in Table~\ref{tab:starform},
where galaxies have been grouped in the same manner as in
Figure~\ref{fig:hb_mgfe50_all}.

About 60 per cent of the 48 galaxies in our sample show roughly flat
age maps of dark-blue to green colour indicating an overall old age
($\ge 9$\,Gyr) over the observed field-of-view. Due to the linear
sampling of the colour map in age, the similar colour in our maps
hides potentially significant age variations in absolute terms
(9-18\,Gyr).  This reflects, however, the increased error in our age
estimates towards older ages, which is dictated by the age sensitivity
of the H$\beta$ index, becoming more uncertain for the same index
measurement error at older ages (see also
Figure~\ref{fig:hb_mgfe50_all}).

In marked contrast, individual galaxies (NGC\,3032, NGC\,3156,
NGC\,3489, NGC\,4150, and NGC\,7457) show age maps indicating a
spatially extended, recent star-formation episode covering large parts
of the observed field-of-view with SSP-equivalent ages of less than
$\sim$3\,Gyr (see Figure~\ref{fig:pop_young}a). Our age maps show that
the youngest stellar populations in these galaxies are found in the
central parts, with typically steep age gradients. Furthermore,
ionized gas, optical dust features or even regular disks are detected
in this group of objects (see Papers~V \& XVI and
Table~\ref{tab:starform}).

All of the above galaxies are also detected in a systematic single
dish CO survey of the \sauron\/ sample \citep{comb07} further
supporting the scenario of recent star-formation. Recently, the
\sauron\/ observations have been extended with auxiliary data in the
NUV with {\em GALEX} imaging (Paper~XIII) and in the mid-IR with {\em
  Spitzer} IRAC imaging and IRS spectroscopy (Paper~XV). The NUV - V
colour may be used to trace young stars (age $<1$\, Gyr) although
conservative limits have to be chosen to avoid confusion with the
UV-upturn phenomenon \citep[e.g.,][Paper~XIII]{ocon99}. Similarly, the
existence of young stars and even ongoing star-formation is closely
connected to the presence of polycyclic aromatic hydrocarbon
(hereafter PAH) emission around 8\,$\mu$m (Paper~XV). Remarkably, but
perhaps not surprisingly, the four galaxies with the youngest
SSP-equivalent ages in our sample (age $<2$\,Gyr within an aperture of
\reb\/) clearly exhibiting signs of recent star-formation in the NUV
and in the mid-IR where data is available (see
Table~\ref{tab:starform}). This suggests residual ongoing
star-formation in this group of galaxies. Evidence for ongoing
star-formation is found in NGC\,3032 even in our own emission-line
analysis of the ionized gas (i.e. a very low [O{\small III}]/\hb\/
ratio) as demonstrated in Paper~XVI.

The next youngest galaxy in our sample NGC\,7457 with a central
SSP-equivalent age of 2.8\,Gyr \re\/ is only marginally detected in
the UV and not detected in CO or by PAH emission in the mid-IR.  We
interpret the slightly older SSP-equivalent age and the absence of PAH
emission in the mid-IR as evidence for an evolved version of the
younger galaxies discussed above. It is interesting to note that all
of the above mentioned galaxies have velocity dispersions of
$\sigma_{\rm e} \le 100$\,\kms\/ ($M_\rmn{dyn} \le 1.6 \times 10^{10}
M_{\sun}$) and thus belong to the low mass end of our sample. Paper~XV
argues that these kinds of galaxies are the product of recent gas-rich
(minor) mergers where the cessation of star-formation occurs in an
'outside-in' manner as the molecular gas is heated or consumed. The
roughly equal distribution of co- and counter-rotating kinematic
structures in these systems supports the external origin of the gas
(Paper~XV).

\begin{table*} 
  \begin{minipage}{175mm} 
    \caption{Summary of star-formation indicators in the \sauron\/ survey.}
    \label{tab:starform}
    \begin{center}
    \begin{tabular}{lccccccccc} \hline
        Name      & Age \re& $\sigma_{\rm e}$& Flatter \mgb & CO & NUV & mid-IR & Optical & Ionized & Rotator    \\
        &  (Gyr)    &    (\kms)  & contours     &    &     & PAH & dust    & gas     & (fast/slow)  \\
	(1) & (2) & (3) & (4) & (5) & (6) & (7) & (8) & (9) & (10) \\  
        \hline 
        \multicolumn{10}{c}{Strong, widespread (post) starburst; see Figure~\ref{fig:pop_young}a} \\
        NGC\,3032 & 0.9 & 90 & N  & Y  & Y  & Y & Y & Y (sf) & F \\
        NGC\,3156 & 0.9 & 65 & N  & Y  & -  & Y & Y & Y & F \\
        NGC\,4150 & 1.2 & 77 & N  & Y  & Y  & Y & Y & Y & F\\
        NGC\,3489 & 1.8 & 98 & N  & Y  & -  & Y & Y & Y & F \\
        \multicolumn{10}{c}{Spatially constrained, central (post) starburst; see Figure~\ref{fig:pop_young}b }\\
        NGC\,4459 & 3.5 &168 & N  & Y  & Y  & Y & Y & Y (sf) & F \\
        NGC\,4382 & 3.7 &196 & Y  & N  & -  & N & N &(Y)& F  \\
        NGC\,4550 & 4.3 &110 & Y  & Y  & Y  & Y & Y & Y &(S) \\
        NGC\,4526 & 6.4 &222 & N  & Y  & Y  & Y & Y & Y (sf) & F \\
        NGC\,5838 & 9.3 &240 & Y  & N  & N  & Y & Y & Y & F \\
        NGC\,4477 & 9.7 &162 & Y  & Y  & N  & Y & Y & Y & F \\
        NGC\,4552 &10.2&252 & N  & N  & N  & N & Y &(Y)& S \\
        NGC\,524  &$>$12&235 & N  & Y  & N  & Y & Y & Y & F \\
        \multicolumn{10}{c}{Spatially extended signs of SSP-equivalent younger age; see Figure~\ref{fig:pop_young}c }\\
        NGC\,7457 & 2.8 & 78 & N  & N  & Y  & N & N &(Y)& F \\
        NGC\,7332 & 2.8 &125 & Y  & -  & -  & N & - & Y & F\\
        NGC\,2549 & 3.7 &145 & N  & N  & -  & - & N & Y & F\\
        NGC\,2685 & 5.1 & 96 & N  & Y  & -  & Y & Y & Y & F \\
        NGC\,3384 & 6.1 &145 & N  & N  & -  & N & N & Y & F \\
        NGC\,2699 & 6.4 &124 & N  & N  & N  & N & N & N & F \\
        NGC\,3377 & 6.7 &138 & Y  & N  & -  & N &(Y)& Y & F \\
        NGC\,4270 & 8.1 &122 & N  & N  & -  & N & N & N & F \\
        NGC\,5831 & 8.1 &151 & Y  & N  & N  & N & N &(Y)& S \\
        NGC\,2768 & 8.9 &216 & N  & Y  & N  & N & Y & Y & F \\
        NGC\,2974 & 8.9 &233 & Y  & N  & Y  & Y & Y & Y & F \\
        NGC\,5982 & 8.9 &229 & Y  & -  & N  & N & N &(Y)& S \\
        NGC\,474  & 9.3 &150 & N  & N  & Y  & N & N & Y & F \\
        NGC\,4387 &10.2 & 98 & N  & N  & N  & N & N & N & F  \\
        NGC\,4564 &10.7 &155 & Y  & N  & N  & N & N & N & F \\
        NGC\,5845 &11.2 &239 & N  & N  & N  &(Y)& Y & N & F \\
        \multicolumn{10}{c}{SSP-equivalent ages consistent with old stellar populations; see Figure~\ref{fig:pop_young}d} \\
        NGC\,4546 &10.7&194 & Y  & N  & N  & N & N & Y& F \\
        NGC\,4458 & 10.7& 85 & N  & N  & N  & N & N & N & S \\
        NGC\,821   & 11.2&189 & Y  & N  & N  & N & N & N & F\\
        NGC\,1023& 11.7&182 & N  & N  &(Y) & N & N & Y & F \\
        NGC\,3608 &11.7 &178 & Y  & N  & -  & N & Y &(Y)& S \\
        NGC\,4473 & 11.7&192 & Y  & N  & N  & N & N & N & F \\
        NGC\,4570 &11.7 &173 & Y  & N  & N  & N & N & N & F \\
        NGC\,5813 & 11.7&230 & N  & N  & N  & N & Y & Y & S \\
        NGC\,2695 &$>$12&188 & N  & N  & N  & N & N & N & F \\
        NGC\,3379 &$>$12&201 & N  & N  & -  & N & Y &(Y)& F\\
        NGC\,3414 &$>$12&205 & N  & N  & -  & N & N & Y& S \\
        NGC\,4262 &$>$12&172 & N  & N  & -  & N & N & Y& F \\
        NGC\,4278 &$>$12&231 & Y  & N  & N  & N & Y & Y & F \\
        NGC\,4374 &$>$12&278 & N  & N  & N  & N & Y & Y & S \\
        NGC\,4486 &$>$12&298 & N  & N  & N  & N & Y & Y & S \\
        NGC\,4621 &$>$12&211 & Y  & N  & N  & N & N &(Y)& F \\
        NGC\,4660 &$>$12&185 & Y  & N  & -  & N & N & N & F \\
        NGC\,5198 &$>$12&179 & N  & N  & N  & - & N &(Y)& S \\
        NGC\,5308 &$>$12&208 & N  & N  & N  & - & N & N & F \\
        NGC\,5846 &$>$12&238 & N  & N  & N  & N & Y & Y & S \\
        \hline 
    \end{tabular} 
    \end{center}
    
        {\em Notes:}\/ This table provides a summary of the star-formation indicators used in the \sauron\/ survey. We group galaxies in four major categories and sort them by increasing SSP-equivalent age in each category. Column 1 lists the galaxy name. Column 2 the SSP-equivalent age over a circular aperture with radius \reb\/ (see also Table~\ref{tab:stellpop}). Column 3 gives the average velocity dispersion over an aperture of $R_{\rmn{e}}$. Column 4 provides information on the shape of the \mgb\/ contours compared to the isophotes (see Paper~VI). Column 5, 6 and 7 indicate detections for molecular gas (CO), NUV star-formation signatures and mid-IR star-formation detections from the studies of \citet{comb07}, Paper~XIII and Paper~XV, respectively. The presence of optical dust and ionized gas emission is indicated in columns 8 and 9 (see Papers V \& XVI). Detections given in brackets ['(Y)'] are marginal, while 'Y (sf)' indicates signs of ongoing star-formation as detected by the analysis of the gas emission in Paper~XVI. Column 10 gives our classification into slow and fast rotators (see Paper IX for details). Since NGC\,4550 features two counter-rotating disks it shows little overall rotation and hence is classified as slow rotator.
  \end{minipage}
\end{table*} 

Another group of galaxies exhibit a more localized central region of
young stars embedded within an otherwise older stellar system (see
maps and Figure~\ref{fig:pop_young}b; NGC\,4459, NGC\,4382, NGC\,4526,
NGC\,4550, NGC\,4552, NGC\,4477, NGC\,5838, and NGC\,524). In all but
two cases (NGC\,4382, NGC\,4552), the recent star-formation event --
with sometimes residual ongoing star-formation seen in the NUV, the
mid-IR and the \sauron\/ gas emission analysis (see
Table~\ref{tab:starform}) -- is located in a thin, embedded disc
clearly visible in the emission line maps and the unsharp-masked
optical dust images of Paper~V. These thin stellar discs are typically
co-rotating with the main body of the galaxy albeit at the distinctly
higher rotational speeds of disks. The metallicity maps (see also
Section~\ref{sec:metal_map}) of these galaxies show a spatially
coincident region of enhanced metallicity, indicating the signs of
ongoing metal enrichment due to disc star-formation. NGC\,524 is
perhaps the least obvious member of this group of galaxies. It shows
only very mild signs of younger stars in the center\footnote{The S/N
  per observed spatial element was particularly low for this galaxy
  and for radii larger than 20\arcsec\/ with large bins the line
  strength indices are of modest quality.}, but it features a
prominent face-on optical dust disk with a clear spiral pattern (see
Paper~V). If there are young stars associated with this thin disc, the
near face-on orientation may dilute their contribution to the
integrated line-of-sight. CO is detected in four out of six cases,
with recent interferometric CO observations of NGC\,4459, NGC\,4526,
NGC\,4550 revealing molecular gas discs co-rotating with the stellar
discs seen in the \sauron\/ observations \citep{young08,crock09}. Also
the {\em Spitzer} IRAC observations (Paper~XV) clearly support the
scenario of star-formation in well defined thin disks for this group
of galaxies with detections for the majority of observed galaxies (see
Table~\ref{tab:starform}.)

The SSP-equivalent ages derived for the regions affected by
star-forming disks in the galaxies above mentioned range from about
3\,Gyr to ages barely below the age of the universe. This, of course,
is a reflection of the luminosity weighting and involved mass
fractions.  Typically these disks represent only a minor mass fraction
of the total galaxy mass \citep[e.g.,][]{crock09}, but contain
contributions of very young ($\sim$ 100-300\,Myr) stars \citep[see
also][]{kav09}. We note that these thin, embedded disks with recent
star-formation typically reside in galaxies of intermediate mass range
with $\sigma_{\rm e} = 160 - 240$\,\kms\/ ($M_\rmn{dyn} = 1 - 25
\times 10^{10} M_{\sun}$; see also Paper~XV).

For the galaxies with circumnuclear star-forming disks or rings,
Paper~XV suggests a scenario where they experience a transient period
of renewed star-formation on top of an underlying much older
population. The origin of the gas may be internal but also very minor
mergers are conceivable. This process is a good candidate for
producing the central, rapidly rotating and metal rich stellar disks
seen in many of our fast rotating galaxies (see Paper~XII and
Section~\ref{sec:metal_map}).

There are two galaxies with a localized region of young stars
(NGC\,4382 and NGC\,4552) lacking a detection of young stars in the
NUV and the mid-IR.  However, optical HST photometry and ground based
imaging provide evidence for a recent merger in the case of NGC\,4382
\citep{lau05}.  In NGC\,4382, the unusual bow-tie like distribution of
the young stars also seen in an enhanced region of the metallicity map
can be matched to the stellar kinematics, which show a marked twist in
these areas with a strong depression in the velocity dispersion map
(see Figure~\ref{fig:maps6} and Paper~III), together with a central
counter-rotating component as seen in the higher spatial resolution
data of Paper~VIII. We interpret this as a kinematically distinct
stellar component (see also Paper~XII) made of stars from a recent,
but more than $\sim$1\,Gyr ago star-formation episode. The ratio of
mass involved in this star-formation event to the total mass of the
galaxy may be significantly higher compared to the mass ratio of the
young stars in the embedded, thin disks seen in the other galaxies of
this group.

For NGC\,4552 the observed, mild increase in \hb\/ absorption
strengths and thus inferred decrease in SSP-equivalent age of the
central region is not supported by any of the other recent
star-formation indicators probed by our survey (see
Table~\ref{tab:starform}). Although this galaxy is known to exhibit a
significant UV-upturn (Bureau et al. in prep.), it is unlikely that
the \hb\/ increase originates from this since none of the other slow
rotator galaxies that show a UV-upturn (NGC\,4374, 4486, 5198. 5846,
5982; Bureau et al. in prep.) exhibit a clear central increase in
\hb\/ absorption strength (see Paper~VI) or corresponding decrease in
SSP-equivalent age. Due to the significant amount of gas emission
present in the center of this galaxy (see Papers~V \& XVI), it is also
possible that the \hb\/ absorption has been somewhat over-corrected.

Typically star-formation and the resulting younger SSP-equivalent ages
are found towards the center of early-type galaxies and star-formation
takes place in a thin disk/ring morphology (see
Figure~\ref{fig:pop_young}a-c and also Paper~XV).  However, there is
also evidence for star-formation outside the nuclear regions in two
galaxies. (1) NGC\,474 shows an age map with a relatively old central
part ($\sim$9\,Gyr) while the SSP-equivalent age becomes younger
towards larger radii ($r \ge 6$\arcsec\/). This observation is
supported by blue NUV-optical colours at the same radii (see
Paper~XIII) and is perhaps connected to the famous shell structures of
NGC\,474 \citep{turn99}.  (2) NGC\,2974 shows significant
star-formation in a ring like structure, both clearly exhibiting signs
of recent star-formation in the NUV and in the mid-IR, outside the
observed field-of-view (hereafter FoV) of the \sauron\/ observations
\citep[][Paper~XV]{jeo07}. There is also HI detected in a ring like
configuration \citep{wei08}. We detect only mild signs of younger
stellar populations.

Besides the galaxies with strong signs of recent star-formation or
spatially constrained star-formation connected to thin, embedded
disks, there is a sizable group of galaxies ($\sim$\,30 per cent of
the sample) that show intermediate SSP-equivalent ages (3 - 11\,Gyr)
over large parts of the observed FoV (in increasing order of central
\reb\/ SSP-equivalent age: NGC\,7457, NGC\,7332, NGC\,2549, NGC\,2685,
NGC\,3384, NGC\,2699, NGC\,3377, NGC\,4270, NGC\,5831, NGC\,2768,
NGC\,5982, NGC\,4387, NGC\,4564, NGC\,5845). With only two exceptions
(the polar ring galaxy NGC\,2768 \citep{crock08a} and NGC\,2685) these
galaxies are not detected in CO, indicating that they have exhausted
their molecular gas supply and are now evolving passively.  Moreover,
neither the dust maps nor the NUV or mid-IR indicators show evidence
for recent ($<$1\,Gyr) star-formation. This is consistent with a
star-formation episode in these galaxies which is long enough ago to
be devoid of dust and very young stars, but recent and massive enough
to be traced by the \hb\/ index and thus our SSP-equivalent age
maps. For these galaxies (see also Figure~\ref{fig:pop_young}c) we
find typically flat age maps or only mildly increasing SSP-equivalent
age as function of radius.

Last, but not least we have the group of galaxies that show line
strength indices consistent with overall old stellar populations, at
least within the sensitivity of our line strength observations and
stellar population analysis (Figure~\ref{fig:pop_young}d). Eight out
of these twenty galaxies are classified as slow rotators in Paper~IX
and it is indeed the group of galaxies with old stellar populations
where we find most of the slow rotators (see also
Section~\ref{sec:central}). Two galaxies (NGC\,2695 for larger radii
and NGC\,4486 for all radii) show \hb\/ line strength which are below
the stellar population model predictions and therefore lead to
non-physically large SSP-equivalent ages. A thorough check of the
data-reduction procedure revealed nothing special for these galaxies
and thus we can offer no explanation to why the \hb\/ line strength is
so weak for these two galaxies.

We note that, when explored at higher spatial resolution, individual
galaxies in the 'old' group can show signs of recent star-formation.
For example, HST imaging of NGC\,4570 \citep{ems98} reveals evidence
for a bar driven ring of young stars ($\sim$2\,Gyr) at a radius of
1\farcs7 at the presumed location of a bar inner Lindblad resonance
(ILR).

It is interesting to explore the stellar populations of the class of
early-type galaxies with kinematically decoupled components (KDCs). We
define them here as sub-components that either rotate around a
different axis to the main galaxy, or that rotate around the same
axis, but with an opposite sense of rotation and thus are distinct
from the young discs discussed above. In Papers~III, VIII and XII, 13
galaxies of the \sauron\ survey were identified to harbor
KDCs. Broadly speaking there appear to be two types of KDCs:
kiloparsec-scale structures which are found almost exclusively in slow
rotating galaxies with predominately old stellar populations
(NGC\,3414, NGC\,3608, NGC\,4458, NGC\,5198, NGC\,5813, NGC\,5831,
NGC\,5982); and more compact KDCs showing a range in SSP-equivalent
ages with a preference towards more recent star-formation episodes
(NGC\,3032, NGC\,4150, NGC\,4382, NGC\,4621, NGC\,7332, NGC\,7457; see
Paper~VIII). While the central, thin discs discussed above show a
prominent connection between young stars, optical dust, nebular
emission, NUV and mid-IR star-formation indicators and enhanced
metallicity, it is remarkable that most of the large scale KDCs do not
show an equivalent, significant signature in the stellar population
maps. NGC\,5831 and NGC\,5982 do exhibit mild signs of a contribution
from young stars, but the spatial location of the KDCs and that of the
young stars is not well correlated. Additionally, the metallicity maps
(see Section~\ref{sec:metal_map}) also do not show a significant
correlation with the rotation direction of the KDC, but rather follow
a normal, smooth radial behavior. In summary, any assembly and
star-formation scenario designed to explain the observed large-scale
KDCs must account for the absence of significant stellar population
parameter signatures (i.e. age, metallicity or abundance ratio).
 
This can be achieved, for example, if the formation process was
completed a long time ago \citep[e.g.,][]{dav01}. However, other
possibilities are being explored too.  A recent dynamical study
\citep{vdB08} applying triaxial orbit-based models to the prominent
KDC galaxy NGC\,4365 \citep[see also][]{ben88a,dav01} shows that the
observed KDC is not physically distinct from the main body, but rather
caused by a superposition of orbits. A more systematic study of the
dynamics of KDC galaxies is in preparation (van den Bosch et al.) but
it may well be that a significant fraction of the observed large scale
KDCs are not dynamically distinct subcomponents, but rather are the
result of a projection of orbits from a more extended structure, thus
naturally explaining the absence of any distinct stellar population
signatures.

\subsubsection{Metallicity and abundance ratio maps}
\label{sec:metal_map} 
For galaxies with old stellar populations, the abundance ratio maps
often show to first order no or only mild positive radial gradients
with values of [$\alpha$/Fe] $\simeq 0.2 - 0.4$ (red to yellow colour
shading). The most significant structures, over and above the
generally rather noisy appearance of the maps, are typically connected
to the presence of (very) young stellar populations, which lead to
lower, more solar values of [$\alpha$/Fe] (see e.g., NGC\,3156,
NGC\,4150, and NGC\,7457). The metallicity maps generally show
negative gradients with increasing radius, often roughly consistent
with the morphology of the light profiles and consistent with many of
the earlier long slit studies \citep[e.g., ][]{dav93,car93,san07}. Two
remarkable outliers from this trend exist: NGC\,3032 and NGC\,4150,
showing a central peak in metallicity with a surrounding ring of
depressed metallicity. These two galaxies are among the most extreme
examples of a recent starburst in our sample (see
Table~\ref{tab:starform}). Due to the degeneracies between
star-formation age and young star mass fractions, as well as the
biases described in Section~\ref{sec:interpretation}, the stellar
population maps of these galaxies are difficult to interpret and a
simple interpretation may be misleading.

Mean metallicity and [$\alpha$/Fe] gradients averaged along isophotes
are described in more detail in Section~\ref{sec:gradients} while we
concentrate in this Section on the two-dimensional aspects of the
maps.

\begin{figure*}
  \includegraphics[height=135mm,viewport=25 0 95 580,clip]{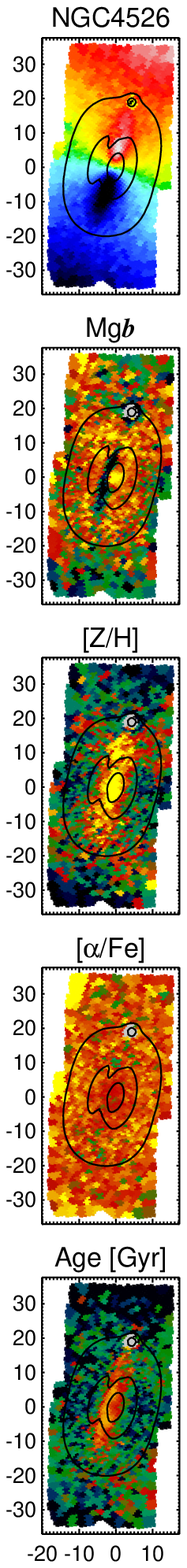}
  \includegraphics[height=135mm,viewport=25 0 120 580,clip]{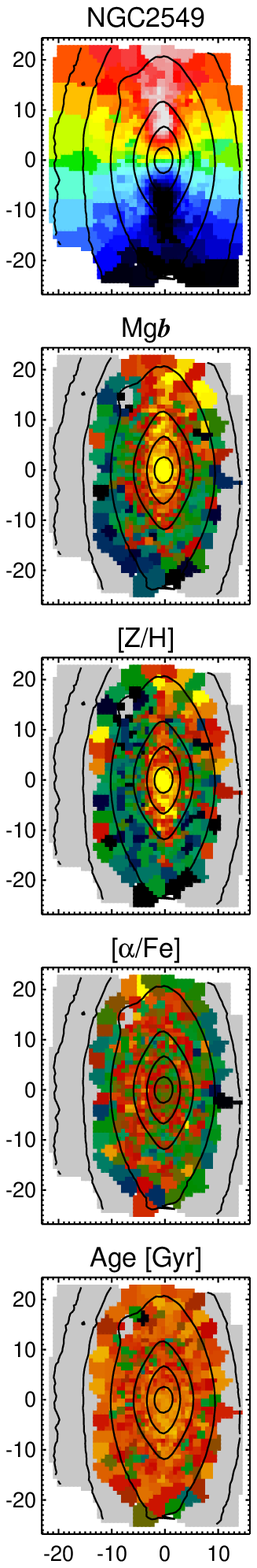}
  \includegraphics[height=135mm,viewport=25 0 150 580,clip]{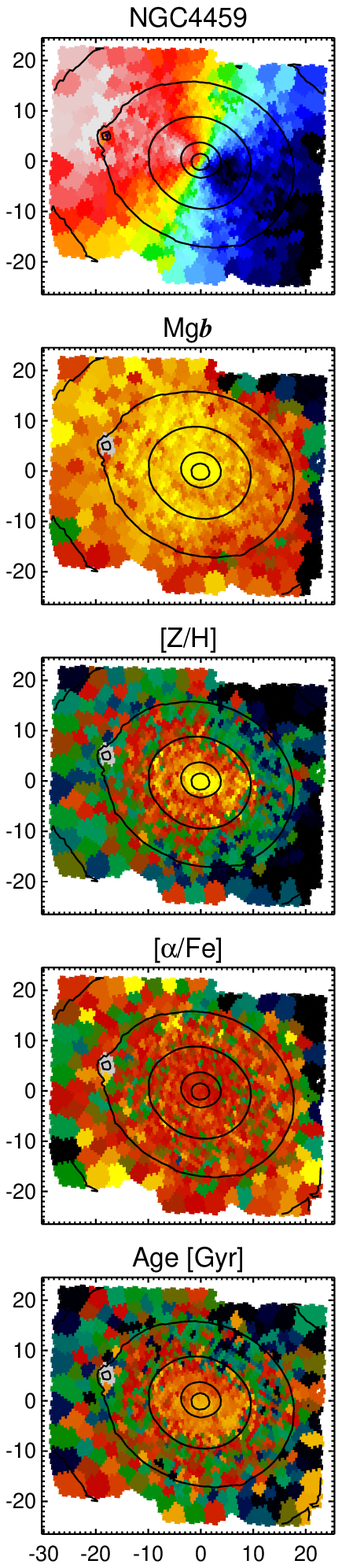}
  \includegraphics[height=135mm,viewport=25 0 115 580,clip]{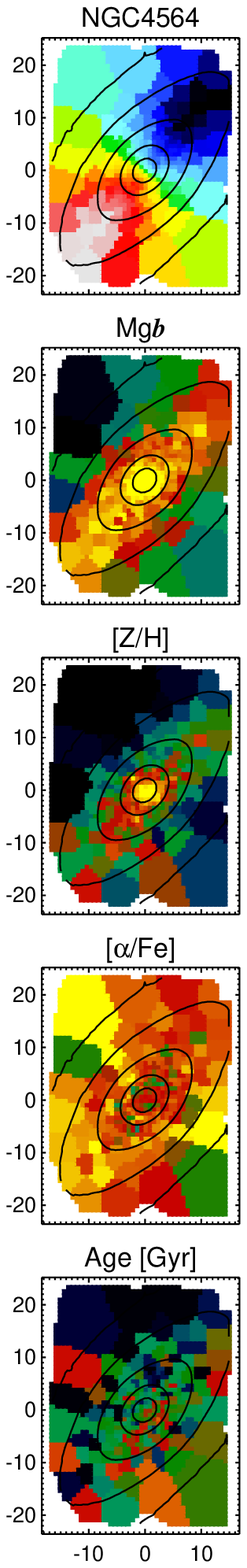}
  \includegraphics[height=135mm,viewport=25 0 140 580,clip]{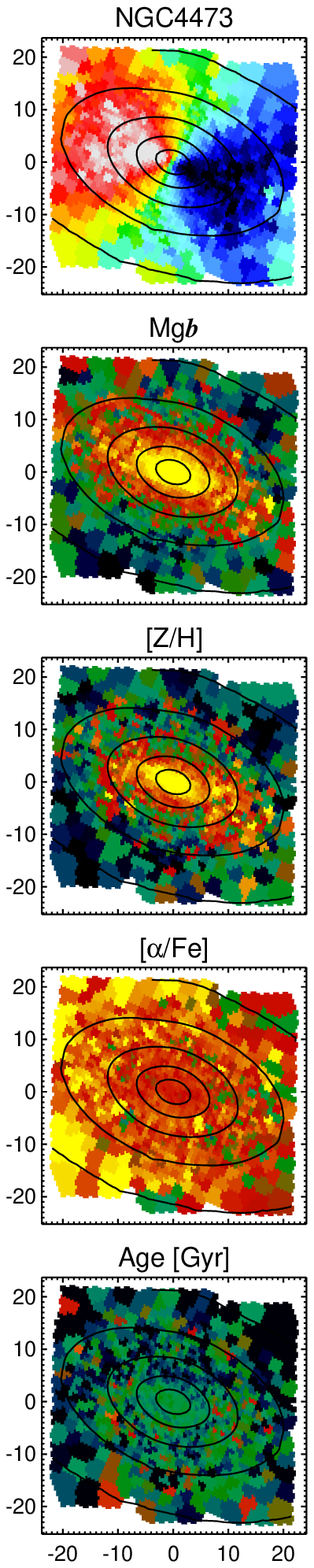}
  \includegraphics[height=135mm,viewport=25 0 150 580,clip]{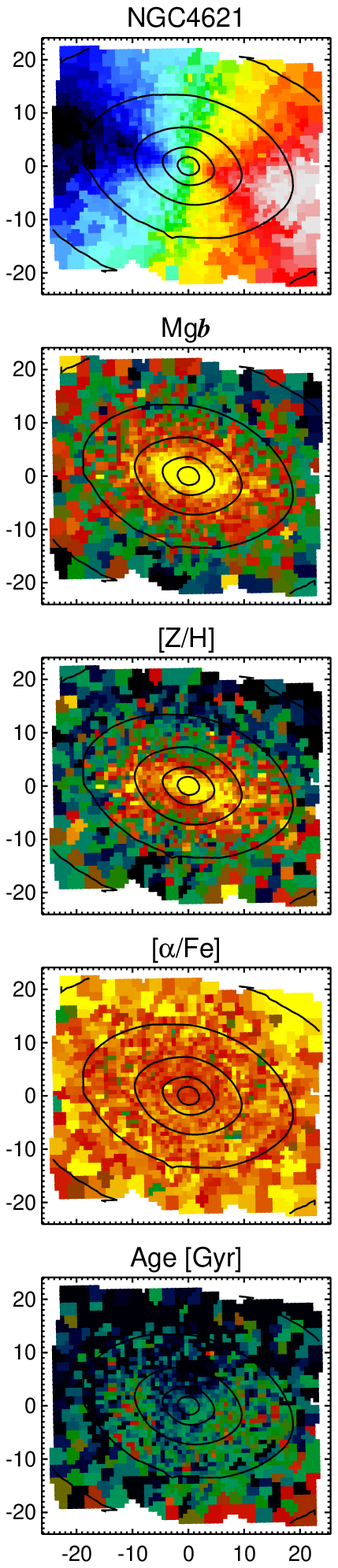}
  \includegraphics[height=135mm,viewport=25 0 115 580,clip]{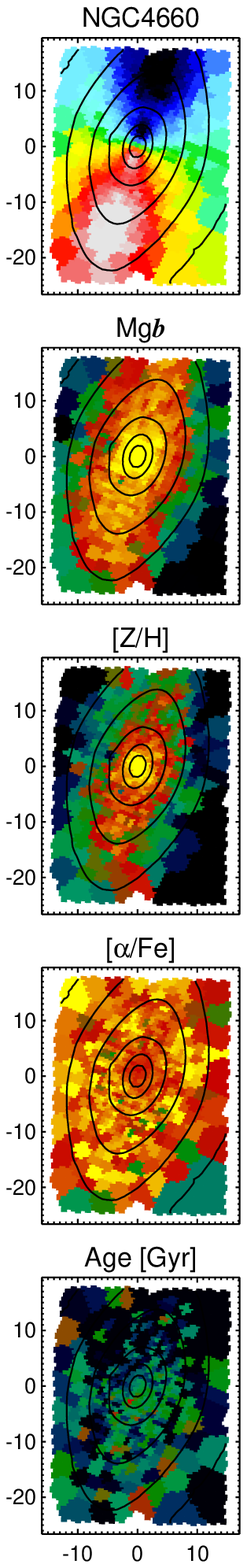} 

  \caption{Prominent examples of fast rotators with a distinct,
    central disk-like component. For each galaxy we show from the top,
    the stellar velocity field, the \mgb\/ line strength map (see
    Paper~VI), the metallicity map, the abundance ratio map
    [$\alpha$/Fe] and the SSP-equivalent age map. The disk-like
    components can be clearly seen in the velocity fields and are
    reflected in enhanced \mgb\/ and [Z/H] regions and mild
    depressions in the [$\alpha$/Fe] maps.  NGC\,4526 and NGC\,4459
    are examples of currently ongoing residual star-formation in a
    disk.  Bins shown in grey colour for NGC\,2549 indicate regions of
    low S/N ratio and thus doubtful line strength measurements. Note
    the varying spatial scales shown for different galaxies; units are
    given in arcsec.}
  \label{fig:disks}
\end{figure*}

In Paper~VI we found that the \mgb\/ isoindex contours appear to be
flatter than the isophotes of the surface brightness for about 40 per
cent of our galaxies without significant dust features (also see
Table~\ref{tab:starform}). This flattening is almost always associated
with significant rotation in the galaxies (see also Paper~XII) and we
inferred from this that the fast-rotating component features a higher
metallicity and/or an increased [$\alpha$/Fe] ratio compared to the
galaxy as a whole. With the stellar populations maps presented in this
study we confirm that the metallicity is enhanced for fast rotating
systems such that the iso-metallicity contours appear more flattened
than the isophotes (prominent examples include NGC\,4564, NGC\,4473,
NGC\,4621 and NGC\,4660; see Figure~\ref{fig:disks}). However, the
region of enhanced metallicity is typically connected with a mild {\em
  decrease}\/ in our abundance ratio maps (see also
Figure~\ref{fig:disks} and~\ref{fig:pop_grads1}).

For individual galaxies such as NGC\,2549 and NGC\,3377 the region of
increased metallicity is connected to the presence of younger stellar
populations in a rotating structure with disk-like kinematics. For
other galaxies there is only mild (e.g., NGC\,4564) or no (e.g.,
NGC\,4660) evidence for younger populations. It is tempting to
associate these flattened regions of enhanced metallicity and somewhat
lower abundance ratios with the more recent star-formation seen in
thin disks in galaxies such as NGC\,4459 and NGC\,4526 (see
Figure~\ref{fig:disks}). Taken together, a scenario emerges in which
the star-formation event that created the fast rotating disc-like
structure is distinct from the formation of the galaxy main body,
which took place at a later stage and on somewhat longer timescales,
leading to an increased metallicity and reduced abundance ratio value.

\citet{nor06} demonstrated in their detailed study of the lenticular
galaxy NGC\,3115, that a simple spheroid + disc formation model, where
the old spheroid has [$\alpha$/Fe]= 0.3 and the somewhat younger disc
shows close to solar abundance ratios, works well to explain the
observed stellar population trends in this galaxy.

In the \sauron\/ sample, we find good evidence for this secondary
star-formation in disk-like structures for low to intermediate mass
($\sigma_{\rm e} = 100 - 160$\,\kms), fast rotator galaxies such as
NGC\,3377 and NGC\,4550 where the young stars are still prominently
seen in the SSP-equivalent age maps affecting large regions of the
galaxy. Can we also find evidence of this type of secondary
star-formation in older galaxies?

\begin{figure}
 \includegraphics[width=84mm]{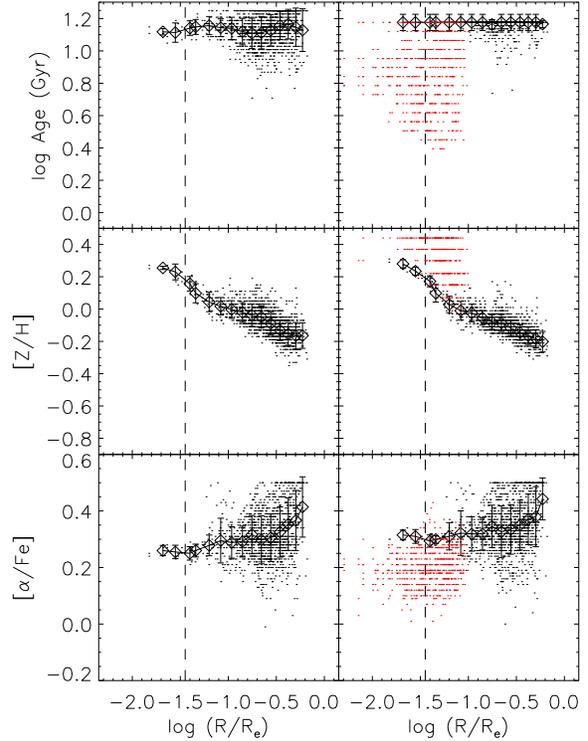}
 \caption{Radial gradients of stellar population estimates for the
   fast rotator galaxy NGC\,4621. The black dots represent individual
   bins while the solid line connects values averaged along isophotes
   where radial bins are indicated by open diamonds. The stellar
   population estimates are shown as derived from \citet[][left
   column]{schia07} and \citet[][right column]{TMB03} models. The red
   dots show the results from high spatial resolution OASIS
   observations (see Paper~VIII) where the stellar population
   estimates were derived with \citet[][]{TMB03} models, however,
   using a larger set of Lick/IDS indices, namely \hb, Fe5015, \mgb,
   Fe5270, Fe5335 and Fe5406.  The vertical dashed lines indicate a
   radius of 2\arcsec. }
 \label{fig:pop_grads_core}
\end{figure}

For more massive galaxies ($\sigma_{\rm e} > 160$\,\kms) the disk
formation and increased metallicity is typically constrained to a
central location involving only a minor fraction of the total mass of
the galaxy (e.g., NGC\,4459, NGC\,4526;
Figure~\ref{fig:disks}). However, examining the [$\alpha$/Fe] maps
more carefully, one can see central depressions in about 25 per cent
of the galaxies (e.g., NGC\,821, NGC\,1023, NGC\,2699, NGC\,4270,
NGC\,4387, NGC\,4473, NGC\,4546, NGC\,4564, NGC\,4570, NGC\,4621,
NGC\,4660, NGC\,5838). All of these galaxies are classified as fast
rotators and exhibit overall old ($> 9$\,Gyr) stellar populations,
while the depression in [$\alpha$/Fe] is highly correlated with an
enhancement of the metallicity estimates in the same spatial position.
A careful examination of the age maps often also reveals a weak
signature of a contribution from a younger stellar population in these
central regions as compared to the outer regions.  However, the most
important connection to mention is that all of the galaxies with
central depressions in [$\alpha$/Fe] show pinched iso-velocity
contours towards the center indicating the existence of a central fast
rotating, disk-like component.\footnote{We note that due to
  inclination effects, edge-on disks in the center are more visible
  than face-on examples.} Remarkably, all galaxies of the above group
that are included in the analysis of HST imaging by \citet[][6/12
galaxies]{lau05} reveal the signature of a stellar disk in the central
regions. Figures~\ref{fig:pop_grads_core} and \ref{fig:pop_grads1}
show radially averaged stellar population gradients for selected
galaxies where the central depressions in [$\alpha$/Fe] are more
apparent.

The kinemetry analysis of our velocity maps presented in Paper~XII
suggested that all fast rotators contain flattened, fast-rotating
components contributing at various levels to the total light of the
galaxy. For example, there are disk dominated galaxies such as
NGC\,3156 and NGC\,2685 and more bulge dominated systems such as
NGC\,821 and NGC4660 (see Paper~XII), the latter showing largely old
stellar populations. Furthermore, 70 per cent of the sample shows
evidence for distinct, multiple kinematic components.

A picture emerges where essentially all fast rotator galaxies have
seen a period (often multiple periods) of secondary star-formation in
kinematically disk-like components on top of an older, spheroidal
component. This picture is perhaps well known and accepted for
classical S0s with small bulge to disk ratios
\citep[e.g.,][]{fis96,nor06}. However, we find sub-components with
disk-like kinematics and distinct stellar populations in all fast
rotators ranging from galaxies classified as Sa to (elongated)
ellipticals: (a) central very young, still star-forming disks in e.g.,
NGC\,4526 (Sa), NGC\,4459 (S0; Papers~XIII and XV), (b) intermediate
age systems devoid of any ongoing star-formation but clear detections
of a younger, metal rich and [$\alpha$/Fe] depressed component with
disk-like kinematics as seen in NGC\,2699 (E) and NGC\,3384 (SB) and
(c) in the systems with largely old stellar populations where the
star-formation event connected to the disk-like kinematics is mostly
evidenced by a raised metallicity in combination with lower
[$\alpha$/Fe] ratios as seen in e.g., NGC\,4621 (E5) and NGC\,4660
(E5). Figure~\ref{fig:disks} presents an overview of selected fast
rotators with metal rich, but [$\alpha$/Fe] depressed disk-like
components ranging from still star-forming disks as found in NGC\,4526
to overall old stellar populations as seen perhaps in NGC\,4660.

The frequent disk-like structures observed at various scales in our
sample of fast-rotating early-type galaxies suggests dissipative
processes play a major role in their formation.  The very central
stellar disks seen in HST imaging and the somewhat larger disk-like
kinematic structures with increased metallicity and depressed
[$\alpha$/Fe] seen in our observations of fast rotators match well
with the recurring but transient circumnuclear star-formation seen in
the mid-IR {\em Spitzer} observations described in Paper~XV. The
rotational support on larger scales for {\em low mass}\/ galaxies,
which are also often showing signs of younger stellar populations is
probably connected to gas-rich (major) mergers adding a widespread,
fast rotating younger component to the galaxies \citep[see
also][]{hof09}. It is not obvious that all fast rotating components on
larger scales in early-type galaxies can be produced in this
way. However, recent IFU observations for $z = 2 \sim 3$ galaxies
revealed striking evidence for a transient phase of clumpy, turbulent
disk formation fed by rapid cold flows \citep{gen08}. The old, {\em
  large-scale}\/ disk-like structures seen in our sample of galaxies
could well be the fossil relic of such a cold flow mode of
star-formation seen at higher redshift. The rotational support and the
relatively short timescale ($\le1$\,Gyr) of the secular evolution in
these ancient turbulent gas-rich disks seems consistent with our
observations \citep[see also][]{elm08}.

The slow rotators in our sample (excluding NGC\,4550 with its two
counter-rotating disks) mostly show flat [$\alpha$/Fe] maps with
metallicity gradients consistent with the isophotes. This supports
their likely different formation path as already argued in Papers~IX,
X and XII. There are however, four slow rotators which show abundance
ratio maps indicating a somewhat depressed [$\alpha$/Fe] ratio towards
the center: NGC\,4552, NGC\,5198, NGC\,5831 and NGC\,5982. This group
of galaxies, which partially also shows weak signs of a contribution
from a younger stellar population, may indicate that the class of slow
rotators is perhaps not as homogeneous in their stellar populations as
their kinematics may suggest.

A few words of caution with regard to the results presented so far are
due.  We note that as shown in Paper~VI, the \mgb\/ index is a very
good {\em observational}\/ tracer of the increased metallicity of the
disk-like structures. However, in the presence of young stellar
populations (SSP-equivalent age $\le 3$\,Gyr) the \mgb\/ index reacts
significantly to age and this causes a decrease in absorption strength
compared to an old population with the same metallicity. For example,
NGC\,4382 (see Paper~VI) shows a central \mgb-index depression which
renders the \mgb\/ index a less useful tool to detect metallicity
enhancements connected to disk-like kinematics in these kind of
galaxies.

Using the Thomas et al. models instead of the Schiavon models does not
alter our main findings and conclusions discussed above. In fact, the
signs of distinct stellar populations in the regions of disk-like
kinematics would be more pronounced in metallicity for galaxies with
old stellar populations. In Figure~\ref{fig:pop_grads_core} we show
results for radially averaged stellar population estimates derived
from both stellar population models for NGC\,4621 (for details of the
procedure see Section~\ref{sec:gradients}).
 
The main limitation of our abundance ratio estimates is the limited
number of indices available for the analysis. New observations
covering large regions of the galaxies as well as a longer wavelength
range, thus including more indices, are needed to confirm the above
findings.

 %
 %
 \renewcommand{\thefigure}{\arabic{figure}\alph{subfigure}}
 \setcounter{subfigure}{1}

 \begin{figure*}
 \begin{center} 
   \includegraphics[draft=false,scale=0.99,trim=0cm 1.cm 0cm 0cm]{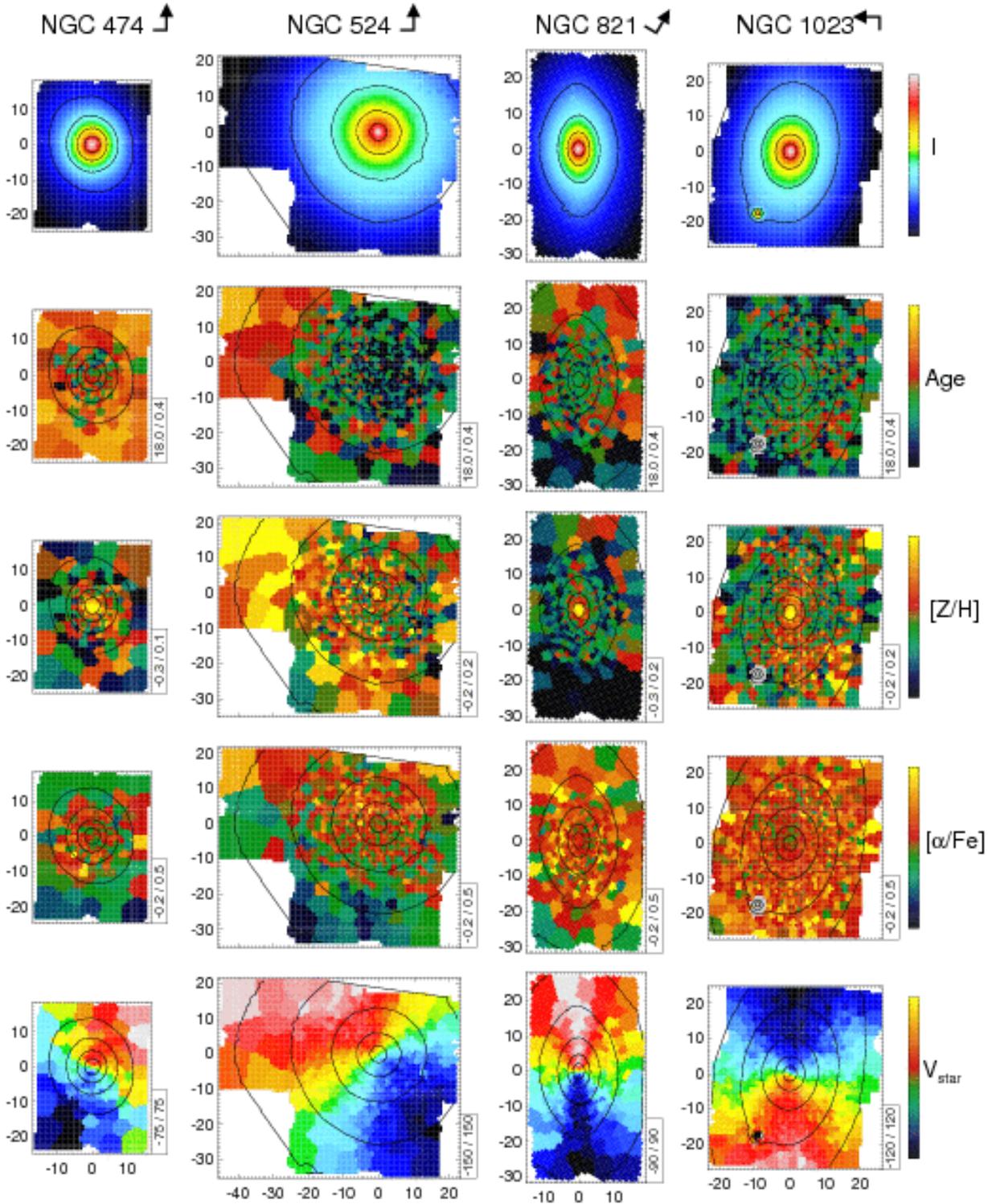} 
 \end{center} 
 \caption[]{Maps of the derived stellar populations of the 48
   early-type galaxies in the \sauron\ representative sample. The
   \sauron\ spectra have been spatially binned to a minimum
   signal-to-noise of 40 by means of the centroidal Voronoi
   tessellation algorithm of \protect\citet{cap03}. All maps have the
   same spatial scale and the units of x- and y-axis are arcsec. From
   top to bottom: i) reconstructed total intensity; ii) SSP-equivalent
   age distribution (the age scale is given in units of Gyr and a
   linear colour scale); iii) metallicity ([Z/H]) distribution and iv)
   [$\alpha$Fe] distribution. The bottom panel in each column
   reproduces the velocity maps from Paper~III.}
 \label{fig:maps1}
 \end{figure*}

 \addtocounter{figure}{-1}
 \addtocounter{subfigure}{1}

 \begin{figure*}
 \begin{center} 
   \includegraphics[draft=false,scale=0.99,trim=0cm 0.cm
     0cm 0cm]{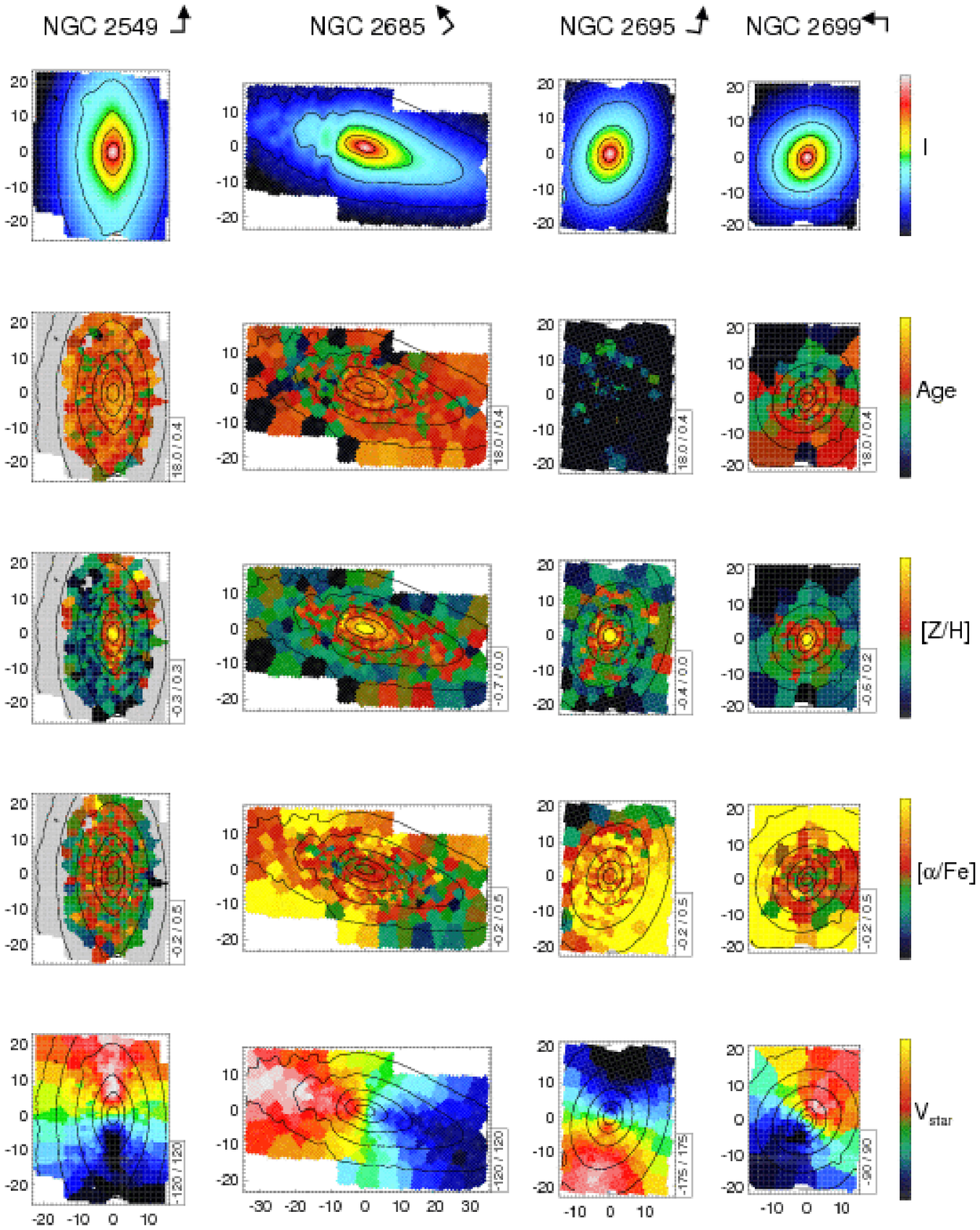} 
 \end{center} 
 \caption[]{}
 \label{fig:maps2}
 \end{figure*}

 \addtocounter{figure}{-1}
 \addtocounter{subfigure}{1}

 \begin{figure*}
 \begin{center} 
   \includegraphics[draft=false,scale=0.99,trim=0cm 0.cm
     0cm 0cm]{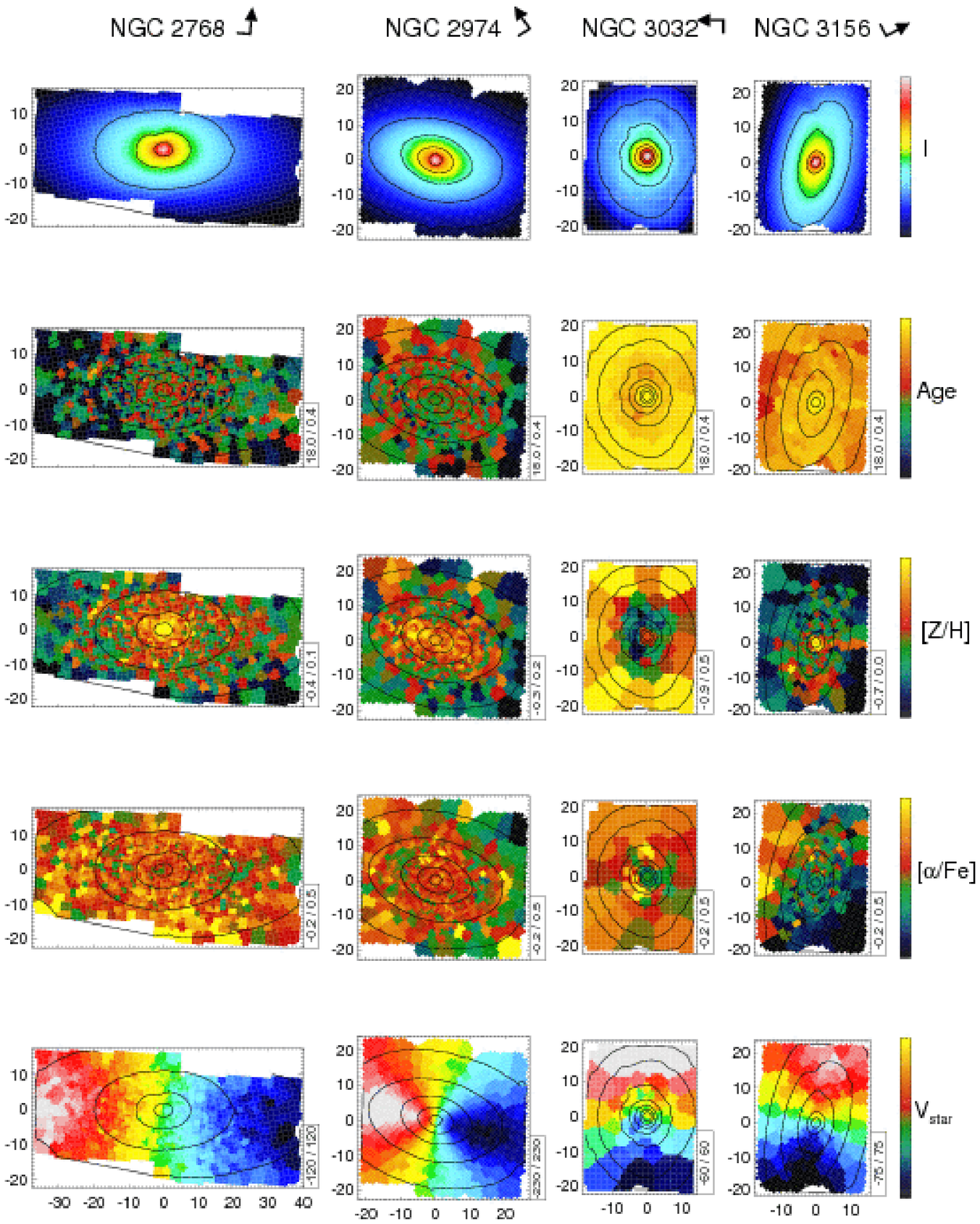} 
 \end{center} 
 \caption[]{}
 \label{fig:maps3}
 \end{figure*}

 \addtocounter{figure}{-1}
 \addtocounter{subfigure}{1}

 \begin{figure*}
 \begin{center} 
   \includegraphics[draft=false,scale=0.99,trim=0cm 0.cm
     0cm 0cm]{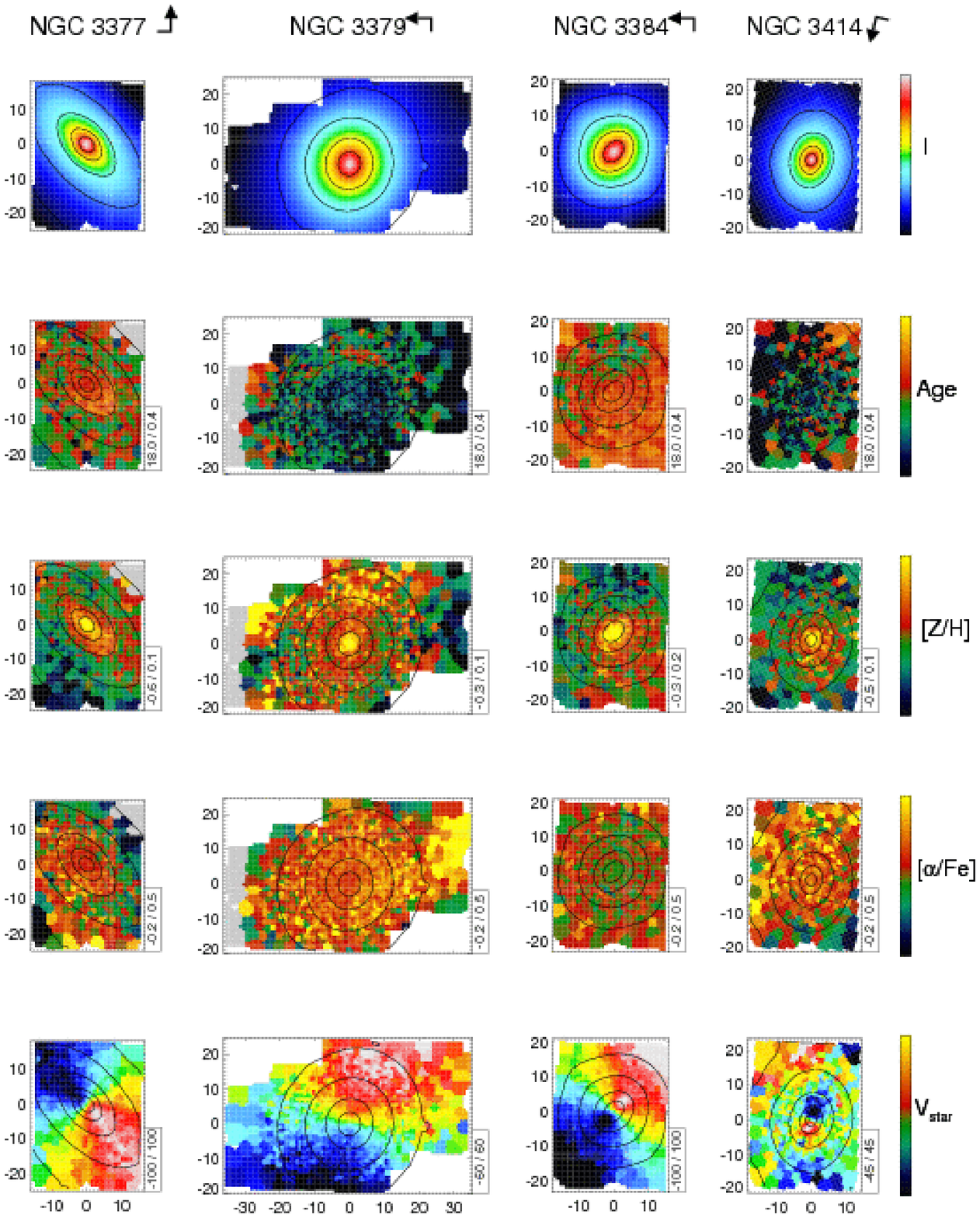} 
 \end{center} 
 \caption[]{}
 \label{fig:maps4}
 \end{figure*}

 \addtocounter{figure}{-1}
 \addtocounter{subfigure}{1}

 \begin{figure*}
 \begin{center} 
   \includegraphics[draft=false,scale=0.99,trim=0cm 0.cm
     0cm 0cm]{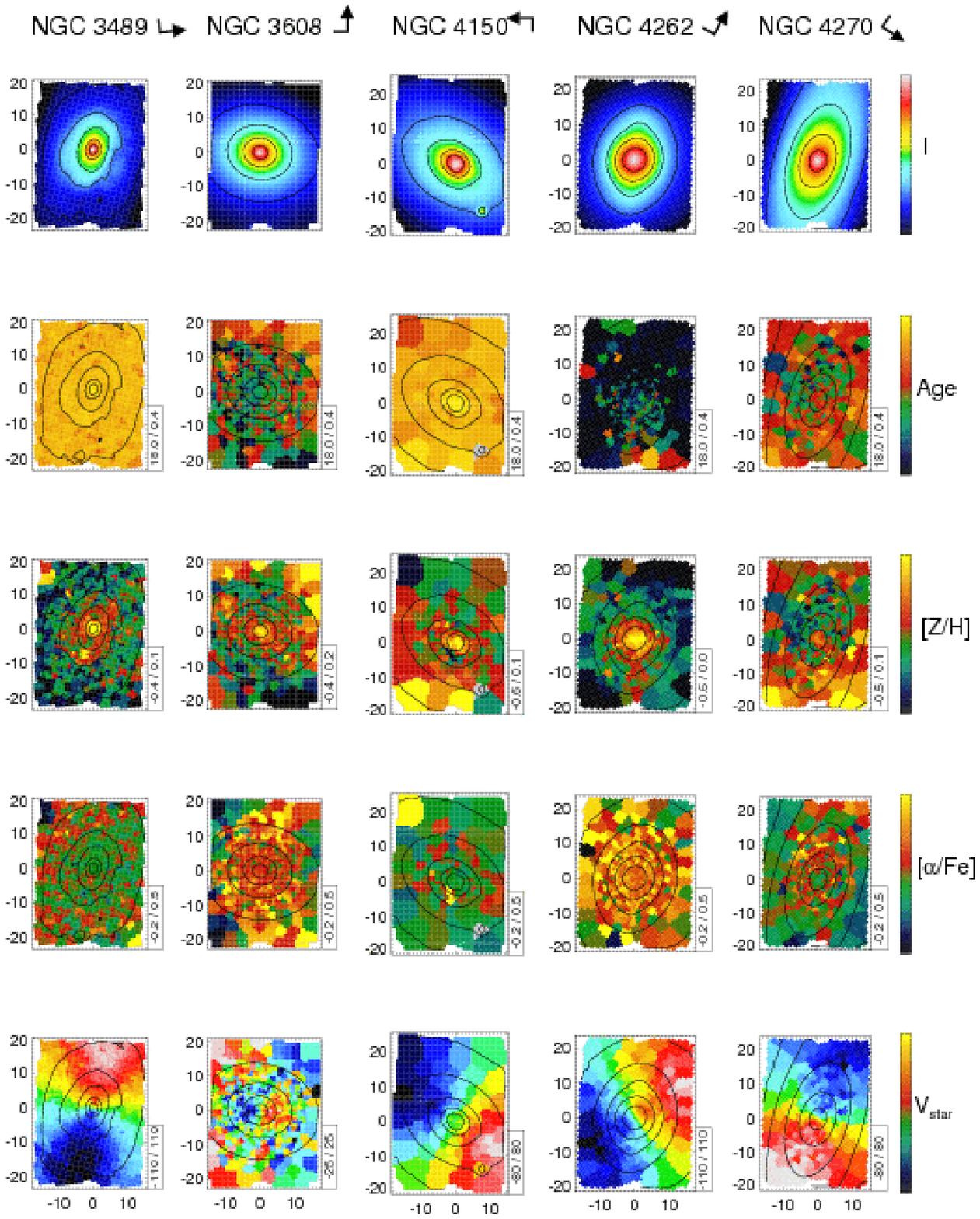} 
 \end{center} 
 \caption[]{}
 \label{fig:maps5}
 \end{figure*}

 \addtocounter{figure}{-1}
 \addtocounter{subfigure}{1}

 \begin{figure*}
 \begin{center} 
   \includegraphics[draft=false,scale=0.99,trim=0cm 0.cm
     0cm 0cm]{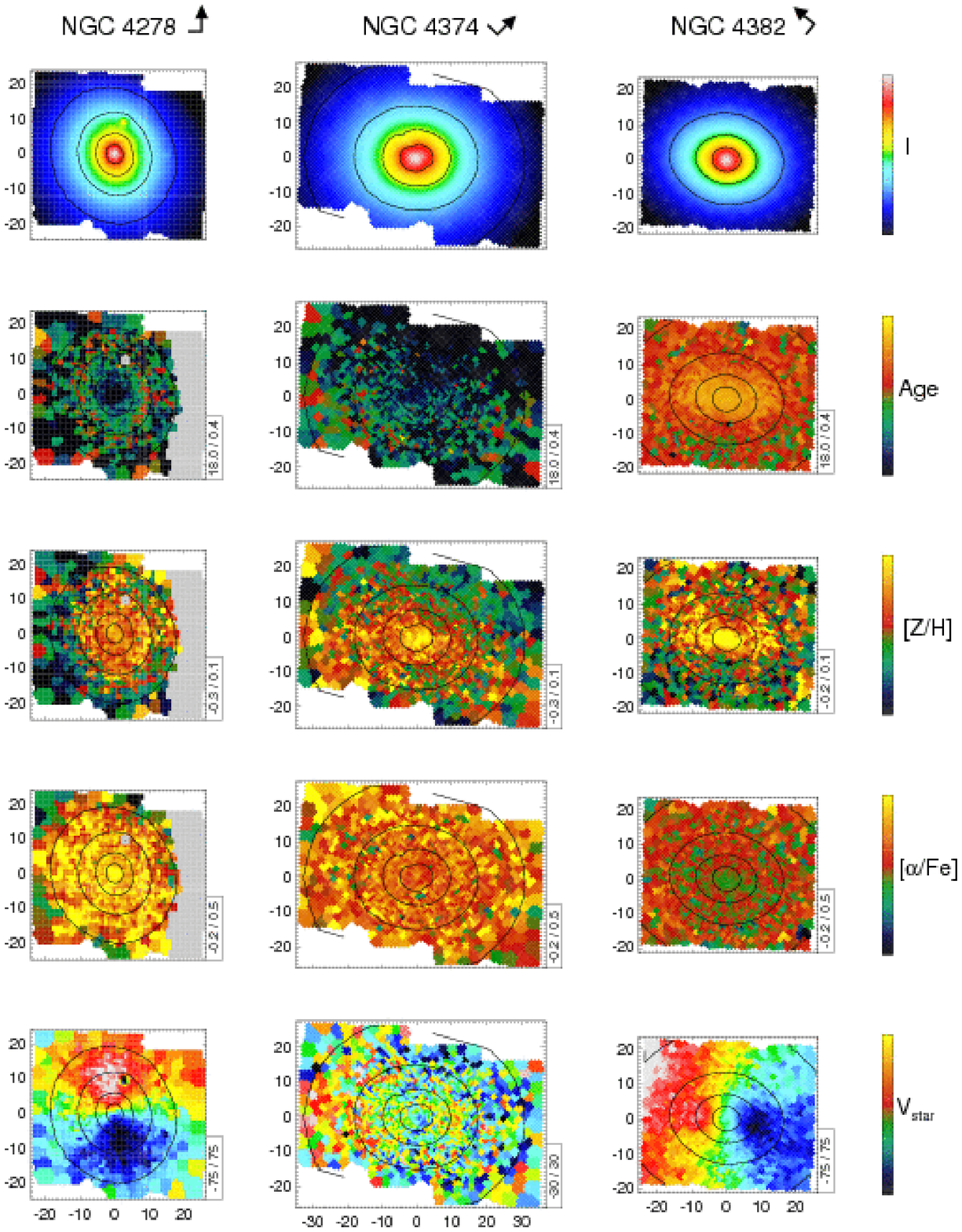} 
 \end{center} 
 \caption[]{}
 \label{fig:maps6}
 \end{figure*}

 \addtocounter{figure}{-1}
 \addtocounter{subfigure}{1}

 \begin{figure*}
 \begin{center} 
   \includegraphics[draft=false,scale=0.99,trim=0cm 0.cm
     0cm 0cm]{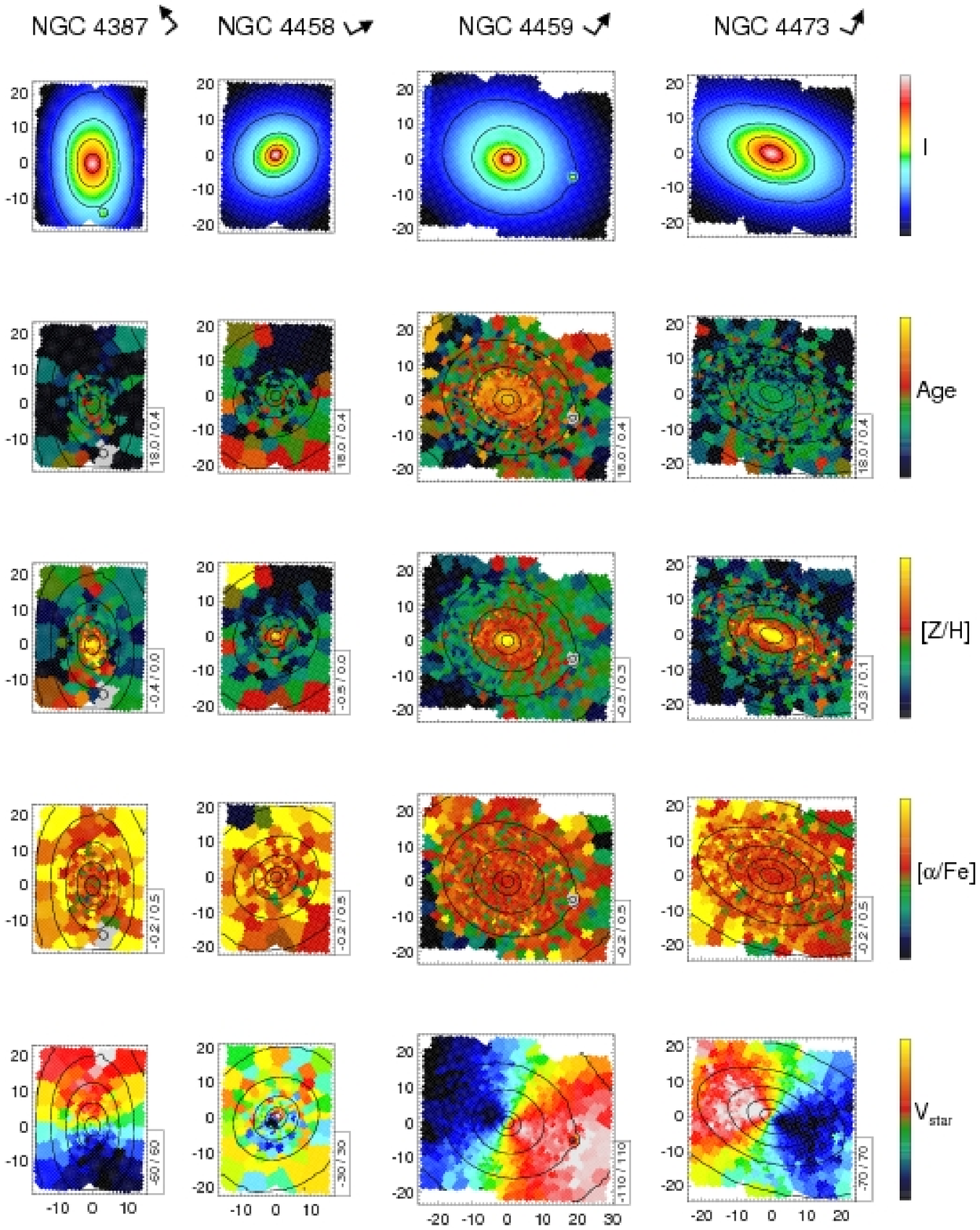} 
 \end{center} 
 \caption[]{}
 \label{fig:maps7}
 \end{figure*}

 \addtocounter{figure}{-1}
 \addtocounter{subfigure}{1}

 \begin{figure*}
 \begin{center} 
   \includegraphics[draft=false,scale=0.99,trim=0cm 0.cm
     0cm 0cm]{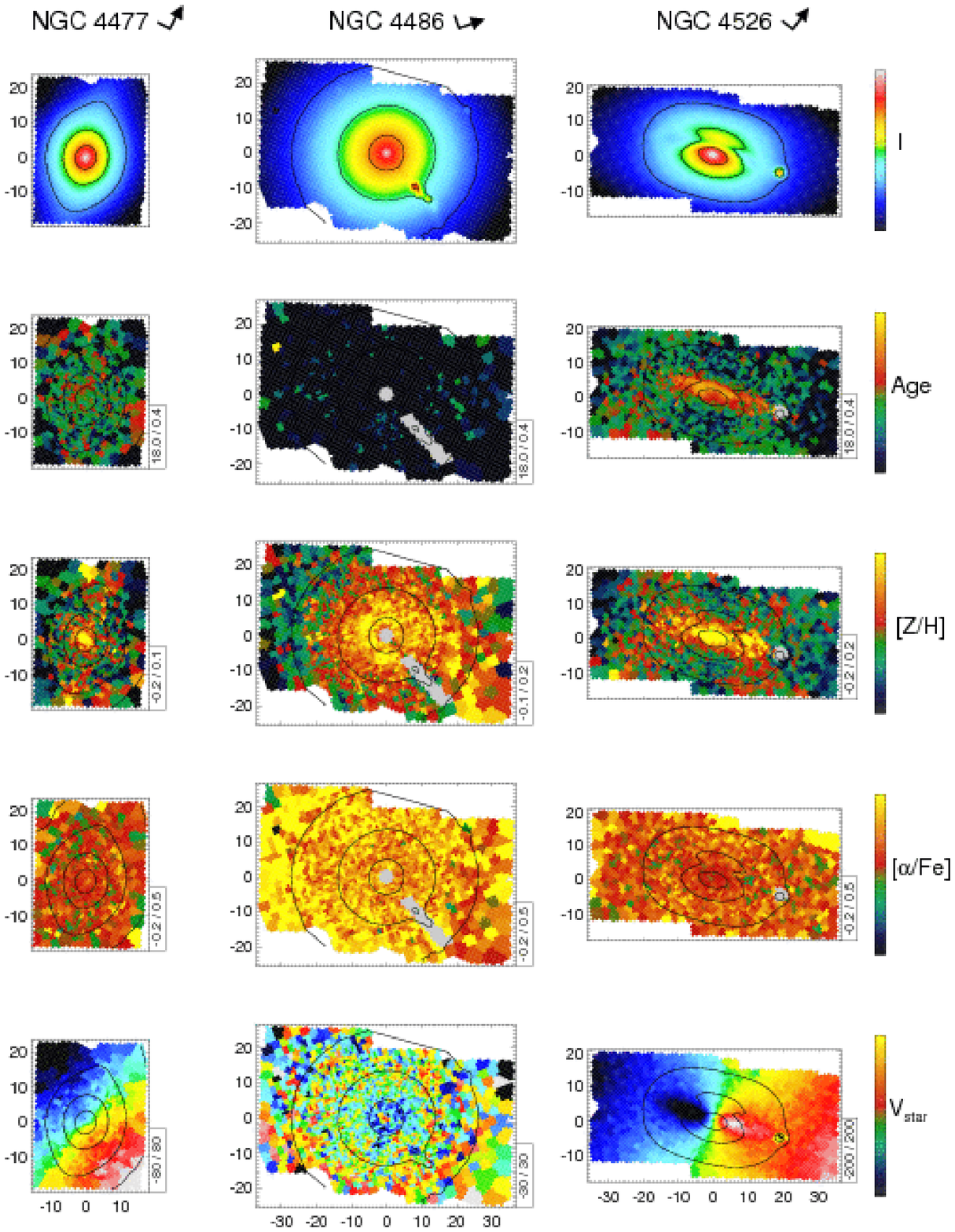} 
 \end{center} 
 \caption[]{}
 \label{fig:maps8}
 \end{figure*}

 \addtocounter{figure}{-1}
 \addtocounter{subfigure}{1}

 \begin{figure*}
 \begin{center} 
   \includegraphics[draft=false,scale=0.99,trim=0cm 0.cm
     0cm 0cm]{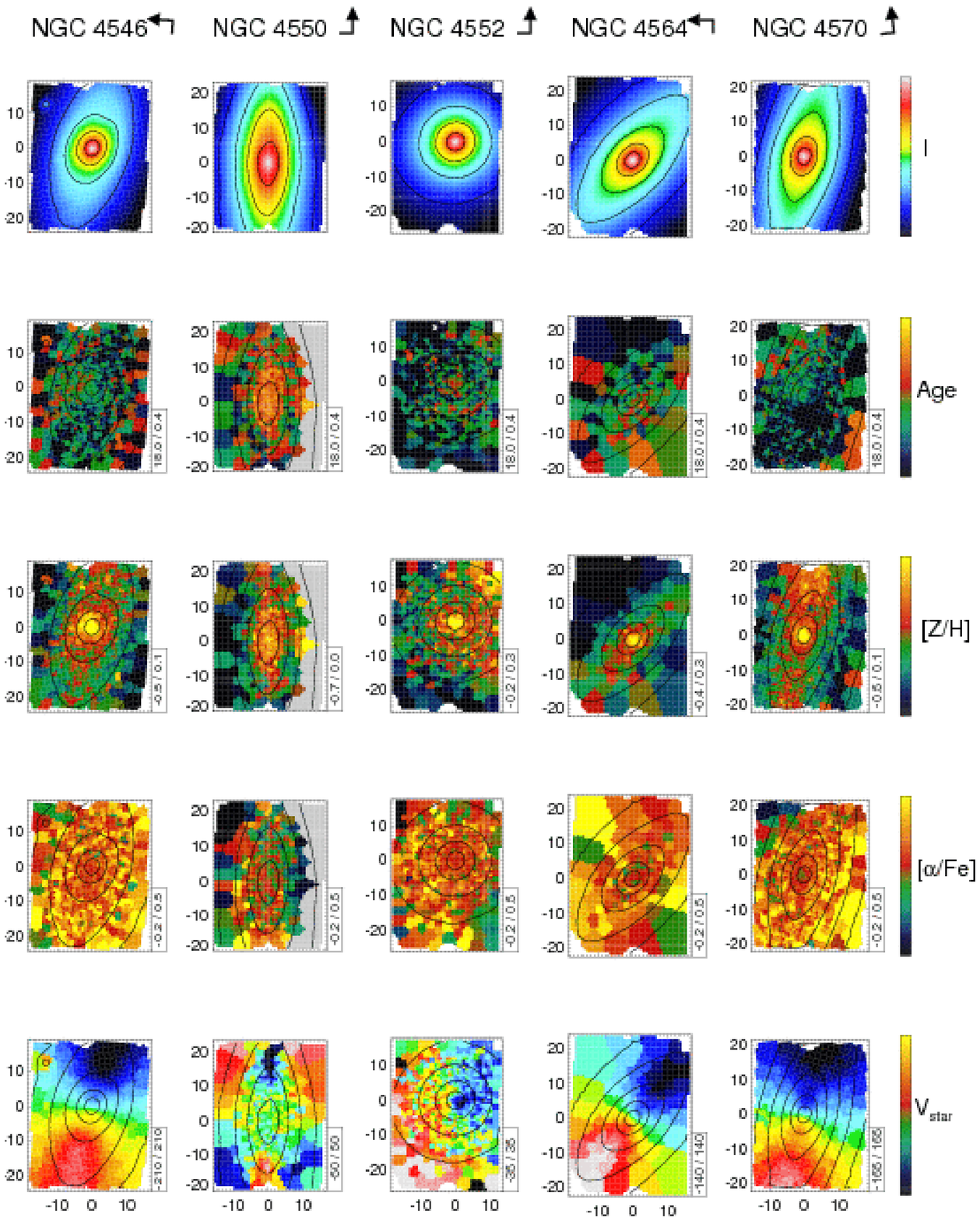} 
 \end{center} 
 \caption[]{}
 \label{fig:maps9}
 \end{figure*}

 \addtocounter{figure}{-1}
 \addtocounter{subfigure}{1}

 \begin{figure*}
 \begin{center} 
   \includegraphics[draft=false,scale=0.99,trim=0cm 0.cm
     0cm 0cm]{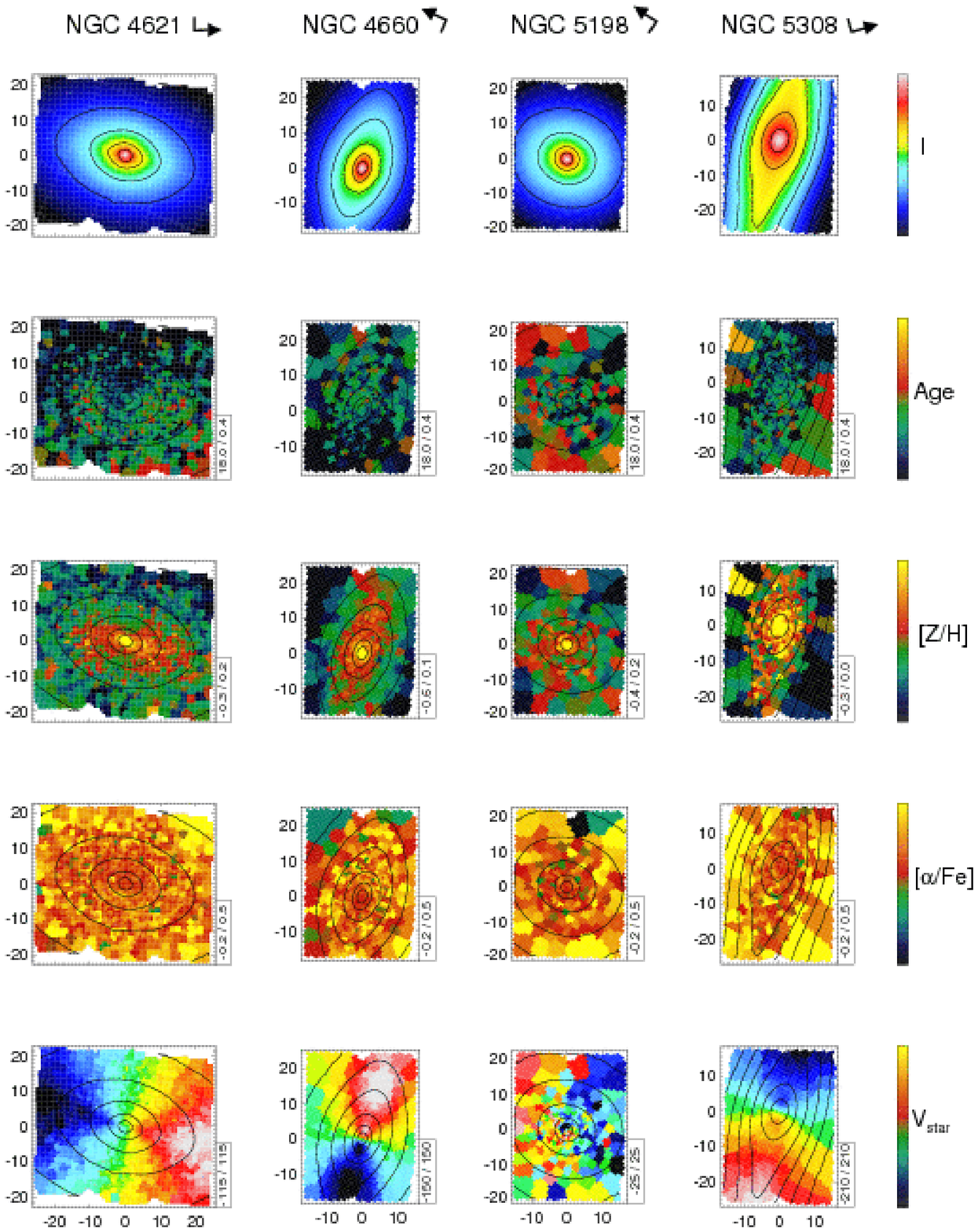} 
 \end{center} 
 \caption[]{}
 \label{fig:maps10}
 \end{figure*}

 \addtocounter{figure}{-1}
 \addtocounter{subfigure}{1}

 \begin{figure*}
 \begin{center} 
   \includegraphics[draft=false,scale=0.99,trim=0cm 0.cm
     0cm 0cm]{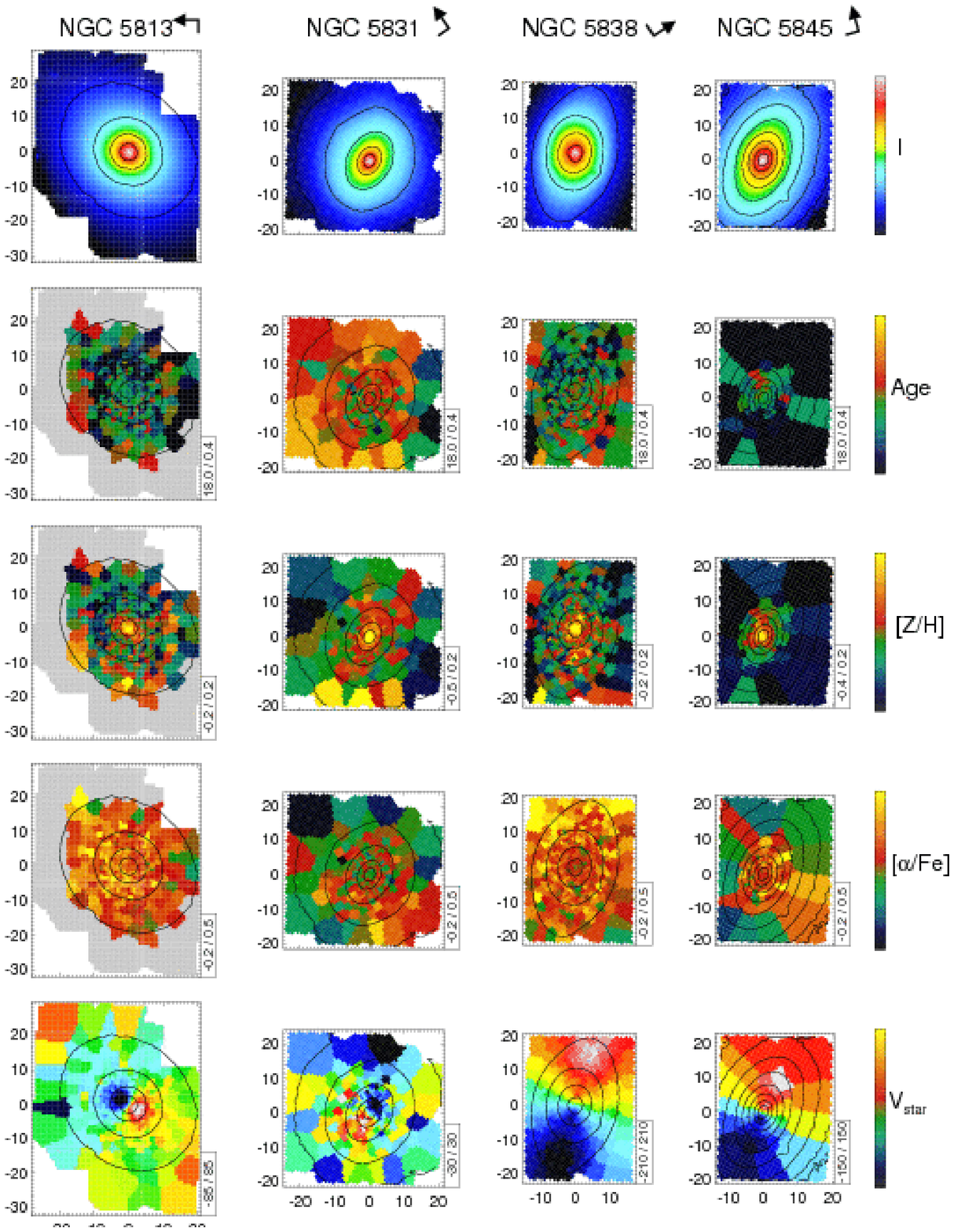} 
 \end{center} 
 \caption[]{}
 \label{fig:maps11}
 \end{figure*}

 \addtocounter{figure}{-1}
 \addtocounter{subfigure}{1}

 \begin{figure*}
 \begin{center} 
   \includegraphics[draft=false,scale=0.99,trim=0cm 0.cm
     0cm 0cm]{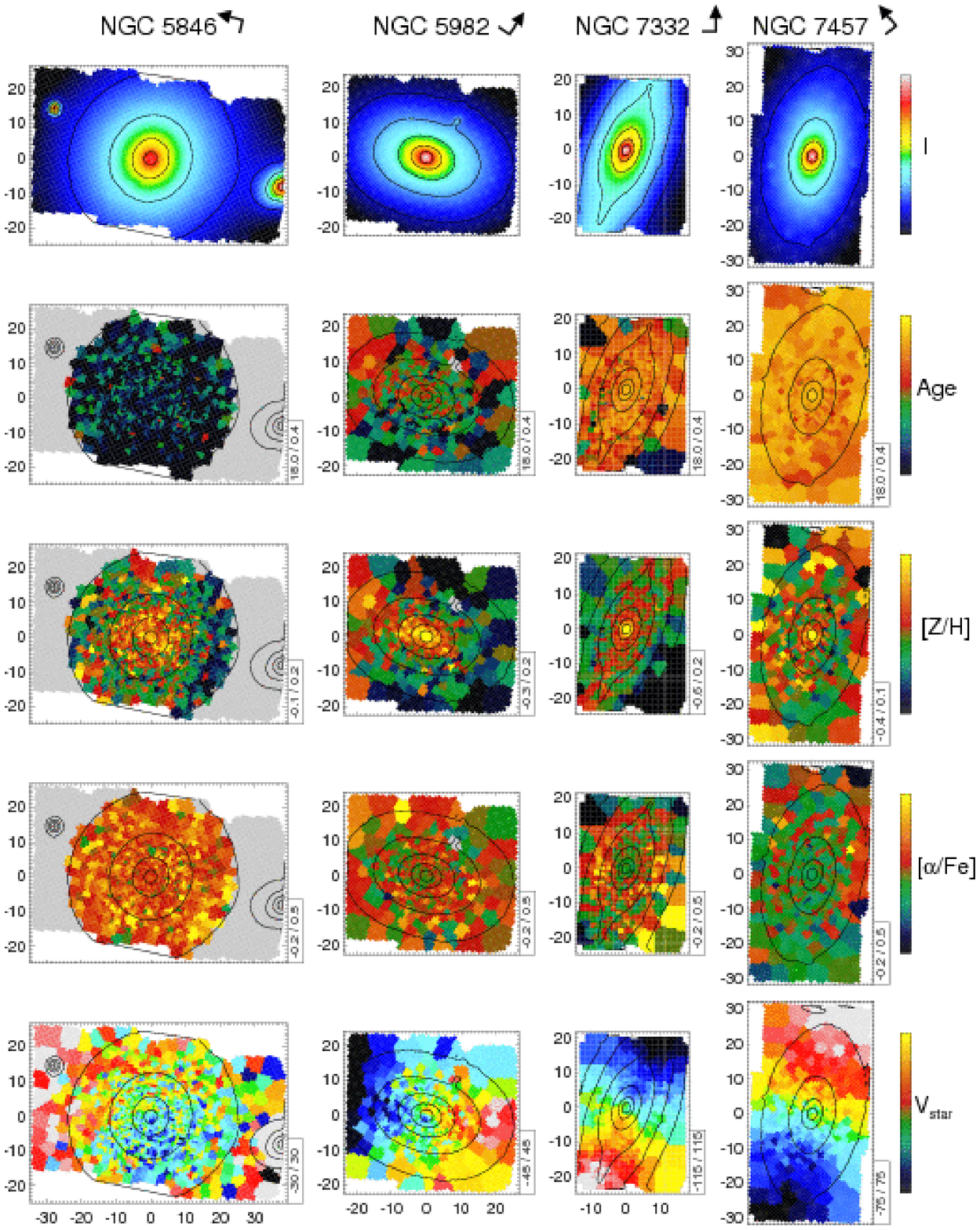} 
 \end{center} 
 \caption[]{}
 \label{fig:maps12}
 \end{figure*}

\renewcommand{\thefigure}{\arabic{figure}}

\subsection{Average stellar population gradients}
\label{sec:gradients}
Average stellar population gradients can be used to study the
formation history of early-type galaxies since different formation
models predict different gradients. In a nutshell, monolithic collapse
models \citep{car84,ari87,pip08} predict steep metallicity gradients
(ranging from $\delta \rmn{[Z/H]}/\delta \log R = \Delta\, \rmn{[Z/H]}
= -0.35$ to $-1.0$) with metal rich centers whereas hierarchical
models, following a merger tree, predict shallower gradients due to
the dilution of any line strength gradients existing in the pre-merger
units \citep{whi80,kob04}. However, adding secondary star-formation,
introduced by gas rich mergers with a range in degree of dissipation,
will typically lead to enhanced metallicity over the inner regions and
thus again steeper metallicity gradients. The metallicity gradients
are predicted to be strongly correlated with the efficiency of the
dissipative process, and to only weakly depend on the remnant mass
(\citeauthor{kob04} 2004; \citeauthor{hop09b} 2009a,\,b).

In this section we derive robust and simple stellar population
gradients for SSP-equivalent age, metallicity and abundance ratio by
averaging the measurements along lines of constant surface brightness
(isophotes) with equal steps in $\log$~flux. This approach does not
take into account the deviations from e.g., metallicity or
[$\alpha$/Fe] contours with respect to isophotes, but does provide
valuable estimates of {\it mean}\/ gradients which can be investigated
throughout the whole sample and compared to the predictions of
simulations. Due to the two-dimensional coverage of our data these
mean radial gradients can be determined with very good precision. The
average value in each radial bin is derived after applying a $3\sigma$
clipping algorithm. The radius for each bin is calculated as the
median major axis radius normalized to the effective radius (see
Table~\ref{tab:LSnew} of the Appendix) along the major axis $a_{\rmn
  e} = R_{\rmn{e}}/\sqrt{1-\epsilon}$, where $R_{\rmn e}$ is the
effective radius and $\epsilon$\/ is the average ellipticity of the
galaxy within the \sauron\/ field.

\begin{figure}
 \includegraphics[width=80mm]{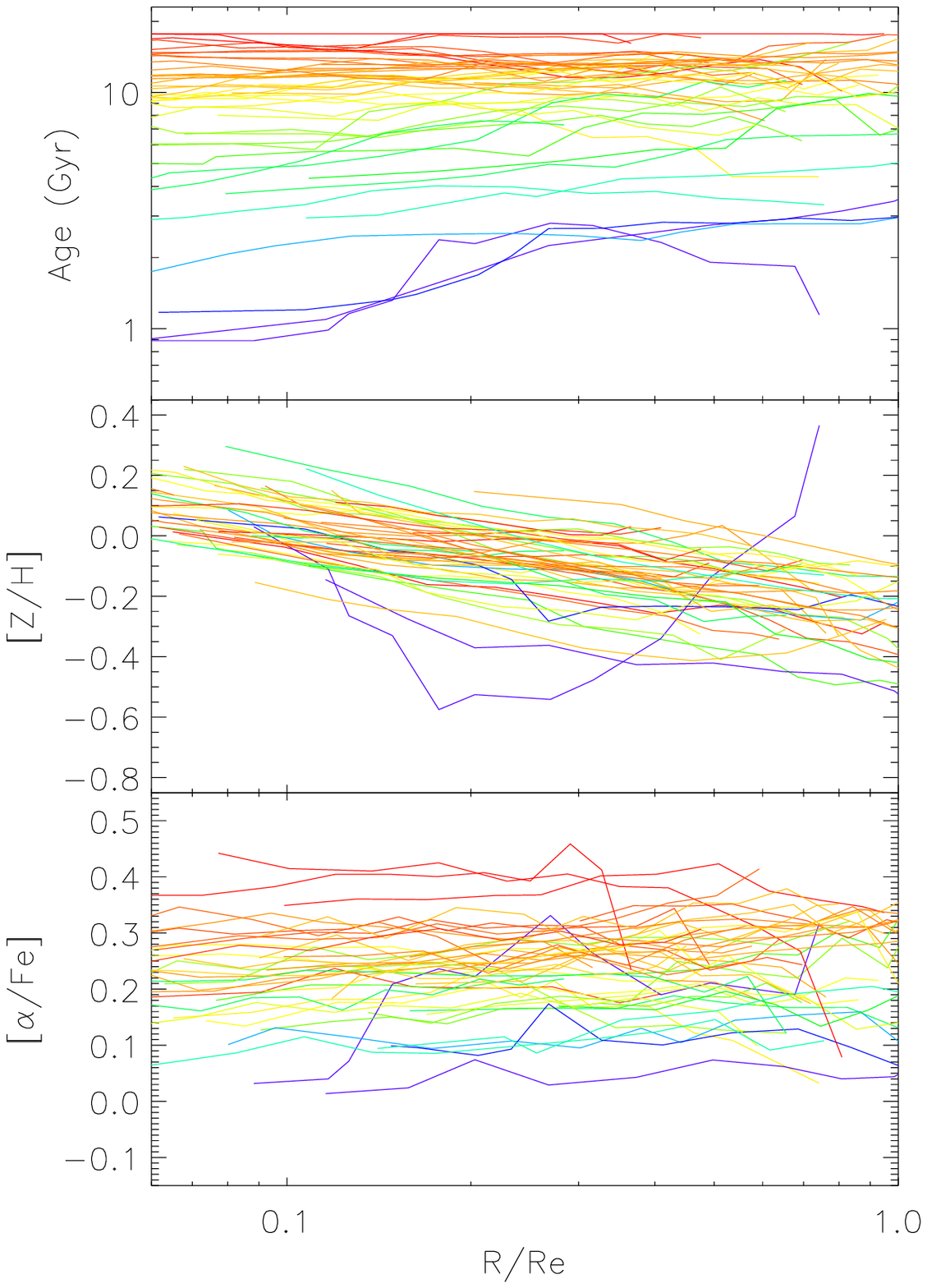}
 \caption{Overview of radial profiles, averaged along isophotes, of
   SSP-equivalent age, metallicity and [$\alpha$/Fe] for the 48
   early-type galaxies in the \sauron\ sample. The colour coding
   reflects central \reb\/ SSP-equivalent age with red referring to
   old stellar populations and blue to young ones (see top plot).}
 \label{fig:pop_grads}
\end{figure}

In Fig.~\ref{fig:pop_grads} we show the radial profiles of
SSP-equivalent age, metallicity and abundance ratio as function of
$\log (R/R_{\rmn e})$ as derived from the data. The following
conclusions can be inferred from the diagram: (1) If the central parts
of a galaxy show old ($>$9\,Gyr) SSP-equivalent ages, then age
gradients remain typically flat. (2) In contrast, the presence of
young stellar populations in the centre typically leads to age
gradients such that older SSP-equivalent ages are found at larger
radii\footnote{A notable exception to this trend is NGC\,474 which
  shows star-formation at larger radii (see also Paper~XIII and
  Section~\ref{sec:agemaps}).}. This observation supports the scenario
of potentially multiple, centrally located star-formation events which
add to the older, underlying stars. (3) Excluding galaxies with
significant recent episodes of star-formation (central age $\le
3$\,Gyr), metallicity gradients appear remarkably homogeneous with a
mean value of -0.28 dex per $\log (R/R_{\rmn e})$ and a rms of
0.12. There is a significant range in central metallicity not
obviously connected to the age of the stellar population. (4)
Abundance ratio gradients are typically consistent with a flat
distribution or are mildly positive gradient (median value of 0.06 dex
per $\log (R/R_{\rmn e})$). Overall, there is a trend in the sense
that the oldest galaxies have the strongest non-solar abundance ratios
(see also Figure~\ref{fig:central_pop_dia}).

\begin{figure*}
 \includegraphics[width=28.1mm]{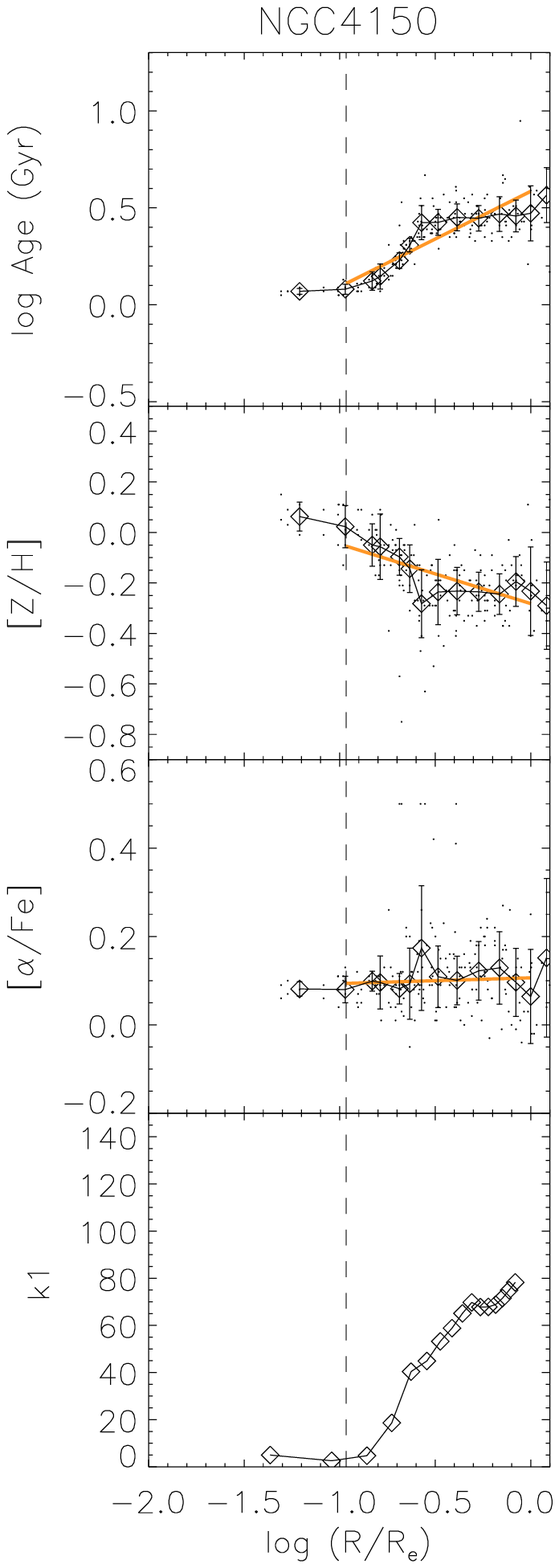}
 \includegraphics*[width=27mm,viewport=8 0 198 566,clip]{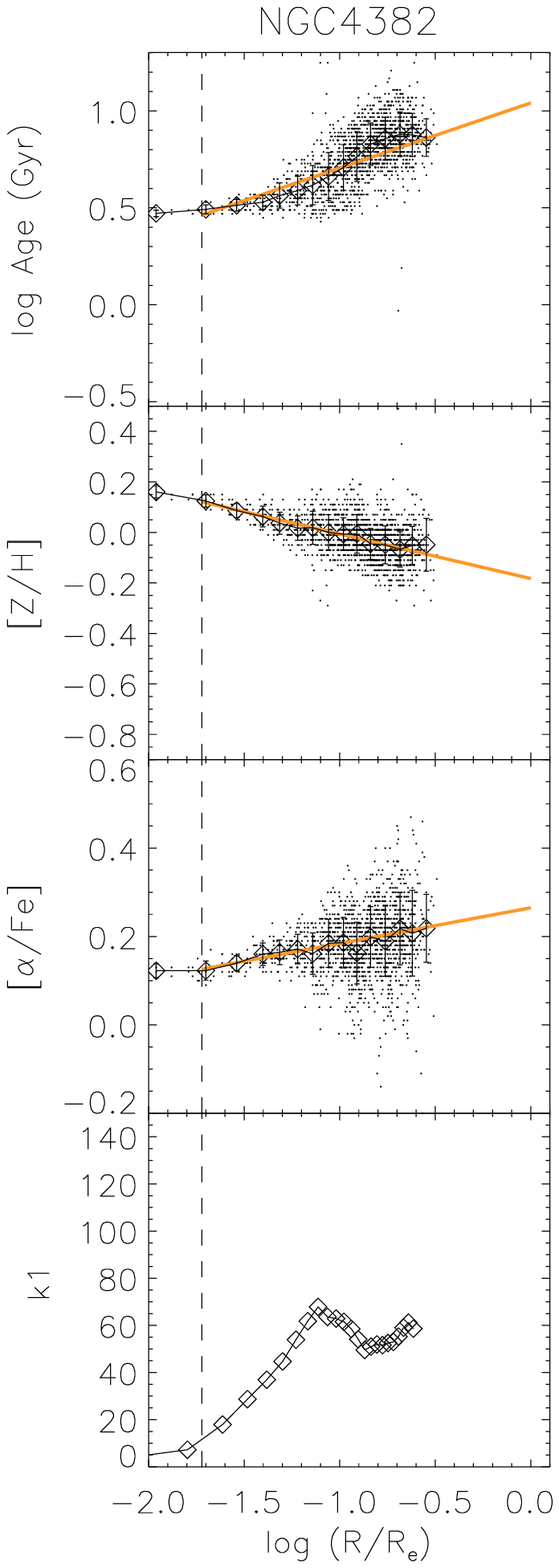}
 \includegraphics*[width=27mm,viewport=8 0 198 566,clip]{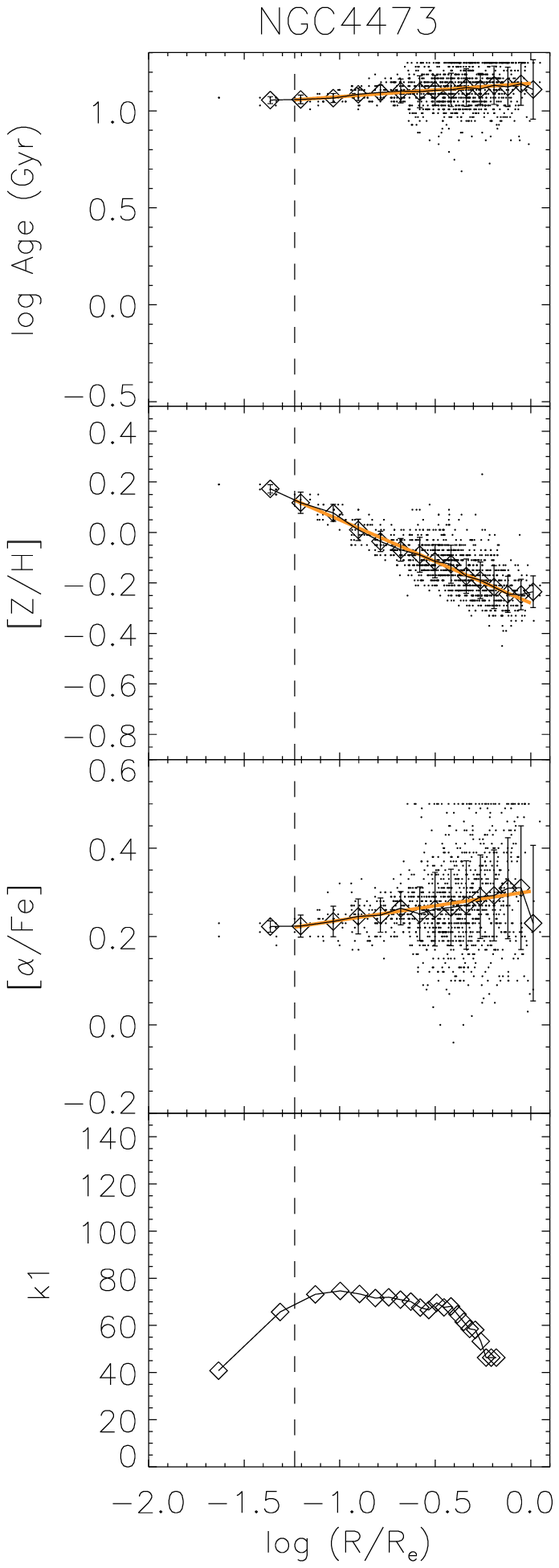}
 \includegraphics*[width=27mm,viewport=8 0 198 566,clip]{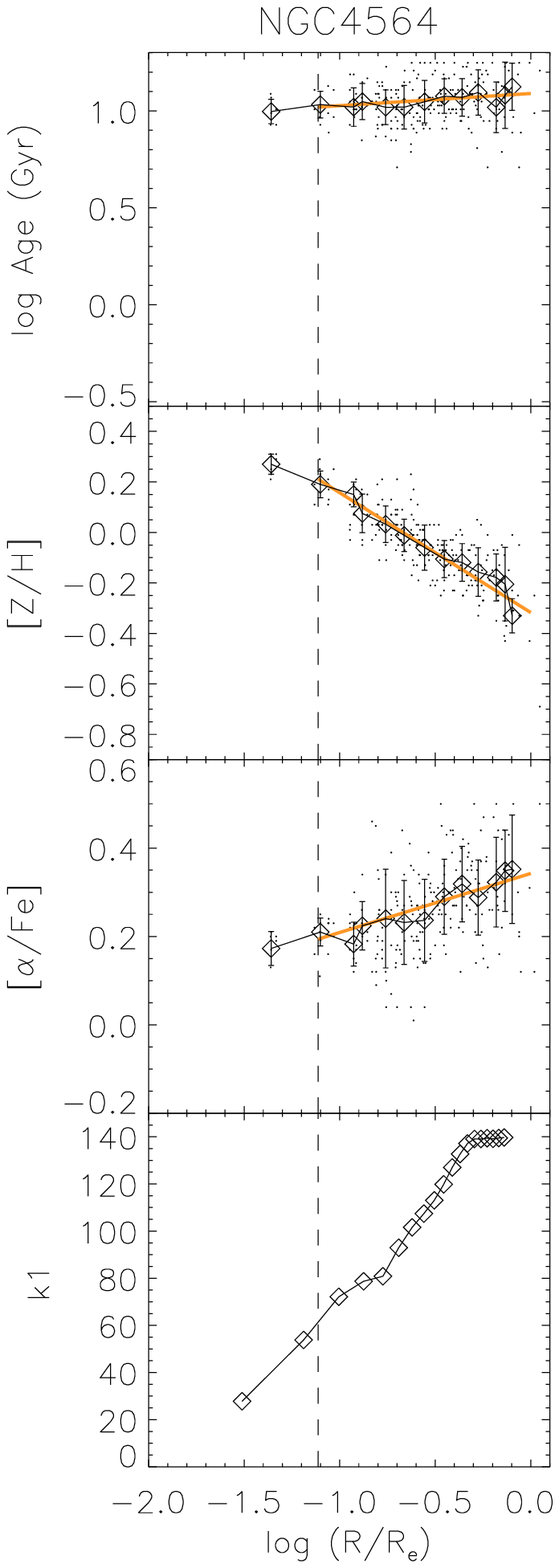}
 \includegraphics*[width=27mm,viewport=8 0 198 566,clip]{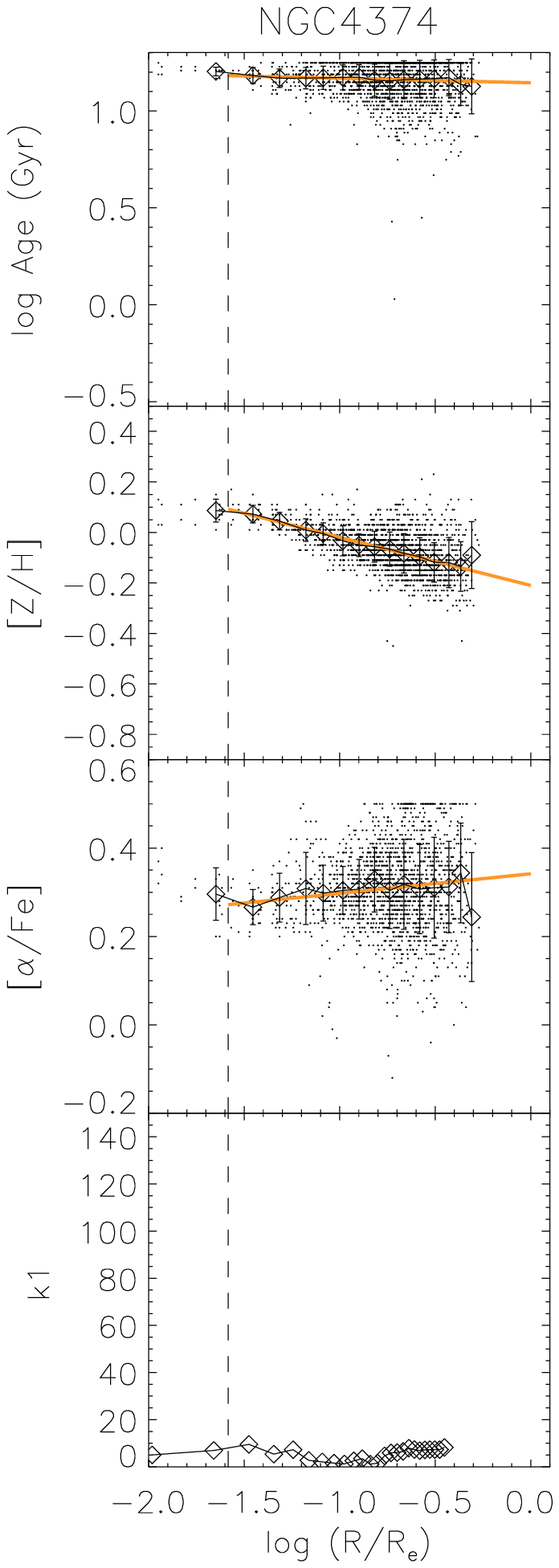}
 \includegraphics*[width=27mm,viewport=8 0 198 566,clip]{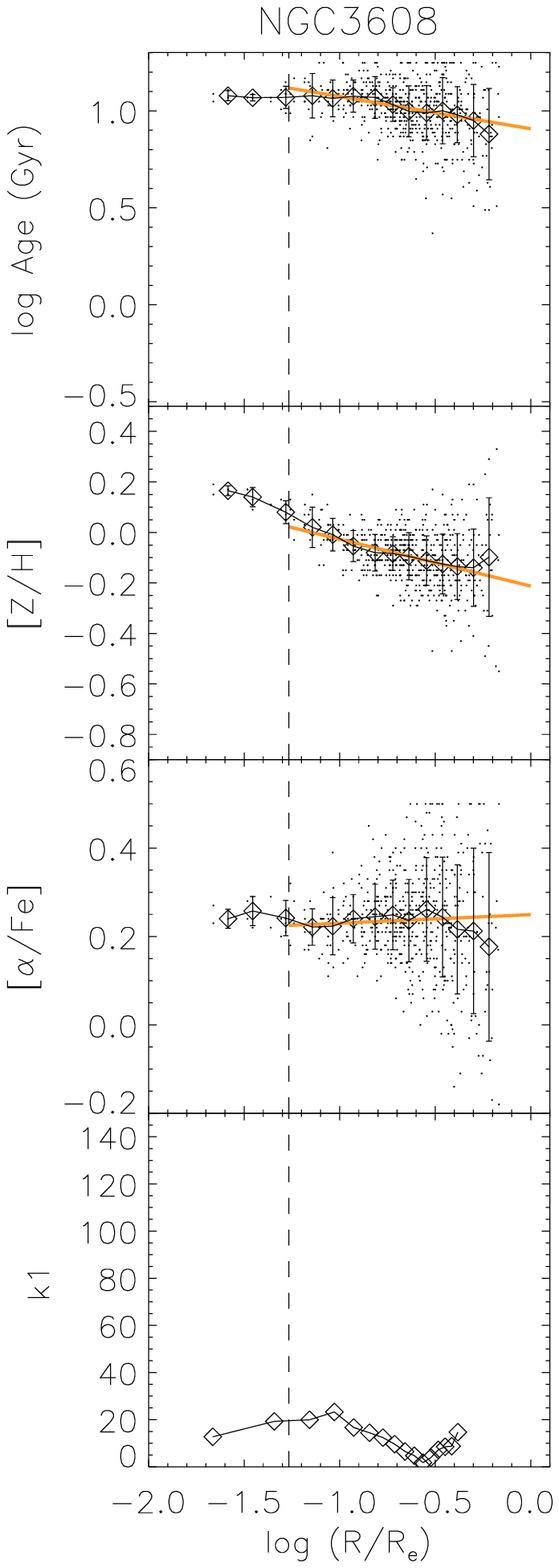}
 \caption{Examples of radial profiles, averaged along isophotes, of
   SSP-equivalent age, metallicity and [$\alpha$/Fe] estimates (open
   diamonds). We fit an error-weighted straight line (orange solid
   line) to each profile for all data available between 2\arcsec\/
   (vertical dashed lines) and $R_{\rmn{e}}$. See text for
   details. Small black points represent individual bins in the
   stellar population maps. The bottom panels show the k1 parameter -
   a measure of the maximum rotation velocity - from the kinemetry
   analysis of Paper~XII.}
 \label{fig:pop_grads1}
\end{figure*}

For each galaxy we fit an error-weighted straight line to the stellar
population values at a given radius such that age gradients are
defined as

\begin{equation}
\Delta\, {\log \rmn{Age}} = \frac{\delta\,{\log \rmn{Age}}}{\delta\, \log (R/R_{\rmn{e}})} \,,
\end{equation}

\noindent where Age is measured in Gyr. A similar definition is used
for measurements of the gradients in [Z/H] and [$\alpha$/Fe], denoted
as $\Delta$\,[Z/H] and $\Delta$\,[$\alpha$/Fe], respectively. Examples
of the fits are shown in Figure~\ref{fig:pop_grads1} for galaxies
representative of the group of fast rotators with young central disks
(NGC\,4150, NGC\,4382), ``old'' fast rotators with evidence for an
embedded disk-like component (NGC\,4473, NGC\,4564) and galaxies from
the group of slow rotators (NGC\,4374, NGC\,3608).

The error of the measurements in each surface brightness bin is taken
as the $1\sigma$ scatter of the data-points within this bin after
applying a $3\sigma$ clipping algorithm. We further restrict the
fitting range to all available data with radii between 2\arcsec\/ and
$R_{\rmn{e}}$. The inner $r < 2$\arcsec\/ regions are excluded from
the fit in order to avoid biases due to seeing effects. The overall
quality of the fits is good and also the assumption of linear
gradients per $\Delta \log R$ is reasonable, with the exception of
some galaxies with contributions from very young stars (see e.g.,
NGC\,4150 and NGC\,4382 in Figure~\ref{fig:pop_grads1}).

\begin{figure*}
 \includegraphics[width=134mm]{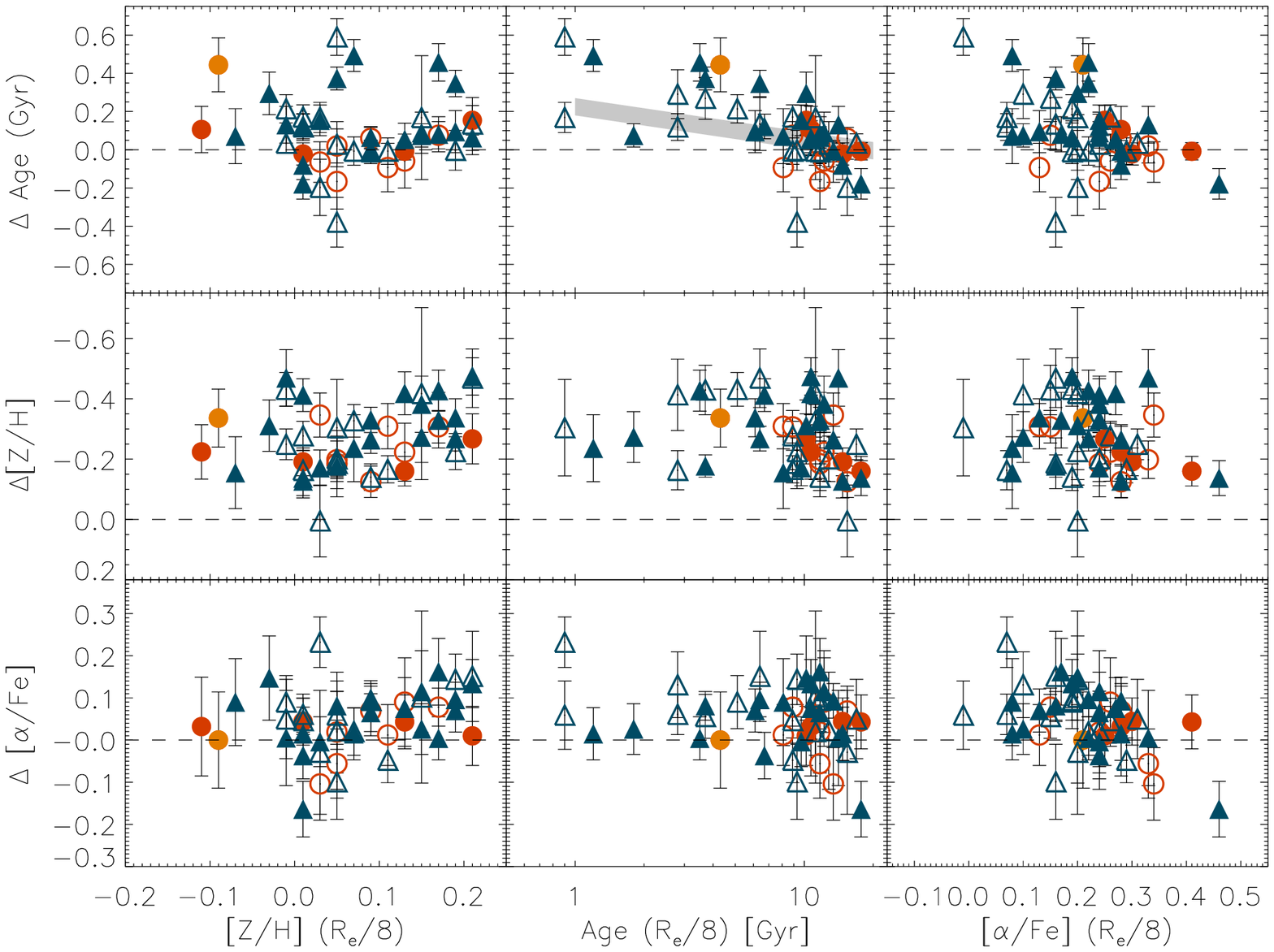}
 \caption{Age gradient, metallicity gradient and [$\alpha$/Fe]
   gradient {\em versus}\/ metallicity, age and [$\alpha$/Fe] measured
   within a circular aperture of \reb. Fast and slow rotators are
   indicated by blue triangles and red circles, respectively.
   NGC\,4550, classified as slow rotator due to two counter-rotating
   disks, is shown as orange filled symbol. Cluster and field galaxies
   are shown as filled and open symbols. In the top middle panel we
   show in grey the region occupied by the simulations of
   \citeauthor[][]{hop09a} (2009a, their Figure 29). }
 \label{fig:pop_grads_age}
\end{figure*}

\begin{figure*}
 \includegraphics[width=134mm]{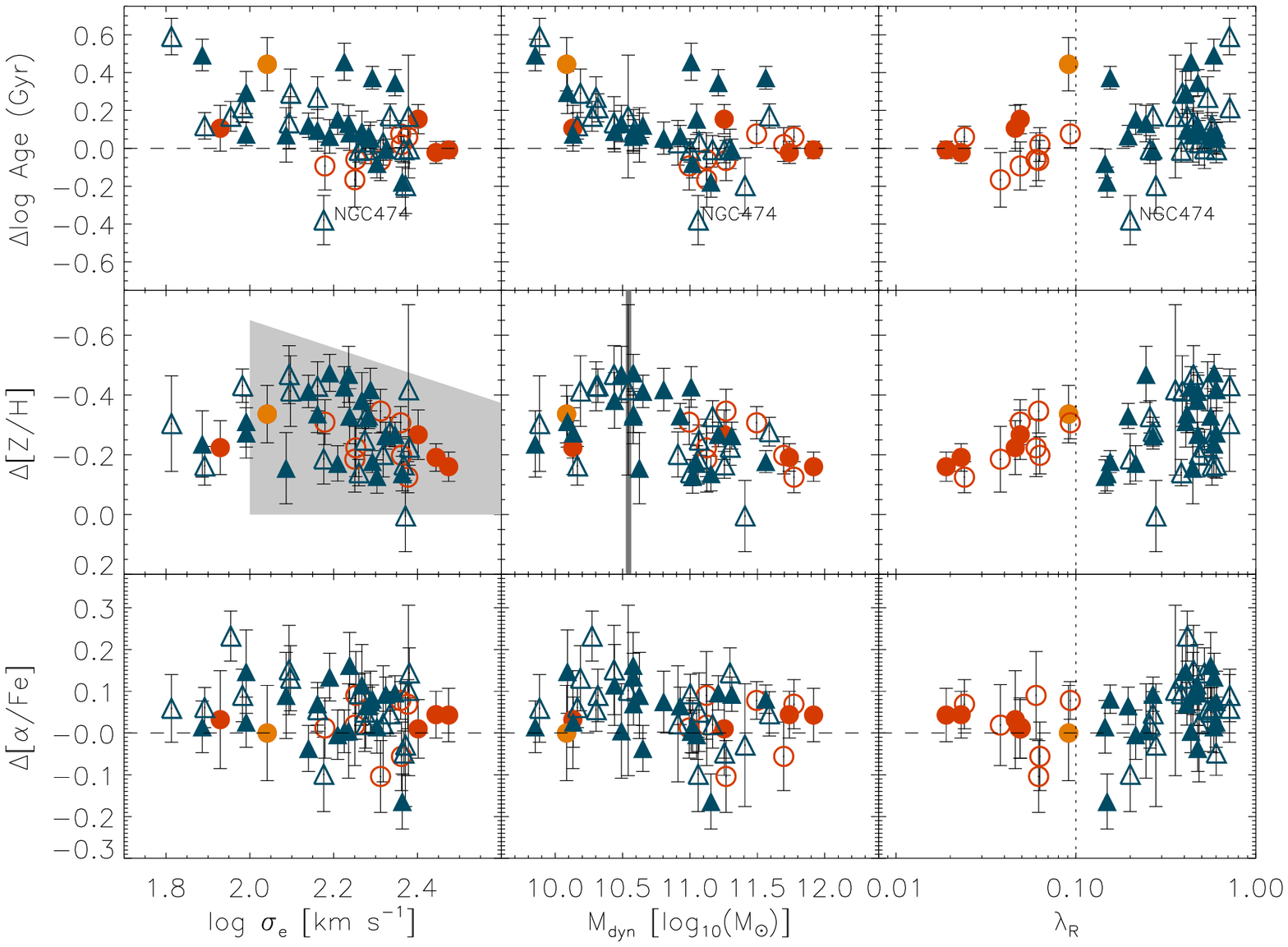}
 \caption{Age gradient, metallicity gradient and [$\alpha$/Fe]
   gradient {\em versus}\/ $\log \sigma_{\rm e}$, dynamical mass
   \mvir\/ and $\log \lambda_R$ for the 48 early-type galaxies in the
   \sauron\ sample. Fast and slow rotators are indicated by blue
   triangles and red circles, respectively. Cluster and field galaxies
   are shown as filled and open symbols.  NGC\,4550, classified as
   slow rotator due to two counter-rotating disks, is shown as orange
   filled symbol. The one slow rotator at small $\log \sigma_{\rm e}$
   is NGC\,4458.  NGC\,474, which shows star-formation at larger radii
   and thus shows a significantly negative age gradient (see also
   Paper~XIII and Section~\ref{sec:agemaps}) is labeled in the Age
   gradient panels.  There is a break in the metallicity gradient vs
   \mvir\/ relation at a mass of $\sim 3.5\times 10^{10} M_{\sun}$
   first detected by \citet{spo09} and indicated by the thick grey
   line.  The grey shaded area in the left middle panel shows the
   approximate region occupied by the simulations of
   \citeauthor[][]{hop09a} (2009a, their Figure 29).}
 \label{fig:pop_grads_lamsig}
\end{figure*}

In Figures~\ref{fig:pop_grads_age} and \ref{fig:pop_grads_lamsig} we
explore the connection between stellar population gradients and
central stellar population estimates within an aperture of \reb, as
well as the connection between stellar population gradients and global
kinematic parameters such as mean velocity dispersion within one
effective radius $\sigma_{\rm e}$, the dynamical mass \mvir\/ (see
Eq.~\ref{equ:mvir}) and the degree of rotational support as measured
by $\lambda_R$.

As was already apparent from Figure~\ref{fig:pop_grads}, old ($\ge
8$\,Gyr) galaxies generally do not show significant age gradients
(mean gradient $0.02\pm0.13$; the error reflects the one sigma
standard deviation of the gradients), whereas young, fast rotator
galaxies typically show positive age gradients (corresponding to a
younger center; mean gradient $0.28\pm0.16$).  Metallicity gradients
are, with two exceptions (NGC\,3032, NGC\,524), all negative with a
median value of $-0.27$. The mean metallicity gradient for old (age
$\ge 8$\,Gyr) galaxies is $-0.25\pm0.11$.  There is, however, a trend
with $\lambda_R$ in the sense that fast rotators have on average the
steeper metallicity gradients with a large scatter at high
$\lambda_R$.

The relation between \sauron\/ line strength index gradients as well
as stellar population gradients with the local escape velocity
$V_{\rmn esc}$ was explored in \citet{sco09}.  A tight correlation
between $V_{\rmn esc}$ and the line strength indices is found. For
\mgb, this correlation exists not only between different galaxies but
remarkably also inside individual galaxies -- it is both a local and
global correlation.  When index measurements are converted to stellar
population estimates the galaxies are found to be confined to a good
approximation to a plane in the four-dimensional parameter space of
SSP-equivalent age, metallicity ([Z/H]), abundance ratio
([$\alpha$/Fe]) and $V_{\rmn esc}$. The observed tight connection
between stellar populations and the gravitational potential, both
locally and globally, presents a strong constraint on galaxy formation
scenarios.

\citet{spo09} noticed a mass-metallicity gradient relation for
early-type galaxies where low mass galaxies ($\lesssim$3.5$ \times
10^{10} M_{\odot}$) form a tight relation of increasing metallicity
gradient with increasing mass. For galaxies above the transition mass
there is a clear downturn in metallicity gradient with increased
scatter. Our analysis is consistent with these observations \citep[see
Figure~\ref{fig:pop_grads_lamsig}, middle panel, but also
see][]{raw2010}. The fraction of slow rotators increases with
increasing mass and the most massive galaxies are slow rotators (see
Paper~IX). Furthermore, there is a weak trend among the slow rotators
in the sense that the galaxies with the smallest $\lambda_R$ exhibit
the shallowest metallicity gradients. This trend clearly needs
confirmation with a larger sample of slow rotators.

These observations suggest that the formation and subsequent evolution
of low-mass early-type galaxies (which are dominated by fast rotators)
leads to a connection between galaxy mass and metal enrichment such
that the steepest metallicity gradients are achieved for masses of
$\sim$3.5$ \times 10^{10} M_{\odot}$.  All these galaxies are
rotationally supported and feature strong disk-components suggesting
gas-rich mergers and accretion to play an important role.  We
emphasize that low-mass galaxies with young {\em and}\/ old
SSP-equivalent ages follow this relation and thus secondary
star-formation due to gas-rich (minor) mergers (see
Section~\ref{sec:agemaps}; Paper~XV) does not destroy the trend.  For
more massive galaxies (above $\sim$3.5$ \times 10^{10} M_{\odot}$) the
relative influence of the various star-formation and assembly
processes which determine the galaxies' properties seem to change. We
observe a large scatter of metallicity gradients with an overall
downturn with mass and increased fraction of slow rotators towards the
most massive systems ($> 10^{11} M_{\odot}$). The observed scatter may
be the result of competing processes such as minor gas-rich mergers
building disk-like structures over extended regions with rather
shallow metallicity gradients, circumnuclear star-formation in rings
and disks leading to steeper metallicity gradients, particularly in
the center, and gas-poor (major) merging removing the signature of
ordered motion and also weakening metallicity gradients in the most
massive remnants. Consistent with this picture, we find the most
massive systems to be slow rotators that exhibit relatively shallow
metallicity gradients and are devoid of any signs of recent
star-formation (see also Paper~XV). Brightest cluster galaxies, which
are underrepresented in our sample, may deviate from the trend
\citep[e.g.,][]{bro07,spo09}. However, the analysis of the projected
axial ratio distribution of quiescent galaxies in the Sloan Digital
Sky Survey \citep{wel09} demonstrates that galaxies above a mass of
$\sim 10^{11} M_{\odot}$ have essentially a spheroidal shape without
major disk contribution whereas the lower mass systems show a large
range in axial ratios.

The average [$\alpha$/Fe] gradients are consistent with zero for the
slow rotators while a significant fraction of the low-mass, fast
rotators show mildly {\em positive}\/ gradients which we interpret as
signs of the contributions from the central, more metal rich and
abundance-ratio depressed regions with disk-like kinematics. The mean
[$\alpha$/Fe] gradients of the fast and slow rotators are
$0.060\pm0.012$ and $0.023\pm0.016$, respectively (the mean was
determined with an outlier resistant bi-weight method and the errors
reflect errors on the mean values).

\begin{figure}
 \includegraphics[width=84mm]{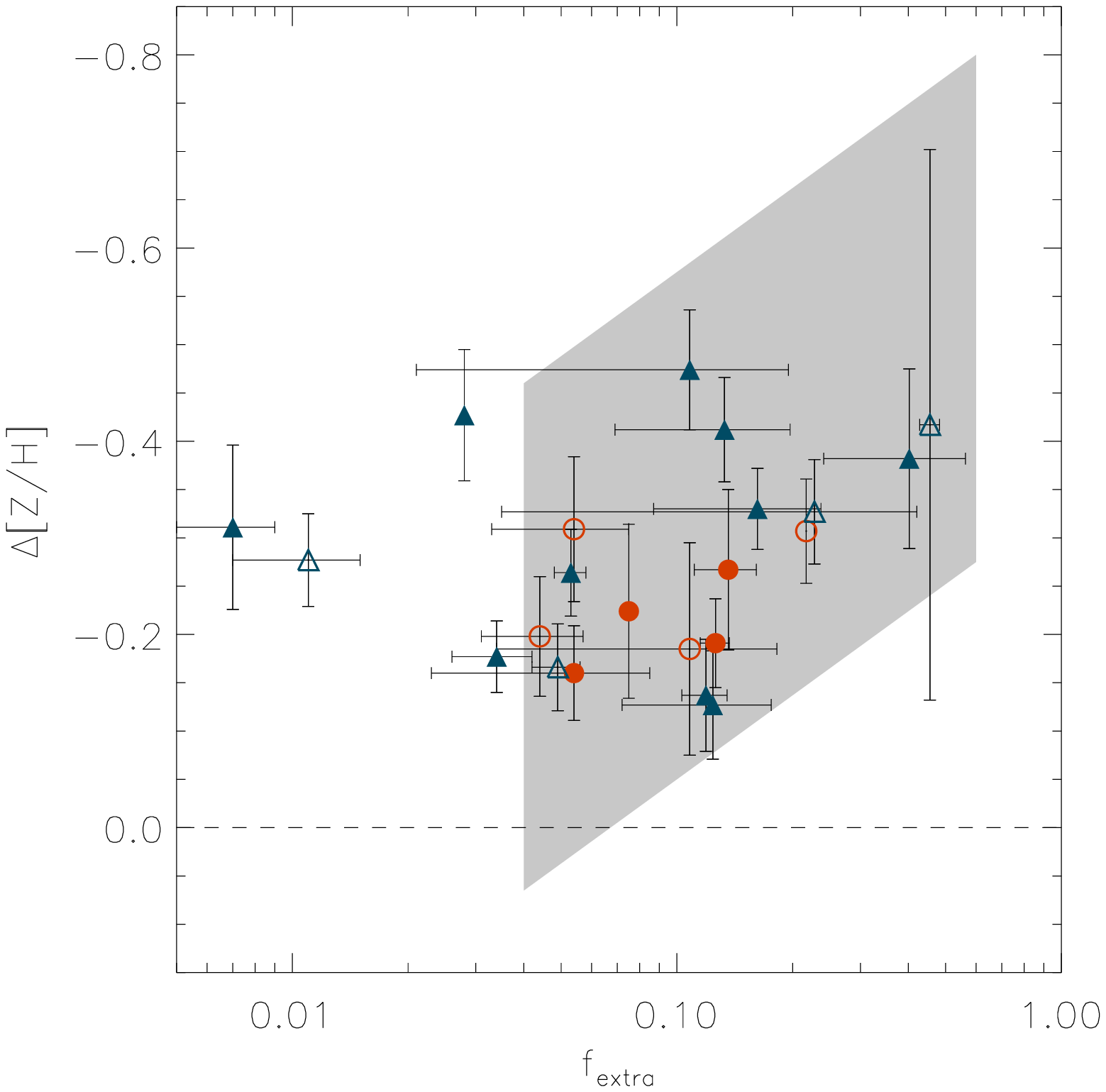}
 \caption{Metallicity gradients as derived from radial averages along
   isophotes {\em versus}\/ f$_\rmn{extra}$, the fraction of extra
   light as determined by \citeauthor[][]{hop09a} (2009a) for galaxies
   in common. The grey shaded area shows the approximate region
   occupied by the simulations of \citeauthor[][]{hop09a} (2009a,
   their Figure 29). Fast and slow rotators are indicated by blue
   triangles and red circles, respectively. Cluster and field galaxies
   are shown as filled and open symbols. NGC\,4550, classified as slow
   rotator due to two counter-rotating disks, is shown as orange
   filled symbol. }
 \label{fig:pop_fextra}
\end{figure}

In the recent simulations of major mergers of gas-rich disks by
\citeauthor[][]{hop09a} (2009a), a large range of metallicity
gradients is found. They emphasize the importance of the degree of
dissipation in the central, merger-induced starburst for the resulting
metallicity gradient. Their predictions (shown as grey shaded regions
in Figures~\ref{fig:pop_grads_age} and \ref{fig:pop_grads_lamsig})
broadly agree with our observations. This includes the overall
downturn in metallicity gradient with increasing mass. For galaxies in
common we specifically explore the correlation between our observed
metallicity gradients with their observed measurements of extra light
f$_\rmn{extra}$ and compare this also with their predictions from
simulations (grey shaded area) in Figure~\ref{fig:pop_fextra}. While
the predictions broadly agree with the observations, a firm conclusion
is not possible with this limited number of galaxies.

\subsection{Scaling relations of integrated stellar population estimates}
\label{sec:central}
After the presentation of the stellar population maps
(Section~\ref{sec:maps}) and the average gradients
(Section~\ref{sec:gradients}) we now discuss more global
SSP-equivalent stellar population parameters. Our integrated
measurements were derived by averaging the (luminosity-weighted)
spectra within a circular aperture with a radius of one $R_{\rmn{e}}$
and re-measuring the line strength indices. Small aperture
corrections, as described in Paper~VI, had to be applied for 28
galaxies. From the resulting line strength values
(Table~\ref{tab:LSnew} of the Appendix), we derived the stellar
population estimates as described in Section~\ref{sec:chi2} and
presented in Table~\ref{tab:stellpop}. We emphasize that these
'global' estimates of SSP-equivalent stellar population parameters
derived from data covering essentially one effective radius (half the
light) for nearby galaxies are much more representative of the
galaxies as a whole compared to the typically central (often $R_{\rmn
  e}$/8) values used in the past \cite[e.g.,][]{tra00b,kun00,tho05}.

\begin{figure*} 
 \includegraphics[width=134mm]{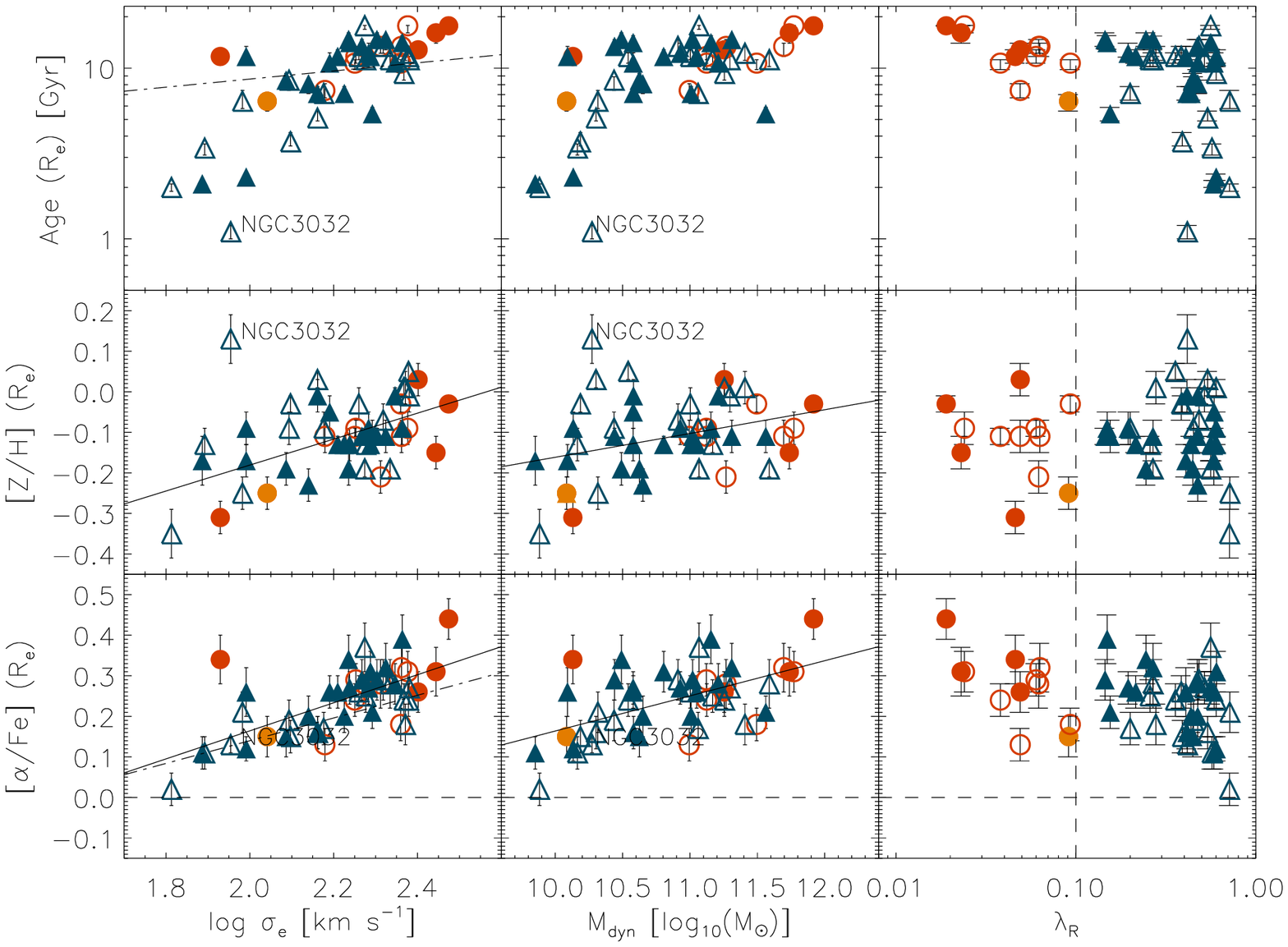}
 \caption{One effective radius, integrated measurements of
   SSP-equivalent age, [Z/H] and [$\alpha$/Fe] versus $\log
   \sigma_{\rm e}$ dynamical mass \mvir\/ and $\log \lambda_R$
   relations. Fast and slow rotators are indicated by blue triangles
   and red circles, respectively. Cluster and field galaxies are shown
   as filled and open symbols. NGC\,4550, classified as slow rotator
   due to two counter-rotating disks, is shown as orange filled
   symbol. The dot-dashed lines in the Age and [$\alpha$/Fe] {\em
     versus}\/ $\log \sigma_{\rm e}$ relations are taken from
   \citet{tho05}, whereas solid lines are linear fits to our data.}
 \label{fig:age_siglam}
\end{figure*}

In Figure~\ref{fig:age_siglam} we show the SSP equivalent age,
metallicity and [$\alpha$/Fe] {\em versus}\/ $\log \sigma_{\rm e}$,
the dynamical mass \mvir\/ and $\log \lambda_R$ relations for the one
$R_{\rmn{e}}$ aperture.  In agreement with \citet{tho05} we find a
correlation of age with velocity dispersion $\sigma_{\rm e}$. This
correlation is driven by the very young, low-mass galaxies
($\sigma_{\rm e} < 100$\,\kms) showing widespread signs of young
stellar populations and the substantial fraction of fast rotators in
the intermediate mass range which are affected to various degrees by
the contribution from younger stars linked to disk-like
kinematics. Since the current sample is not complete, particularly at
the low mass end, it is difficult to assess how many low-mass, old
systems exist. Perhaps more importantly, the age versus dynamical mass
plot shows that a strong contribution from young stellar populations
is confined to the low mass end ($\lesssim 2 \times 10^{10} M_{\odot}$). With
increasing mass the relative fraction of 'young' galaxies decreases
because any recent star-formation contributes relatively less to the
total mass of the system. The slow rotators appear to be the oldest
galaxies in the sample with a weak trend of decreasing $\lambda_R$ and
increasing SSP-equivalent age.  Confirmation of this trend awaits the
analysis of a larger more complete sample of slow rotators. However,
there is also a substantial number of {\em fast}\/ rotators for which
the last star-formation event was either long ago or very minor in
mass fraction so that they show old SSP-equivalent ages.

In agreement with previous investigations
\citep[e.g.,][]{kun02,tho05,ber06,col06,tho2010} there is mild
evidence for early-type galaxies residing in low density regions to
have slightly younger ages on average compared to the cluster
environment. For the \sauron\/ sample, we find mean ages of
$10.8\pm0.8$ and $9.7\pm0.9$\,Gyr, for cluster and field galaxies,
respectively (the errors are given as errors on the mean).

In summary, massive slow rotators have typically old SSP-equivalent
ages while low-mass fast rotators make up the most prominent examples
of young SSP-equivalent ages. In the intermediate mass range, galaxies
show a range of SSP-equivalent ages of greater than 5\,Gyr. The
contributions from younger stellar populations are linked to secondary
star-formation in disk- or ring-like structures (see
Section~\ref{sec:agemaps}).

Despite the various, distinct components and structures we find in the
stellar population maps and discuss in Section~\ref{sec:maps}, the
global, integrated metallicity estimates over one effective radius
show a significant, positive correlation with $\log \sigma_{\rm e}$. A
linear fit, excluding NGC\,3032, gives (see
Figure~\ref{fig:age_siglam}):
$$ [\rmn{Z/H}] (R_{\rmn{e}}) = 0.32(\pm0.07)  \log \sigma_{\rm e} - 0.82(\pm0.15)\,. $$ 

\noindent The metallicity -- $\log \sigma_{\rm e}$ relation is in
agreement with previous investigations
\citep[e.g.,][]{kun00,tho05}. On average we find smaller metallicities
compared to the study of e.g., \citet{tho05}, which is caused by the
larger aperture we use in our analysis (one $R_{\rmn e}$ {\em
  versus}\/ $R_{\rmn{e}}/10$). The most extreme case of a post
star-burst galaxy in our sample, NGC\,3032, deviates from the
relation. However, we ascribe this to the extreme luminosity weighting
of young and old stars in this galaxy and therefore consider the
SSP-equivalent estimate of the metallicity as not meaningful. There is
no evidence for an environmental dependence of the
metallicity-$\sigma_{\rm e}$ relation in our sample.

The correlation of our global (one $R_{\rmn e}$) metallicity estimates with
dynamical mass \mvir\/ is significant at the 3$\sigma$ level but shows a
significant scatter in metallicity at a given mass (rms = 0.08 dex):
$$ [\rmn{Z/H}] = 0.06(\pm0.02)  \log M_\rmn{dyn} - 0.75(\pm0.25)\,. $$

\noindent \citet{tra00b} showed in a study of 50 local elliptical
galaxies that in the four-dimensional space of metallicity,
$\log$~age, $\log \sigma$ and [$\alpha$/Fe] only a two-dimensional
plane is occupied.  Furthermore, velocity dispersion seems to be the
only structural parameter that determines the stellar population of a
galaxy. One of the projections from this plane is the remarkably tight
correlation between [$\alpha$/Fe] and $\log \sigma$. We show our
[$\alpha$/Fe] estimates as function of central velocity dispersion in
the bottom left panel of Figure~\ref{fig:age_siglam}. In good
agreement with the literature, we find a rather tight relation in the
sense that more massive galaxies exhibit larger $\alpha$/Fe ratios
\citep[see also][]{jor99,kun01,proc02,tho05,tho2010}. We do not detect
a significant difference between low and high density
environments. The best fitting relation for the full sample is:
$$ [\alpha/\rmn{Fe}] = 0.35(\pm0.06) \log \sigma_{\rm e} - 0.53(\pm0.13)\,. $$ 

\noindent The best fitting relation of [$\alpha$/Fe] with dynamical mass
is:
$$ [\alpha/\rmn{Fe}] = 0.09(\pm0.02)  \log M_\rmn{dyn} - 0.70(\pm0.20)\,. $$

\noindent What drives the [$\alpha$/Fe] -- $\log \sigma_{\rm e}$ (or
mass) relation?  Taking our results from Section~\ref{sec:maps} into
account we find that the fast rotator galaxies with intermediate
velocity dispersions ($\sigma_{\rm e} = 100 - 160$\,\kms) are the ones
with contributions of stellar populations formed in a component with
disk-like kinematics and more solar [$\alpha$/Fe] values. The
luminosity-weighted averages of the disk-like component with the main
body of the galaxy results in intermediate values of
[$\alpha$/Fe]. The group of galaxies with SSP-equivalent ages of $<
4$\,Gyr have the strongest relative contribution of a young disk and
SSP-equivalent estimates of [$\alpha$/Fe] $\simeq 0.1$ and thus
determine the low-mass point of the relation. We conclude that in our
sample there is evidence that the young stars with more solar-like
[$\alpha$/Fe] ratios, created in fast-rotating disk-like components in
low and intermediate mass galaxies, reduce the global [$\alpha$/Fe]
and thus significantly contribute to the apparent [$\alpha$/Fe] --
$\log \sigma_{\rm e}$ relation. However, we also find at least one low
mass galaxy (NGC\,4458; $\sigma_{\rm e} < 100$\,\kms), which has a
[$\alpha$/Fe] value comparable with the most massive systems in our
sample.

For the slow rotators, there is evidence for a trend of increasing
[$\alpha$/Fe] with decreasing $\lambda_R$.  Overall, however, the
metallicity and [$\alpha$/Fe] estimates of the slow rotators appear
surprisingly inhomogeneous. Although the current sample of slow
rotators is limited, the significant spread in the detailed stellar
population properties may indicate that there could be more than one
formation path resulting in slow rotators.

\section{Concluding remarks}
\label{sec:conclusions}
We caution the reader at this point that the \sauron\/ sample is a
representative sample of limited size and not complete; therefore
results for the population of early-type galaxies as a whole await
confirmation with larger samples.

Nevertheless, the SSP-equivalent stellar population maps presented in
this paper show that the 48 early-type galaxies in the \sauron\/
sample display a significant and varied structure in their stellar
populations. Up to 50 per cent of the sample galaxies show signs of a
contribution from a young stellar population. The young stars,
dominating the light, are very apparent in the low mass ($\lesssim 3
\times 10^{10} M_{\odot}$) post-starburst systems, showing signs of
residual star-formation in the NUV and mid-IR and even optical gas
emission lines consistent with star-formation in the most extreme case
(NGC\,3032). For these systems we observe SSP-equivalent ages of
$\le$3\,Gyr over the full field-of-view, typically covering one
effective radius $R_\rmn{e}$.  Spatially well localized cases of young
stars in circumnuclear disks and rings are found in eight intermediate
mass systems. The latter star-formation can be linked to thin, dusty
disks/rings also seen by star-formation signatures in the NUV or
mid-IR and optical gas emission lines consistent with star-formation
for two galaxies (NGC\,4459, NGC\,4526).
  
About 30 per cent of the sample galaxies show intermediate
SSP-equivalent ages (3-9\,Gyr) where ongoing star-formation is rarely
detected by NUV or mid-IR signatures, suggesting that the
star-formation process was concluded more than $\sim$1\,Gyr
ago. Common to most of these galaxies is the ubiquitous presence of
ordered stellar motion, classifying them as fast-rotators. We show
that the flattened components with disk-like kinematics identified in
all fast rotators (Paper~XII) are connected to regions of distinct
stellar populations.  This is clearly seen in the examples of the very
young, still star-forming circumnuclear disks and rings with increased
metallicity preferentially found in intermediate-mass fast rotators
(e.g., NGC\,4526).  However, signs of increased metallicity and mildly
depressed [$\alpha$/Fe] ratios compared with the main body of the
galaxy are seen in many fast rotators extending to apparently old
galaxies with extended disk-like kinematics (e.g.,
NGC\,4660). Additionally, intermediate mass fast-rotators show often
evidence for nuclear stellar disks with increased metallicity.  A
galaxy formation picture emerges where essentially all fast rotators
have experienced, often multiple, periods of secondary star-formation
in kinematically disk-like components on top of an older, spheroidal
component.
  
In contrast, the slow rotators, often harboring kinematically
decoupled components in their central regions, generally show no
stellar population signatures over and above the well known
metallicity gradients in early-type galaxies and are largely
consistent with old ($\ge$10\,Gyr) stellar populations.
  
Using our radially averaged stellar population gradients we find that
low mass fast rotators form a sequence of increasing metallicity
gradient with increasing mass, consistent with the observations of
\citet{spo09}. For more massive systems (above $\sim$3.5 $\times
10^{10} M_{\odot}$) there is an overall downturn such that metallicity
gradients become shallower with increased scatter at a given mass
leading to the most massive systems being slow-rotators with
relatively shallow metallicity gradients. The shallower metallicity
gradients and increased scatter could be a consequence of the
competition between different star-formation and assembly scenarios
following a general trend of diminishing gas fractions and more equal
mass mergers with increasing mass, leading to the most massive systems
being devoid of ordered motion and any signs of recent star-formation.
  
We also use our observations to compute more global stellar population
parameters integrated over circular apertures of one-eighth and one
effective radius, and compare the resulting relations of stellar
population parameters versus velocity dispersion with previous
results. In agreement with previous investigations we find significant
metallicity -- $\sigma$ and [$\alpha$/Fe] -- $\sigma$ relations. The
metallicity - dynamical mass relation is shallow and shows a large
scatter at a given mass possibly reflecting the many different
formation and assembly histories which can lead to an early-type
galaxy of a given mass.  We argue that the [$\alpha$/Fe] -- $\sigma$
relation is largely driven by an increasing relative contribution from
metal rich and mildly [$\alpha$/Fe] depressed disk-like components in
galaxies with decreasing mass.

\renewcommand{\thefigure}{\arabic{figure}}

\section*{Acknowledgments}
We would like to thank Ricardo Schiavon for providing the custom built
stellar population models with adjusted [Ti/Fe] ratios and Claudia
Maraston and Daniel Thomas for interesting discussions on the use of
stellar population models. We are very grateful to the referee for
comments that improved the quality of this manuscript. HK is grateful
to Michael J. Williams for advice on english language.

The \sauron\/ project is made possible through grants 614.13.003,
781.74.203, 614.000.301 and 614.031.015 from NWO and financial
contributions from the Institut National des Sciences de l'Univers,
the Universit\'e Lyon I, the Universities of Durham, Leiden and
Oxford, the Programme National Galaxies, the British Council, PPARC
grant 'Observational Astrophysics at Oxford 2002 - 2006' and support
from Christ Church Oxford, and the Netherlands Research School for
Astronomy NOVA. NS is grateful for the support of an STFC
studentship. MC acknowledges support from a STFC Advanced Fellowship
(PP/D005574/1).  RLD is grateful for the award of a PPARC Senior
Fellowship (PPA/Y/S/1999/00854), postdoctoral support through PPARC
grant PPA/G/S/2000/00729 and from the Royal Society through a Wolfson
Merit Award. The PPARC Visitors grant (PPA/V/S/2002/00553) to Oxford
also supported this paper. GvdV acknowledges support provided by NASA
through Hubble Fellowship grant HST-HF-01202.01-A awarded by the Space
Telescope Science Institute, which is operated by the Association of
Universities for Research in Astronomy, Inc., for NASA, under contract
NAS 5-26555. This paper is based on observations obtained at the
William Herschel Telescope, operated by the Isaac Newton Group in the
Spanish Observatorio del Roque de los Muchachos of the Instituto de
Astrofísica de Canarias. It is also based on observations obtained at
the 1.3m Mcgraw-Hill Telescope at the MDM observatory on Kitt Peak,
which is owned and operated by the University of Michigan, Dartmouth
College, the Ohio State University, Columbia University and Ohio
University. This project made use of the HyperLeda and NED data bases.
Part of this work is based on HST data obtained from the ESO/ST-ECF
Science Archive Facility.

%
%
%

\bibliographystyle{mn2e}
\bibliography{references}
%
%
%


\appendix

\section{Revised line strength measurements for circular apertures of $R_{\rm e}/8$ and $R_{\rm e}$}
\label{sec:appendix1}

We present revised line strength measurements on the Lick/IDS system
over circular apertures of $R_{\rm e}/8$ and $R_{\rm e}$ in
Table~\ref{tab:LSnew}.  This table supersedes the data presented in
Table~5 of Paper~VI. The details of the revised data-reduction are
given in Section~\ref{sec:revised}. We further note that the effective
radius values were revised using a $R^{1/n}$ growth curve analysis of
our wide-field MDM (1.3m) imaging survey of the early-type galaxy
sample (Falc{\'o}n-Barroso in preparation).

\begin{table*}
\begin{center}
  \begin{minipage}{175mm}
 \caption{List of revised line strength measurements within a circular aperture of $R_{\rm e}/8$ and $R_{\rm e}$.}
 \label{tab:LSnew}
 \begin{center}
 \begin{tabular}{lrrccccrrcc}
  \hline
Name & $R_{\rmn{e}}$ &  $\sigma$  & \hb  &Fe5015 & \mgb & \fes  &  $\sigma_{\rm e}$  & \hb  & Fe5015  & \mgb \\
     & (\arcsec)    &  (\kms) & (\AA) & (\AA)  & (\AA) & (\AA) &  (\kms) & (\AA) & (\AA)& (\AA)\\
Aperture     &              & \reb  & \reb &  \reb & \reb  & \reb &  $R_{\rmn{e}}$ &  $R_{\rmn{e}}$ & $R_{\rmn{e}}$ & $R_{\rmn{e}}$\\
 (1) & (2)          &  (3)  & (4) & (5) & (6) & (7) & (8) & (9) & (10) & (11)\\
\hline
 %
 %
   NGC\,474 & 28.0 & 166 &  1.66 &  5.33 &  4.08 &  --   & 142 &  1.82 &  4.79 &  3.44 \\
  NGC\,524 & 35.4 & 251 &  1.34 &  5.43 &  4.50 &  --   & 225 &  1.48 &  5.35 &  4.22 \\
  NGC\,821 & 31.3 & 200 &  1.52 &  5.31 &  4.38 &  --   & 182 &  1.59 &  4.67 &  3.75 \\
 NGC\,1023 & 48.7 & 204 &  1.49 &  5.46 &  4.45 &  2.33 & 165 &  1.51 &  4.96 &  4.12 \\
 NGC\,2549 & 13.9 & 148 &  2.04 &  5.93 &  4.12 &  2.44 & 140 &  1.96 &  5.15 &  3.59 \\
 NGC\,2685 & 23.6 &  90 &  1.99 &  4.94 &  3.52 &  2.17 & 100 &  1.91 &  4.20 &  3.00 \\
 NGC\,2695 & 18.7 & 222 &  1.31 &  5.03 &  4.58 &  --   & 184 &  1.26 &  4.33 &  3.95 \\
 NGC\,2699 & 12.1 & 150 &  1.79 &  5.79 &  4.29 &  2.42 & 123 &  1.75 &  4.81 &  3.60 \\
 NGC\,2768 & 68.0 & 205 &  1.68 &  4.96 &  4.06 &  --   & 200 &  1.60 &  4.43 &  3.62 \\
 NGC\,2974 & 28.3 & 237 &  1.64 &  5.16 &  4.42 &  2.27 & 227 &  1.66 &  5.02 &  4.05 \\
 NGC\,3032 & 19.3 &  94 &  4.77 &  4.47 &  1.71 &  1.61 &  90 &  3.86 &  4.08 &  2.20 \\
 NGC\,3156 & 14.8 &  66 &  4.44 &  3.88 &  1.54 &  1.58 &  66 &  2.98 &  3.78 &  1.84 \\
 NGC\,3377 & 38.3 & 144 &  1.83 &  4.88 &  3.81 &  2.08 & 126 &  1.80 &  4.37 &  3.18 \\
 NGC\,3379 & 44.9 & 216 &  1.39 &  5.12 &  4.51 &  2.19 & 190 &  1.43 &  4.74 &  4.10 \\
 NGC\,3384 & 28.5 & 161 &  1.85 &  5.81 &  4.14 &  2.46 & 141 &  1.82 &  5.07 &  3.68 \\
 NGC\,3414 & 32.0 & 232 &  1.42 &  4.99 &  4.51 &  2.17 & 191 &  1.49 &  4.46 &  3.70 \\
 NGC\,3489 & 21.5 & 105 &  2.81 &  4.93 &  2.82 &  2.01 &  99 &  2.60 &  4.37 &  2.58 \\
 NGC\,3608 & 33.6 & 191 &  1.51 &  5.25 &  4.39 &  --   & 167 &  1.61 &  4.74 &  3.77 \\
 NGC\,4150 & 15.9 &  83 &  3.53 &  4.22 &  2.16 &  1.69 &  77 &  2.82 &  4.13 &  2.35 \\
 NGC\,4262 & 10.6 & 196 &  1.41 &  4.94 &  4.45 &  2.06 & 164 &  1.45 &  4.41 &  3.85 \\
 NGC\,4270 & 13.7 & 138 &  1.75 &  5.15 &  3.49 &  --   & 125 &  1.75 &  4.60 &  3.26 \\
 NGC\,4278 & 30.6 & 252 &  1.28 &  4.69 &  4.70 &  2.11 & 217 &  1.44 &  4.51 &  4.17 \\
 NGC\,4374 & 70.2 & 292 &  1.38 &  5.08 &  4.52 &  2.24 & 261 &  1.40 &  4.61 &  4.07 \\
 NGC\,4382 & 94.4 & 187 &  2.11 &  5.06 &  3.45 &  2.11 & 178 &  1.99 &  4.60 &  3.27 \\
 NGC\,4387 & 11.0 & 100 &  1.62 &  5.07 &  3.95 &  2.22 &  98 &  1.54 &  4.54 &  3.69 \\
 NGC\,4458 & 19.9 & 106 &  1.61 &  4.60 &  3.82 &  1.92 &  83 &  1.61 &  4.05 &  3.33 \\
 NGC\,4459 & 41.0 & 178 &  2.14 &  5.25 &  3.85 &  2.26 & 155 &  1.84 &  4.60 &  3.37 \\
 NGC\,4473 & 26.8 & 192 &  1.49 &  5.37 &  4.51 &  2.31 & 186 &  1.50 &  4.78 &  3.98 \\
 NGC\,4477 & 46.5 & 170 &  1.62 &  5.06 &  4.14 &  2.22 & 147 &  1.55 &  4.67 &  3.81 \\
 NGC\,4486 &106.2 & 311 &  1.15 &  5.15 &  5.11 &  2.25 & 268 &  1.15 &  4.59 &  4.59 \\
 NGC\,4526 & 35.7 & 232 &  1.84 &  5.51 &  4.31 &  2.32 & 214 &  1.58 &  4.91 &  4.16 \\
 NGC\,4546 & 22.0 & 220 &  1.54 &  5.34 &  4.60 &  2.28 & 189 &  1.54 &  4.56 &  3.87 \\
 NGC\,4550 & 11.6 &  88 &  2.11 &  4.54 &  3.17 &  2.00 & 103 &  1.92 &  4.35 &  2.96 \\
 NGC\,4552 & 33.9 & 268 &  1.55 &  5.65 &  4.83 &  2.29 & 233 &  1.45 &  5.14 &  4.40 \\
 NGC\,4564 & 19.3 & 173 &  1.52 &  5.87 &  4.77 &  2.42 & 150 &  1.57 &  4.84 &  4.00 \\
 NGC\,4570 & 12.8 & 197 &  1.45 &  5.80 &  4.67 &  2.33 & 167 &  1.45 &  4.75 &  3.97 \\
 NGC\,4621 & 46.0 & 225 &  1.40 &  5.29 &  4.68 &  2.27 & 200 &  1.44 &  4.67 &  4.11 \\
 NGC\,4660 & 11.5 & 221 &  1.43 &  5.56 &  4.76 &  2.30 & 181 &  1.47 &  4.71 &  4.02 \\
 NGC\,5198 & 18.0 & 206 &  1.46 &  5.44 &  4.73 &  --   & 173 &  1.55 &  4.72 &  3.97 \\
 NGC\,5308 &  9.9 & 244 &  1.46 &  5.28 &  4.46 &  --   & 201 &  1.46 &  4.83 &  4.16 \\
 NGC\,5813 & 55.9 & 226 &  1.49 &  5.00 &  4.47 &  2.12 & 210 &  1.49 &  4.63 &  4.03 \\
 NGC\,5831 & 29.2 & 168 &  1.72 &  5.55 &  4.13 &  2.38 & 148 &  1.81 &  4.83 &  3.34 \\
 NGC\,5838 & 20.6 & 283 &  1.59 &  5.68 &  4.60 &  2.38 & 232 &  1.57 &  4.96 &  4.17 \\
 NGC\,5845 &  4.3 & 269 &  1.51 &  5.62 &  4.64 &  2.30 & 237 &  1.51 &  5.23 &  4.36 \\
 NGC\,5846 & 76.8 & 236 &  1.33 &  5.37 &  4.80 &  --   & 213 &  1.30 &  4.83 &  4.34 \\
 NGC\,5982 & 24.9 & 260 &  1.62 &  5.74 &  4.41 &  --   & 223 &  1.59 &  5.11 &  3.95 \\
 NGC\,7332 &  9.2 & 137 &  2.27 &  5.87 &  3.67 &  2.28 & 125 &  2.14 &  4.85 &  3.19 \\
 NGC\,7457 & 33.2 &  70 &  2.37 &  4.98 &  2.92 &  2.09 &  75 &  2.25 &  4.56 &  2.80 \\
\hline
 \end{tabular}

\end{center}

{\em Notes:}\/ This table supersedes the values published in Table~5
of Paper~VI. (1) NGC\, number. (2) Effective (half-light) radius
$R_{\rmn e}$ measured with a $R^{1/n}$ growth curve analysis from our
wide-field MDM (1.3m) imaging survey of the early-type galaxy sample
(Falc{\'o}n-Barroso in preparation). (3) Velocity dispersion of the
luminosity weighted spectrum within a circular aperture of \reb. Only
the first two moments, $v$ and $\sigma$ are used in the fit to the
spectrum.  (4)~--~(7) Fully corrected line strength index measurements
on the Lick/IDS system of the luminosity weighted spectrum within
\reb\/ for the \hb, Fe5015, \mgb\/ and \fes\/ indices. Due to limited
field coverage we cannot determine the \fes\/ indices for 11
galaxies. (8) Velocity dispersion of the luminosity weighted spectrum
within a circular aperture of one $R_{\rmn e}$ . Only the first two
moments, $v$ and $\sigma$ are used in the fit to the spectrum.
(9)~-~(11) Fully corrected line strength index measurements on the
Lick/IDS system of the luminosity weighted spectrum within
$R_{\rmn{e}}$ for the \hb, Fe5015, and \mgb\/ indices. For galaxies
with less than one $R_{\rmn{e}}$ coverage we applied the aperture
corrections given in Paper~VI. Formal errors of the line strength
indices are below 0.1\,\AA. We note, however, that there are
systematic errors which we estimate to be of the order of 0.06, 0.15,
0.08, 0.06\,\AA\/ for the \hb, Fe5015, \mgb, and \fes\/ indices,
respectively. For the velocity dispersion we adopt an error of 5 per
cent.

\end{minipage}

\end{center}
\end{table*}

\label{lastpage}
\end{document}